\documentclass{article}

\usepackage{arxiv}
\usepackage{amsmath}
\usepackage{fullpage}
\usepackage{natbib}
\usepackage{graphicx}
\usepackage[toc,page]{appendix}
\usepackage{float}
\usepackage{xcolor}
\usepackage{authblk}
\usepackage{setspace}
\usepackage{amsfonts}
\usepackage{caption}
\usepackage{url}

\begin{document}

\title{EnsAI: An Emulator for Atmospheric Ammonia Ensembles}

\author{
 Michael Sitwell \\
 Air Quality Research Division \\
 Environment and Climate Change Canada \\
 Toronto, Ontario, Canada \\
 \texttt{michael.sitwell@ec.gc.ca} \\
}

\maketitle

\begin{abstract}
Ensemble-based methods for data assimilation and emissions inversions are a popular way to encode flow-dependency within the model error covariance. While most ensemble methods do not require the use of an adjoint model, the need to repeatedly run a geophysical model to generate the ensemble can be a significant computational burden. In this paper, we introduce \textit{EnsAI}, a new AI-based ensemble generation system for surface ammonia concentrations. When trained on an ensemble for ammonia generated by the GEM-MACH air quality model, it was shown that the ensembles produced by EnsAI can accurately reproduce the meteorology-dependent features of the original ensemble, while generating the ensemble 3,300 times faster than the original GEM-MACH ensemble. While EnsAI requires an upfront cost for generating the ensemble used for training, as well as the training itself, the long term computational savings can greatly exceed these initial computational costs. When used in an emissions inversion system, EnsAI produced similar inversion results to those in which the original GEM-MACH ensemble was used while using significantly less computational resources.
\end{abstract}

\section{Introduction}

Ensemble forecasting of atmospheric models has become an invaluable tool in atmospheric modeling over the past few decades. An ensemble of atmospheric states has the ability to represent the uncertainty associated with an atmospheric model and can encode information about both the level of uncertainty of particular state variables as well as the correlation between the uncertainties of different variables. Quantifying the uncertainty of atmospheric forecasts is not only useful when interpreting the results of the forecast, but is also essential for data assimilation and other inversion methods that blend the forecast with observational information.

The goal of an assimilation process is to produce what is known as an \textit{analysis}, which is a combination of forecast (or \textit{background}) and observational information in a way that is optimal (or close to optimal) according to some metric. The uncertainties ascribed to both the model and observations help determine this optimal combination. The error covariances associated with the model and observations set both the level of impact that the observations have on the analysis as well as the spread of the observational information via the error correlations. The error correlations not only spread observational information spatially, but can also impose balance between different variables in multivariate applications \citep{bannister2008}. Furthermore, as the analysis increment lies in the subspace
spanned by the background error covariance matrix, using a background error covariance with sufficient rank is essential.

While having an accurate estimate of a model's error covariance is of high importance in many applications, actually obtaining such an estimate is often difficult. Earlier numerical weather prediction (NWP) assimilation systems often used static forecast error covariances as this was more computationally feasible at the time. But as more computational resources became available, more weather centers were able to incorporate flow-dependent error covariances within their systems. One popular method for doing this was by using four-dimensional variational (4DVar) assimilation \citep{thepaut1991,zupanski1993,rabier2000}. However, 4DVar's need for a model adjoint, which can be taxing to develop and maintain, has increased the popularity of adjoint-free methods, in particular algorithms that use ensembles to encode flow-dependent information. The most common ensemble-based algorithms currently employed at weather centers are the ensemble Kalman filter (EnKF) \citep{evensen1994,lorenc2003,houtekamer2016} and ensemble–variational assimilation (EnVar) \citep{buehner2010,desroziers2014,fairbairn2014}. 

While ensemble-based assimilation methods offer many benefits over other methods, a significant downside of using these methods is the computational cost of creating these ensembles, which often requires running an atmospheric model tens to hundreds of times for each time step. The computational cost often limits the number of ensemble members, which can result in spurious correlations in the covariance.

The assimilation of atmospheric chemical observations has been an important component of air quality modeling \citep{miyazaki2012a,robichaud2017,huijnen2020,park2022,garrigues2022}, stratospheric ozone and UV forecasting \citep{lahoz2007,inness2013,rochon2019}, climate modeling \citep{peters2010,miyazaki2011,bruhwiler2014,agusti2023}, and emissions inversions \citep{miyazaki2012a,miyazaki2012b,zhu2013,cao2020}. The computational challenges associated with ensemble generation are compounded when applied to atmospheric chemistry models, which typically require more computational resources than NWP models. While many operational NWP ensemble systems are run at weather centers with ensemble sizes on the order of tens to hundreds, the higher computational requirements of running atmospheric chemistry models make running a similarly designed chemical data assimilation system difficult, especially in an operational setting. When chemical ensembles are feasible to run, it typically has a much smaller ensemble size than in NWP systems, often with less than a hundred members \citep{pendergrass2023}. In contrast to most modern NWP systems, it is not uncommon for chemical data assimilation systems to use static background error covariances. For example, at Environment and Climate Change Canada (ECCC), the two operational systems that produce analyses for chemical fields, the air quality surface analysis \citep{menard2018a,menard2018b,menard2018c} and the stratospheric ozone forecasting system \citep{rochon2019,buehner2024}, both use static background error covariances. ECCC's air quality forecasting system uses the Global Environmental Multiscale – Modelling Air quality and Chemistry (GEM-MACH) chemical weather model, which takes approximately four to five times more computing time and up to nine times more disk space than the Global Environmental Multiscale (GEM) NWP model that GEM-MACH is built upon, illustrating the challenges with using ensemble methods with atmospheric chemical models.

The recent explosion in machine learning–based weather prediction (MLWP) \citep{pathak2022,bi2023,chen2023a,lam2023,chen2025} offers the potential of atmospheric forecasting that requires a fraction of the run time as compared to traditional physically-based models. The drastic reduction in computational run time is due to both the artificial intelligence (AI) algorithms used as well as running on specialized hardware such as graphics processing units (GPUs). Soon after the initial release of MLWP models, new MLWP systems were developed to produce ensembles of forecasts \citep{price2023,li2024a}, combine data assimilation methods into MLWP models \citep{xiao2023,li2024b}, and end-to-end forecasting/assimilation systems \citep{chen2023b,vaughan2024}. The ability of these MLWP models to meet or exceed the performance (at least with respect to some metrics) of traditional physically-based models that run on central processing unit (CPU) computing clusters, while requiring far less computation time, is opening up a new frontier in the atmospheric sciences.

Currently, most MLWP systems are primarily trained using the ERA5 data set \citep{hersbach2020}. For atmospheric chemistry applications, additional data sets are needed to provide information on chemical constituents during the training of the AI model. The closest analogue at present to the ERA5 reanalysis for atmospheric chemistry is the Copernicus Atmosphere Monitoring Service (CAMS) \citep{inness2019} reanalysis, which only dates back to 2003. Recently, \citet{bodnar2025} developed an AI geophysical system that is pretrained on NWP data and fine-tuned for atmospheric chemistry using the CAMS reanalysis as well as the CAMS analysis for more recent years. However, at present, the atmospheric chemistry data that could potentially be used for training is much more limited as compared to NWP, particularly for ensemble information for chemicals in the troposphere important for air quality forecasting.

\citet{sitwell2022, sitwell2026} detail the development of an ensemble-based ammonia emissions inversion system that was used to update the ammonia emissions provided to the GEM-MACH air quality model. While these studies have shown that this ensemble-based inversion system can significantly improve the parameters used for generating ammonia emissions, comprehensive assessments of the ammonia emissions can only be done occasionally when using the GEM-MACH generated ensemble due to the high computational cost of ensemble generation.

The aim of this work is to develop an AI-based model for efficient ensemble generation of atmospheric ammonia surface concentrations for the purpose of air quality forecasting, which we will refer to as \textit{EnsAI}. An existing ensemble of atmospheric ammonia concentrations that was used for ammonia emissions inversions will be leveraged for training. For our application, the ensemble of emissions perturbations is computationally inexpensive to create, but translating the ensemble of emissions perturbations into an ensemble of ammonia concentration perturbations using a physically-based atmospheric model is usually very computationally demanding. EnsAI uses a U-Net convolutional neural network to map ammonia emissions perturbations to ammonia surface concentration perturbations, using surface wind and temperature fields as auxiliary inputs to the neural network.

As EnsAI can generate ensembles of concentrations without needing to run computationally expensive physically-based atmospheric models, it can potentially serve as an integral component of an ammonia emissions inversion system that can be run frequently without an excessive computational burden. Although EnsAI has the upfront cost from training (as well as generating the ensemble training data if not already available), these initial costs will be much smaller than the long-term computational savings over the intended use period.

\section{Emissions Inversions}
\label{sec:emissions_inversions}

This study develops an AI-based ensemble generation system to be used in an ammonia emissions inversion system. As in \citet{sitwell2022, sitwell2026}, ammonia retrievals from the satellite-borne CrIS instrument are used to refine the ammonia emissions parameters, which are then used in the GEM-MACH air quality model. While the inversion methods are described in full in \citet{sitwell2022, sitwell2026}, we summarize the main features in this section, as well as a few modifications made for this study.

\subsection{The GEM-MACH Air Quality Model}
\label{sec:gemmach}

GEM-MACH \citep{moran2010,gong2015,pavlovic2016} is a chemical transport model used in ECCC’s operational air quality forecasting system that is embedded within ECCC’s operational GEM weather forecasting model \citep{cote1998a,cote1998b,girard2014}. GEM-MACH is a physically-based model of atmospheric gas-phase, aqueous-phase, and heterogeneous chemistry. Version 3.1.0 of GEM-MACH was used for this study, which has 85 vertical levels that extend from the surface to 0.1 hPa. GEM-MACH can be run with both global and limited area horizontal grids. For this work, a limited area horizontal grid with spacing of $0.09^{\circ}$ ($\sim$10 km) was used. The meteorology is refreshed at the beginning of each forecast using analyses from ECCC’s Global Deterministic Prediction System \citep{buehner2015}.

\begin{figure}
\begin{center}
\noindent\includegraphics{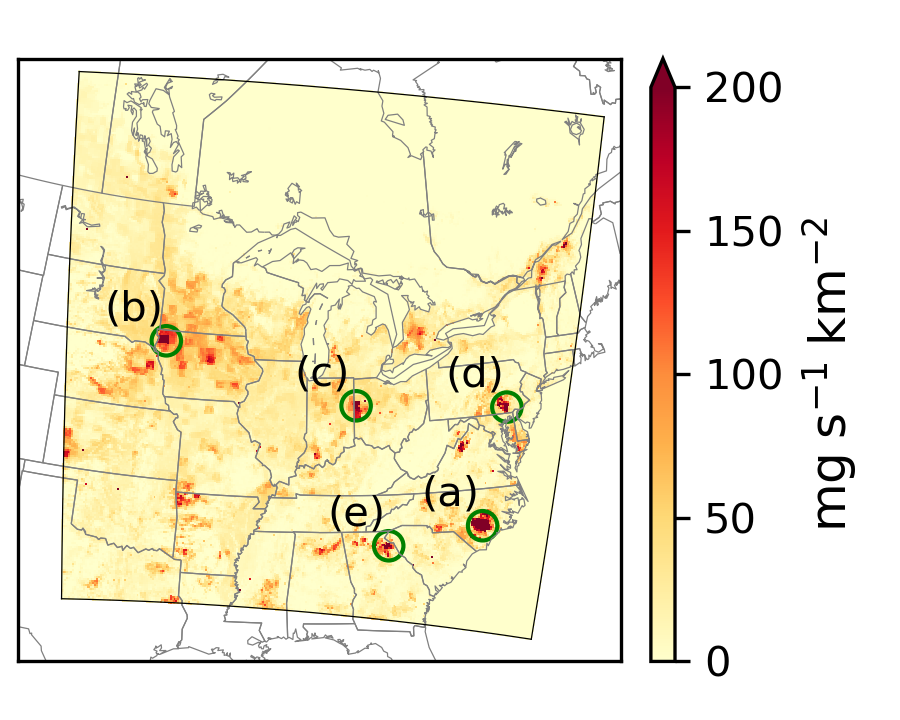}
\end{center}
\caption{Annual mean ammonia emissions from the (unperturbed) inventory. Emissions are only displayed within the domain of the EnsAI model. Particular locations examined in this work are also displayed, which are located (in order of descending emissions) in (a) North Carolina, (b) Iowa, (c) Ohio, (d) Pennsylvania, and (e) Georgia.}
\label{fig:mean_emissions}
\end{figure}

This version of GEM-MACH also includes a bidirectional flux model for ammonia \citep{sitwell2026}. The bidirectional flux model combines monthly-mean ammonia emission potentials with meteorological information to produce hourly meteorologically-dependent ammonia emissions. The emissions for all other species are computed by translating monthly or annual emissions inventories into hourly emissions through the use of sector-dependent time profiles. The unperturbed set of emissions was version 3.1.2 of the operational GEM-MACH emissions data set. The annual mean of this emissions inventory for ammonia is shown in Figure \ref{fig:mean_emissions}.

\subsection{CrIS Ammonia Retrievals}

The CrIS instrument, which is aboard the Suomi National Polar-orbiting Partnership (SNPP), NOAA-20 and NOAA-21 satellites, is a Fourier transform spectrometer that measures the infrared spectrum in three different spectral bands. The CrIS Fast Physical Retrieval (CFPR) algorithm \citep{shephard2015a, shephard2020}, based on the TES ammonia retrieval algorithm \citep{shephard2011}, used for the ammonia retrievals makes use of the longwave band between 960 and 967 $\mathrm{cm}^{-1}$. As we will focus on the years 2014 to 2016, we use retrievals from the CrIS instrument aboard the SNPP satellite. The SNPP satellite, which is in a sun-synchronous orbit, has local overpass times at approximately 01:30 and 13:30, though only the day-time retrieval is used in the inversions. CrIS has a resolution at nadir of around 14 km. The retrievals are made on 14 pressure levels, with the averaging kernel typically peaking at or near the surface.

Retrievals are from version 1.6.4 of the CFPR algorithm that includes non-detects when the signal is below the detection limit \citep{white2023}. Only retrievals over land with a signal-to-noise ratio of at least one and have at least 0.1 degrees of freedom are used in the inversions.

The goal of this work is to develop an efficient ensemble generation model that runs on GPUs, which typically have access to less memory than traditional CPU clusters. As such, to reduce memory requirements, only the retrieval at the surface will be used in the inversions, instead of all 14 levels that were used in the inversions in \citet{sitwell2022, sitwell2026}. This should be a reasonable simplification since the retrievals are mainly sensitive to the surface and lower atmospheric levels.

\subsection{Inversion Methods}
\label{sec:inversion_methods}

Inversion algorithms are used to create a statistical blend of observational and a priori (background) information by solving an inverse problem, the result of which is known as the analysis. Both the background model state and observations are weighted by their error covariances. The error variances determines the level of impact the observations have on the analysis, while the correlations can (among other things) spread the observational information to other variables in the model state. Most algorithms solve this problem by finding an additive state vector, known as the \textit{increment}, to add to the a priori state vector. 

For our current application, a set of emissions $\textbf{e}$ is used as input into GEM-MACH to generate atmospheric concentrations $\textbf{c}$, where only $\textbf{c}$ is observed. As mentioned in Section \ref{sec:gemmach}, because we will only be utilizing the lowest level of the model output, $\textbf{c}$ will denote only the surface concentrations. If our model state vector $\textbf{x}$ is represented in block form as
\begin{equation}
\textbf{x} = \begin{bmatrix}
\textbf{c} \\ \textbf{e}
\end{bmatrix} ,
\end{equation}
then the background error covariance $\textbf{B}$ can be expressed as
\begin{equation}
\textbf{B} = \begin{bmatrix}
\textbf{B}_{cc} & \textbf{B}_{ec}^{\rm{T}} \\
\textbf{B}_{ec} & \textbf{B}_{ee}
\end{bmatrix} ,
\label{eq:bmatrix}
\end{equation}
where $\textbf{B}_{cc}$ and $\textbf{B}_{ee}$ represent the univariate covariances for the surface concentrations and emissions, respectively, while $\textbf{B}_{ec}$ represents their cross-covariance. Although the uncertainties in the atmospheric concentration of ammonia are likely due to a number of different factors including uncertainties in emissions, meteorology, initial conditions, and chemistry, for these inversions only the uncertainty due to the emissions is considered.

Ensemble methods model the background error covariance, in full or in part, by use of an ensemble of model states, where each member of the ensemble has a different perturbed state. An ensemble $\{ \textbf{x}_1, \ldots, \textbf{x}_N \}$ of $N$ model states can be used to form a background error covariance $\textbf{B}^{\rm{ens}}$ as
\begin{equation}
\textbf{B}^{\rm{ens}} = \frac{1}{N-1} \mathcal{L} \circ \left( \sum_{i=1}^N \delta\textbf{x}_i \delta\textbf{x}_i^{\rm{T}} \right) ,
\label{eq:bens}
\end{equation}
where $\delta\textbf{x}_i$ is the model state perturbation of ensemble member $i$. Physically-based atmospheric models are usually computationally costly to run, so ensembles typically consist of tens to hundreds of members. These small ensemble sizes produce error covariances with low rank and spurious long-distance correlations. A localization scheme is usually included to mitigate these effects, which we implement through an element-wise multiplication (denoted by $\circ$) of the ensemble error covariance with a localization matrix $\mathcal{L}$.

\begin{figure}
\begin{center}
\noindent\includegraphics{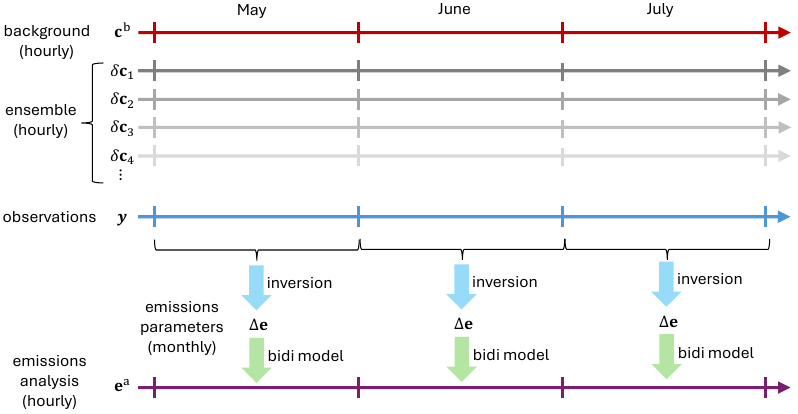}
\end{center}
\caption{Diagram of the ammonia emissions inversion process. The hourly background concentrations ($\textbf{c}^{\rm{b}}$) and ensemble concentration perturbations ($\delta\textbf{c}$) are used with observations ($\textbf{y}$) within a month to produce monthly increments to the emissions parameters $\Delta\textbf{e}$. GEM-MACH's bidirectional flux model (represented by the green arrows in the diagram) combines the monthly emissions parameters with meteorological fields to yield hourly updated emissions ($\textbf{e}^{\rm{a}}$).}
\label{fig:inversion_diagram}
\end{figure}

A diagrammatic representation of the emissions inversion process is shown in Figure \ref{fig:inversion_diagram}. As individual CrIS ammonia retrievals have relatively large uncertainties \citep{shephard2015a}, each inversion uses a month of retrievals so that a sufficient amount of observational information is used in each inversion. The hourly fields from an unperturbed GEM-MACH run are used as the a priori (background) model state. The ensemble used in the inversions is comprised of a set of hourly fields produced by perturbed GEM-MACH runs. The emissions inversions use a calendar month's worth of retrievals with the hourly background and ensemble model fields to produce updated emissions model parameters for a particular month. While the product of each inversion is a single field of the emissions parameters for a particular month, the posterior ammonia emissions vary within the inversion time window as the ammonia emissions are dependent on meteorological conditions, most notably the surface temperature. Further details of the inversion algorithm can be found in Section \ref{sec:inversion_details} of the Appendix.

\subsection{GEM-MACH Ensemble Generation}
\label{sec:gm_ensemble}

Ensemble generation is often the most computationally demanding part of an ensemble forecasting system. EnsAI was trained on a previously existing ammonia ensemble, which removed the need to generate new training data. This section briefly summarizes the processes that were used to generate this ensemble.

The ensemble generation process consists of two steps: First, for each ensemble member $i$, a Gaussian random field is drawn that is used to construct the time-mean emissions $\textbf{e}_i$. These emissions are then combined with the unperturbed model input, collectively denoted by $\boldsymbol\theta$, which includes the emissions from all chemical species other than ammonia as well as initial conditions for both meteorological and chemical concentration fields. The GEM-MACH model, here represented by the operator $\mathcal{M}^{\rm{GM}}$, is then run to produce the atmospheric concentrations $\textbf{c}_i$ for that ensemble member, which is represented symbolically as
\begin{equation}
\textbf{c}_i=\mathcal{M}^{\rm{GM}}(\textbf{e}_i,\boldsymbol\theta) .
\label{eq:gm_model}
\end{equation}

The initial step of generating the emissions ensemble is relatively computationally inexpensive. A random field is repeatedly drawn from an isotropic normal distribution and are used to perturb the ammonia emissions. The standard deviations of the emissions perturbation field was set to 50\% of their monthly mean values, with a minimum standard deviation value of 10 $\mathrm{mg}\ \mathrm{s}^{-1}\ \mathrm{km}^{-2}$ imposed so that all locations have a non-negligible ensemble variance. The horizontal spatial correlations of this distribution were chosen to have a half-width at half-maximum of 40 km. The second step of repeatedly running GEM-MACH with the perturbed emissions fields accounts for most of the computational cost of this inversion method and is the motivation for exploring AI-based ensemble generation methods that can potentially produce an ensemble at much lower computational costs.

\section{AI Ensemble Generation}

As repeatedly running GEM-MACH is very computationally demanding for ensembles of any reasonable size, we wish to find an AI-based model $\mathcal{M}^{\rm{EnsAI}}$ that can be used in Eq. (\ref{eq:gm_model}) in place of $\mathcal{M}^{\rm{GM}}$ that emulates the GEM-MACH model but is able to run much faster than GEM-MACH.

In the original ensemble generation procedure described in Section \ref{sec:gm_ensemble}, we note from Eq. (\ref{eq:gm_model}) that the full GEM-MACH model $\mathcal{M}^{\rm{GM}}$ is used to produce the \textit{perturbed} concentration fields $\textbf{c}_i$, which in our case are the ammonia surface concentrations. A full GEM-MACH emulator would aim to reproduce the full forecasts for multiple chemical species. However, as our goal is only to reproduce ensemble perturbations, $\mathcal{M}^{\rm{EnsAI}}$ will only predict the ensemble perturbations of the ammonia surface concentrations $\delta\textbf{c}_i$ given the perturbations to the ammonia emissions $\delta\textbf{e}_i$. Since $\mathcal{M}^{\rm{EnsAI}}$ only needs to simulate the ensemble perturbations of certain model fields (only surface ammonia in our case), $\mathcal{M}^{\rm{EnsAI}}$ can potentially be much simpler than an emulator that aims to reproduce the full GEM-MACH output.

\begin{figure}
\begin{center}
\noindent\includegraphics{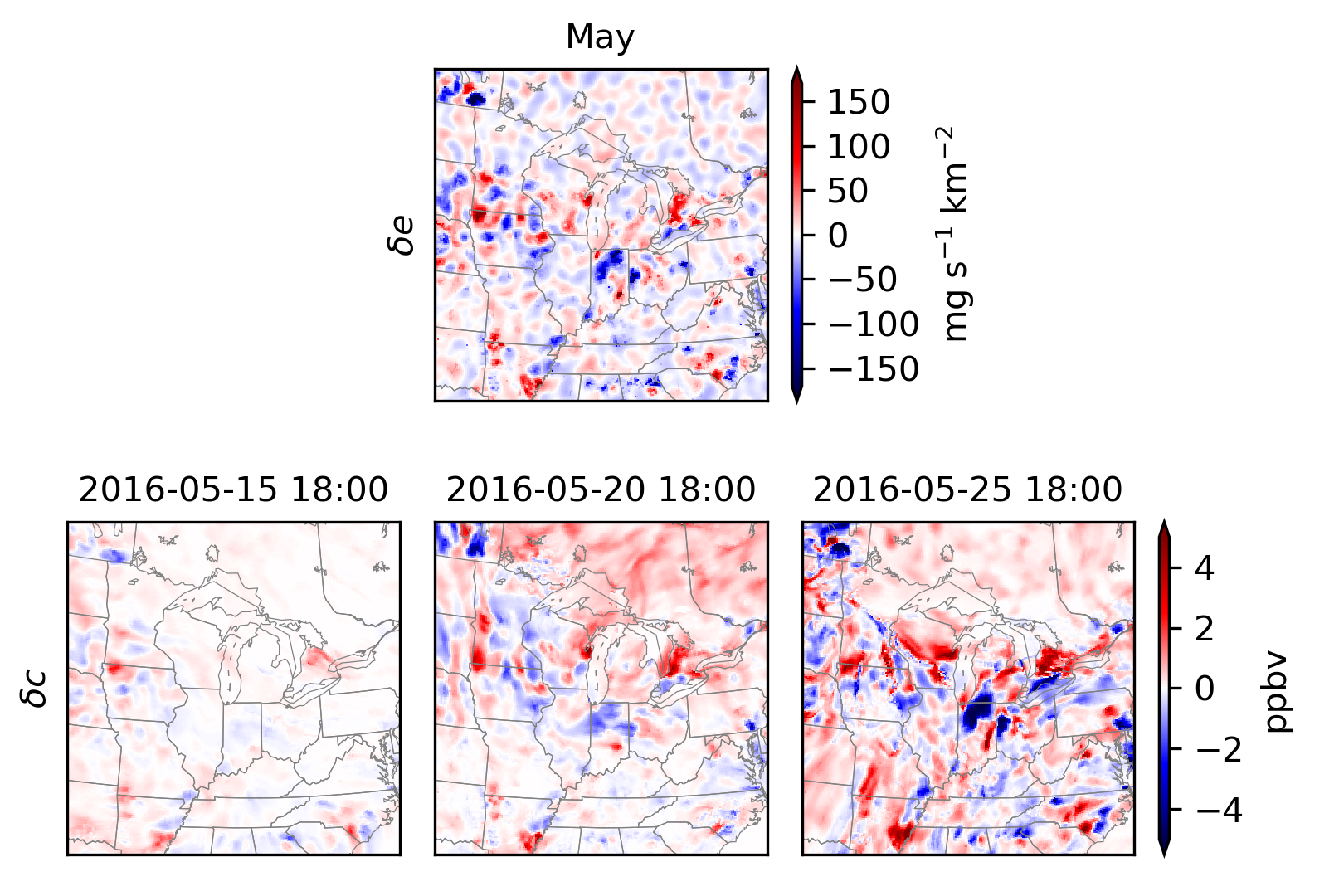}
\end{center}
\caption{Monthly-mean ammonia emissions perturbations for May (top row) and the resulting ammonia surface concentration perturbations at three different times in May (bottom row) for one member of the GEM-MACH ensemble.}
\label{fig:perturbations}
\end{figure}

For application to ammonia emissions inversions, the same perturbation to the time-mean ammonia emissions can result in different perturbations to the atmospheric ammonia concentration at different times due to the dependence on the meteorological conditions. The strength of the ammonia emissions are strongly dependent on temperature, while the local spatial distribution of its atmospheric concentration is dependent on advection patterns. This is illustrated in Figure \ref{fig:perturbations}, which displays the ammonia emission and surface concentration perturbations for one particular member of the ensemble. The monthly-mean ammonia emission perturbations for May is shown in the top row, while the bottom row shows the surface concentration perturbations resulting from this emissions perturbation at three different times (all at the same time, 18 UTC, but for different days). While all three surface concentration perturbations originate from the same emissions perturbation, the different meteorological conditions at each time have a clear effect on both the magnitude and spatial pattern of surface concentration perturbations.

For the ensemble generation described in Eq. (\ref{eq:gm_model}), the input to GEM-MACH consists of a randomly perturbed ammonia emissions field $\textbf{e}_i$, as well as a large number of unperturbed fields $\boldsymbol\theta$. In addition to meteorological fields, GEM-MACH requires hundreds of chemical constituent fields as well as other auxiliary fields such as those describing land-use and soil hydrology. As GEM-MACH is typically run on large CPU clusters that have access to much larger amounts of memory than is currently available to GPUs, EnsAI was designed to use drastically fewer unperturbed input fields.

As ammonia emissions have a strong temperature dependence and the local spatial patterns of ammonia concentrations depend on its advection, the meteorological input fields to EnsAI were limited to the surface temperature $\textbf{T}$ and horizontal wind fields $\textbf{u}$ and $\textbf{v}$. Ammonia has a relatively short atmospheric lifetime and individual perturbations to the ammonia concentrations are short-lived. We exploit these properties by neglecting concentration or meteorological fields from previous time steps to further reduce the size of the input. Hence, EnsAI models a diagnostic, rather than a prognostic, relationship. The EnsAI model can now be summarized as
\begin{equation}
\delta\textbf{c}_i=\mathcal{M}^{\rm{EnsAI}}(\delta\textbf{e}_i, \textbf{u}, \textbf{v}, \textbf{T}) .
\label{eq:AI_model}
\end{equation}
As EnsAI describes a diagnostic relationship between variables, $\textbf{u}$, $\textbf{v}$, $\textbf{T}$, and $\delta\textbf{c}$ in the equation above are all valid at the same time (while $\delta\textbf{e}$ is a monthly mean parameter). We emphasize that the simplification to a diagnostic relationship will only be a reasonable approximation for short-lived species. Additionally, the original $772 \times 642$ horizontal grid of the GEM-MACH regional model, which covers all of Canada, the United States, and northern Mexico, was reduced to a $256 \times 256$ grid that cover only the central and eastern United States and Canada (displayed in Fig. \ref{fig:mean_emissions}) to further reduce memory requirements. 

It is worth noting that generating the ensemble of emissions is simple, as they are constructed from a Gaussian random field as described in Section \ref{sec:gm_ensemble}. In Eq. (\ref{eq:gm_model}), the GEM-MACH model $\mathcal{M}^{\rm{GM}}$ translates the emissions ensemble into an atmospheric concentration ensemble, and the desire is for $\mathcal{M}^{\rm{EnsAI}}$ to emulate this process. This is in contrast to other AI-based geophysical systems, such as those using diffusion-based generative models \citep{price2023,li2024a}, that construct a latent space that is easy to sample in. In diffusion-based generative models, a diffusion process is used to add noise to an image (which in our application would correspond to an atmospheric field) until the image is dominated by noise. The goal of the noising process is for the distribution of noised images to be well described by a probability distribution that is easy to sample from. In these systems, random samples are drawn in the latent space and are then denoised using the reverse diffusion process \citep{song2020a,song2020b,ho2020} to transform the random samples into the space of the atmospheric model. In our application, sampling from the emissions distribution is already relatively easy, so constructing and transforming into a latent space is not necessary. In both cases, samples are randomly drawn in a space where the probability distribution is relatively simple and the AI model transforms that sample to the state space of the atmospheric model. However, EnsAI can potentially be simpler than diffusion-based (or other similar) generative models since a diffusion process to denoise an image does not need to be modeled.

\section{Neural Network Architecture}

\begin{figure}
\begin{center}
\noindent\includegraphics{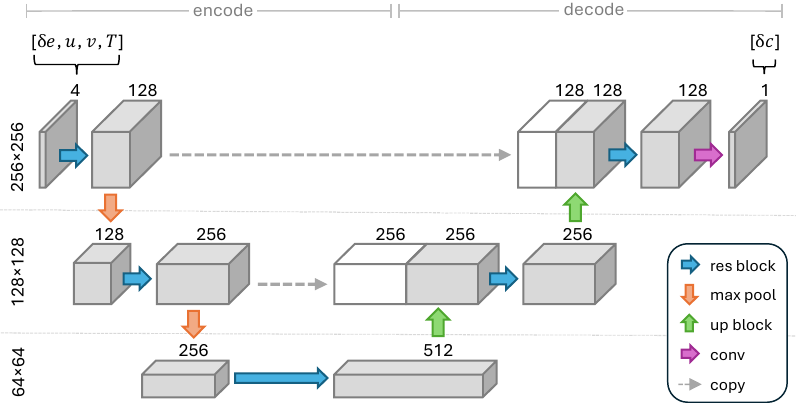}
\end{center}
\caption{U-Net architecture used for EnsAI. Each box represents an image or feature map, where the $x, y$ dimensions are displayed on the left and the number of channels are displayed above each box. White boxes denote feature maps that have been copied during the encoding sequence and concatenated with feature maps produced in the decoding sequence. The blue, orange, green, and purple arrows represent residual blocks, max pooling, upscaling blocks, and two-dimensional convolution, respectively. This figure is adapted from the depiction of the original U-Net architecture shown in Fig. 1 of \citet{ronneberger2015}.}
\label{fig:unet}
\end{figure}

EnsAI is based on the popular U-Net convolutional neural network architecture \citep{ronneberger2015} that was originally developed for image segmentation, but has since been widely used for other applications such as diffusion-based image generation \citep{ho2020}. The U-Net used for this study is displayed in Figure \ref{fig:unet}, where each box represents either the model state at the input or output, which in the language of image processing can be referred to simply as an \textit{image}, or the image transformed into an intermediate state often referred to as a \textit{feature map}. The U-Net broadly consists of a series of convolutions and max pooling that encodes images into various feature maps, followed by a series of convolutions and upsampling to decode the feature maps back to an image. The dimensions of the horizontal grid at each stage within the U-Net is given on the left of the diagram, while the number of fields, usually referred to as \textit{channels} in the current context, is displayed above each box.

The input fields to EnsAI are displayed in the upper-left corner of Fig. \ref{fig:unet}, which consists of the four input fields $[\delta e, u, v, T]$ on the $256\times 256$ horizontal grid. In the encoding branch of the network, the input image is sequentially transformed into feature maps with more channels and a reduction in spatial resolution to encode different image features. The decoding branch decreases the number of channels and increases the spatial resolution, with the final output having the original spatial resolution and a single channel for the atmospheric surface concentration perturbation $\delta c$ (shown in the upper-right corner of Fig. \ref{fig:unet}). Details on each component of the U-Net can be found in Appendix \ref{sec:unet_details}.

The U-Net used for EnsAI has two downsampling layers, in contrast to many other U-Net designs that have three. Ammonia has a relatively short atmospheric lifetime which limits the scale of its advection. Consequently, its background error covariance matrix is sparse and concentrated within $\sim 100$ km of each location. After EnsAI's two downsampling layers, at the bottleneck of the network, the network's receptive field is about 320 km (or 160 km in either direction), which should be adequately large to properly capture the spatial correlations of the error covariance matrix. Adding additional downsampling layers would increase the number of tunable model parameters beyond what is required for this particular application, although we note that longer lived species would likely require additional downsampling layers. As mentioned in Section \ref{sec:inversion_methods}, localization is applied to the ensemble covariance to remove spurious long-distance correlations due to the limited ensemble size. As the horizontal localization length scale for the ammonia emission inversions had half-width at half-maximum of 120 km (see Section \ref{sec:inversion_details} of the Appendix), any additional longer-scale features created by adding more downsampling layers would be removed after localization in any case.

\section{Training the Network}
\label{sec:training}

An existing GEM-MACH ammonia ensemble for January to December 2016 was used for training the neural network. As mentioned in Section \ref{sec:inversion_methods}, the ensemble consisted of hourly GEM-MACH model fields. The ensemble consisted of 60 members, so the training set contained $366 \;\rm{days} \times 24 \;\rm{h}\;\rm{day}^{-1} \times 60 \;\rm{members} = 527,040 \;\rm{samples}$.

The mean squared error (MSE) was used as the loss function for training. While the MSE loss function can lead to smoothing at smaller scales, no significant smoothing was observed in our experiments. More details of the training can be found in Section \ref{sec:unet_details} of the Appendix.

Ensembles for January to December 2015 and July 2014 were generated to use as test and validation data sets. We note that some GEM-MACH files during January and December 2015 were either unavailable or contained erroneous values and were excluded from the test data set. As such, the test set covers slightly less than a full year's worth of data, with January 1-4, January 9-31, and December 5-11 were excluded from the test set. As the change in total atmospheric ammonia in North America is small between these years \citep{vandamme2021}, evaluating on the year prior to the year used for training is unlikely to be an issue.

\section{Run Time Comparison}
\label{sec:runtime}

One of the main motivations of this work is to develop a method for ensemble generation that can run significantly faster than traditional physically-based systems. The original GEM-MACH ensemble took roughly 6.5 hours of wall time to generate a week's worth of simulated data per ensemble member when run on a CPU cluster using 720 computing nodes. In contrast, EnsAI when run on a single GPU took only 7 seconds of wall time to generate a week's worth of simulated data per ensemble member, or about 3,300 times faster than GEM-MACH (see Section \ref{sec:inference_details} of the Appendix for more details).

The run time comparisons above for EnsAI do not include the upfront cost of generating the GEM-MACH ensemble used for training as well as the time for training. The intent is to run EnsAI for time periods much longer than the training period, so that the long-term computational savings of running EnsAI instead of GEM-MACH for ensemble generation becomes much larger than these initial costs. For operational implementation, the desire is to only retrain EnsAI occasionally when there are significant changes in the operational version of GEM-MACH.

It is important to note that while EnsAI runs significantly faster than GEM-MACH, GEM-MACH produces forecasts for hundreds of chemical fields on 85 vertical levels, while EnsAI only produces a single surface concentration field. However, while the wealth of information produced by GEM-MACH can have many uses, the full volume of this information is not always needed for every application. Generating an ensemble of ammonia concentrations at or near the surface to be used for emissions inversions is one example where not having as much information generated is not necessarily a drawback, but where the long run times of the physically-based model may limit how often these inversions can be performed.

\section{Static Error Covariances}

To provide a point of reference for the differences between the GEM-MACH and EnsAI ensemble covariances, we also examine the differences between the GEM-MACH ensemble error covariances and the model error covariances derived in a similar manner to those used operationally at ECCC for surface pollutants \citep{menard2018c,menard2020}. ECCC's operational air quality system produces hourly surface analyses of $\rm{O}_3$, $\rm{NO}$, $\rm{NO}_2$, $\rm{SO}_2$, $\rm{PM}_{2.5}$, $\rm{PM}_{10}$ using error statistics that have a diurnal variation, but are otherwise constant within a two-month period. We refer to these flow-independent error statistics as the `static' error covariances, although this is a bit of a misnomer as they do contain diurnal patterns. Including these comparisons will help quantify the effect of using flow-dependent error covariances over flow-independent error covariances that mirror what is currently used in operational air quality systems at ECCC.

While the ECCC operational error statistics for surface pollutants are formed using the time-average departures of a single forecast from its time-mean and adjusting the error variances using innovation statistics, to isolate the effects of including flow-dependent error covariances, the static error covariances used for this study were formed simply by computing the two-month mean of the GEM-MACH ensemble covariance for the training set (done separately for every hour of the day to retain the mean diurnal variation).

\section{Surface Observations}

In addition to direct comparisons of the covariances, EnsAI's performance will also be evaluated by comparing the posterior ammonia surface fields to ammonia surface observations from the US's Ammonia Monitoring Network (AMoN; \url{http://nadp.slh.wisc.edu/amon}) and Canada's National Air Pollution Surveillance (NAPS; \url{https://www.canada.ca/en/environment-climate-change/services/air-pollution/monitoring-networks-data/national-air-pollution-program.html}) network. Ammonia observations from the AMoN network are made using Radiello\textsuperscript\textregistered passive diffusion samplers and measure ammonia concentrations over two-week periods. The NAPS stations that monitor ammonia use citric acid-coated denuders to measure ammonia concentration over a 24-hour period every three or six days. A total of 55 AMoN stations and 6 NAPS stations were within the model domain and had valid observations during the test period, the locations of which are shown in Figure S2 of the Supplement.

\section{Direct Comparisons of the Covariances}

In this section, the different covariances will be directly compared (in contrast to Section \ref{sec:inversions} that compares products of inversions that use these covariances). Although EnsAI was trained on a 60 member ensemble, it is possible for EnsAI to generate ensembles with more than 60 members. While the speed at which EnsAI can run may allow for much larger ensemble sizes, new strategies for memory management may be required for very large ensembles. In any case, to simplify the comparison between covariances, for this study the results presented for EnsAI will be for an ensemble of the same size as the original GEM-MACH ensemble.

In this section, uncertainties on quantities derived from the GEM-MACH ensemble due to the limited size of the ensemble will be estimated via bootstrapping, in which the GEM-MACH ensemble members were resampled (with replacement) 50 times. 1--$\sigma$ confidence intervals are illustrated on line plots by lightly-shaded region around the curve and by error bars for scatter plots. We note that for some curves the uncertainties are small enough that the confidence regions are not easily distinguished from the main curve.

\subsection{Covariance Parametrization}
\label{sec:cov_param}

For the direct comparison between covariances, each covariance is decomposed into variance and correlation components. The error variances can be constructed from the emissions standard deviations $\sigma_e$ and the surface concentrations standard deviations $\sigma_c$, while the correlations are comprised of the univariate emissions correlations $\rho_{ee}$, the univariate surface concentration correlations $\rho_{cc}$, and the emissions/surface concentration cross-correlations $\rho_{ec}$. As EnsAI uses the same emissions ensemble $\{ \delta\textbf{e}_i \}$ as the GEM-MACH ensemble, no comparisons of $\sigma_e$ or $\rho_{ee}$ are necessary. 

While the ensemble standard deviations $\sigma_c$ can be represented by a vector and is straightforward to display its values (for example on top of a 2D map), it is more difficult to display the correlations $\rho_{cc}$ and $\rho_{ec}$ in full, as these represent matrices, although a single row/column can easily be shown. To aid in summarizing the information in the correlation matrices, we introduce the isotropic correlation length $L^{\rm{iso}}$, which we define at location $\textbf{\textit{x}}$ as
\begin{equation}
L^{\rm{iso}}(\textbf{\textit{x}}) = \sqrt{\frac{2}{\pi}} \int \varrho(\textbf{\textit{x}}, \textbf{\textit{x}}') |\textbf{\textit{x}}'| d\textbf{\textit{x}}' ,
\end{equation}
where $\varrho$ is the normalized spatial correlation function defined by
\begin{equation}
\varrho(\textbf{\textit{x}}, \textbf{\textit{x}}') \equiv \frac{\rho(\textbf{\textit{x}}, \textbf{\textit{x}}')}{\int \rho(\textbf{\textit{x}}, \textbf{\textit{x}}') d\textbf{\textit{x}}'} .
\end{equation}
Defined in this manner, the isotropic correlation length is the expectation value of distance weighted by $\varrho$. The factor of $\sqrt{2/\pi}$ in this definition is included so that for a 2D isotropic Gaussian correlation model, $L^{\rm{iso}}$ coincides with the Daley length-scale \citep{daley1993} (as well as the standard deviation of the Gaussian model). We can extend this definition to anisotropic modes $m$ of the correlation length by introducing a multipole expansion of the correlation length as
\begin{equation}
L[m](\textbf{\textit{x}}) \equiv \sqrt{\frac{2}{\pi}} \int e^{im\phi'} \varrho(\textbf{\textit{x}}, \textbf{\textit{x}}') |\textbf{\textit{x}}'| d\textbf{\textit{x}}' ,
\label{eq:lcftr_def}
\end{equation}
where $\phi'$ is the angle between $\textbf{\textit{x}}'$ and the $x$-axis. With this definition, we have $L[0]=L^{\rm{iso}}$, $L[1]$ is proportional to the dipole moments of the function $\varrho$, and $L[2]$ is related to its quadrupole moments. More details of this expansion can be found in Section \ref{sec:corrlen_details} of the Appendix.

\begin{figure}
\begin{center}
\noindent\includegraphics{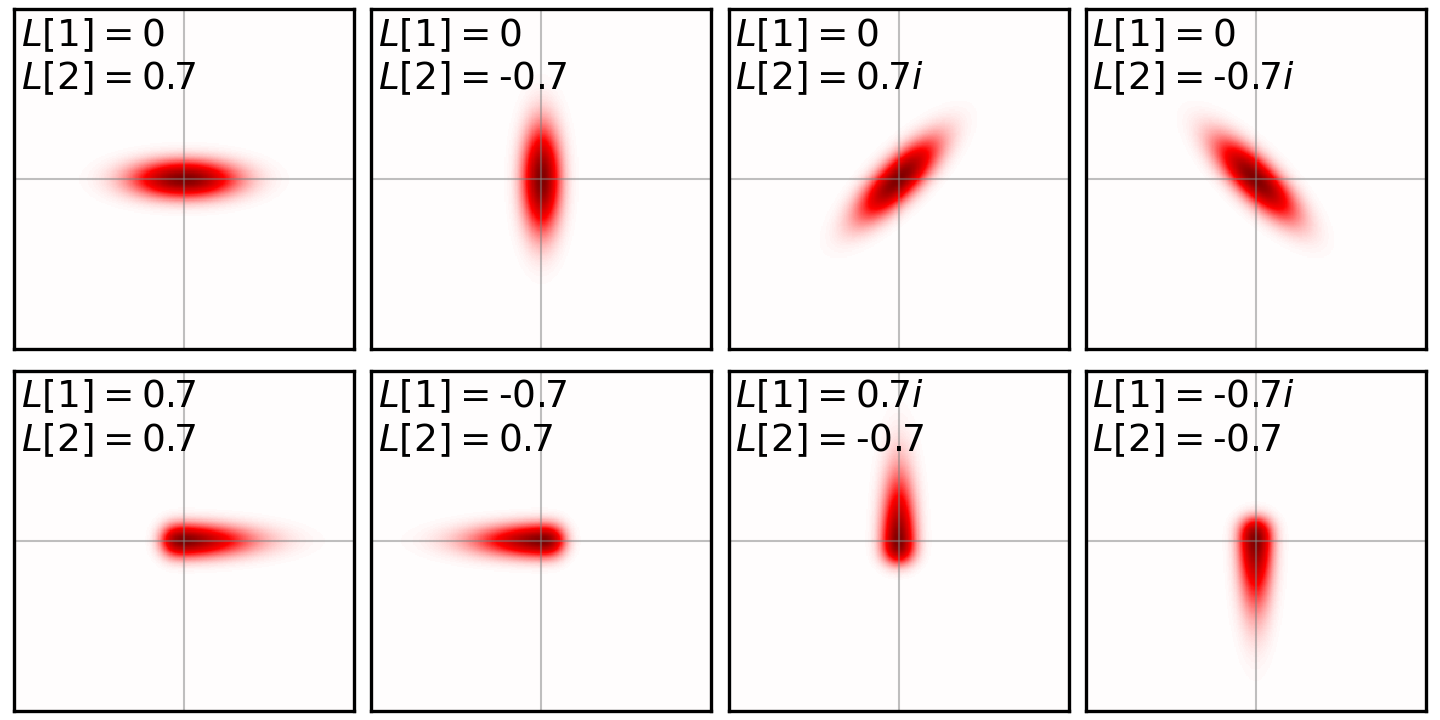}
\end{center}
\caption{Examples of correlations and their corresponding $L[1]$ and $L[2]$ values, which are displayed in the upper left corner of each panel. The correlation length modes are taken with respect to the origin. The values for $L[0]$ are the same for all panels and units are selected so that $L[0]=1$.}
\label{fig:corr_examples}
\end{figure}

To better understand this parameterization of the correlation length, example correlations are given in Figure \ref{fig:corr_examples}, with the values for $L[1]$ and $L[2]$ displayed in the upper left corner of each panel. In these plots, the correlation length modes are taken with respect to the origin. The isotropic correlation length $L[0]$ is the same in each panel and the units of $L[m]$ are set such that $L[0]$ is equal to one. In the top row, all correlations have $L[1] = 0$ but have varying values for $L[2]$, which results in anisotropic correlations centered on the origin and its major axis pointing in different directions. The correlations in the bottom row of Fig. \ref{fig:corr_examples} have $L[2]$ values with the same magnitude as those in the top row, but also have nonzero values for $L[1]$, which skews the correlations in a particular direction.

The set of covariance parameters to compare is then $\{\sigma_c, L_{ec}[m], L_{cc}[m]\}$, where $L_{ec}[m]$ and $L_{cc}[m]$ are the correlation length modes for the emissions/surface concentration cross-correlation and surface concentration univariate correlations, respectively. For each of these covariance parameters, we will examine the anomaly correlation coefficient ($ACC$) and the root mean square error ($RMSE$), which are defined in Section \ref{sec:comparison_statistics} of the Appendix. Lastly, since the comparisons of $L_{ec}[m]$ and $L_{cc}[m]$ yielded fairly similar conclusions, our analysis will focus on $L_{ec}[m]$ for brevity and only some comparisons for $L_{cc}[m]$ will be discussed.

\subsection{Comparison Results for Error Variances}
\label{sec:results_variances}

\begin{figure}
\begin{center}
\noindent\includegraphics[width=0.7\textwidth]{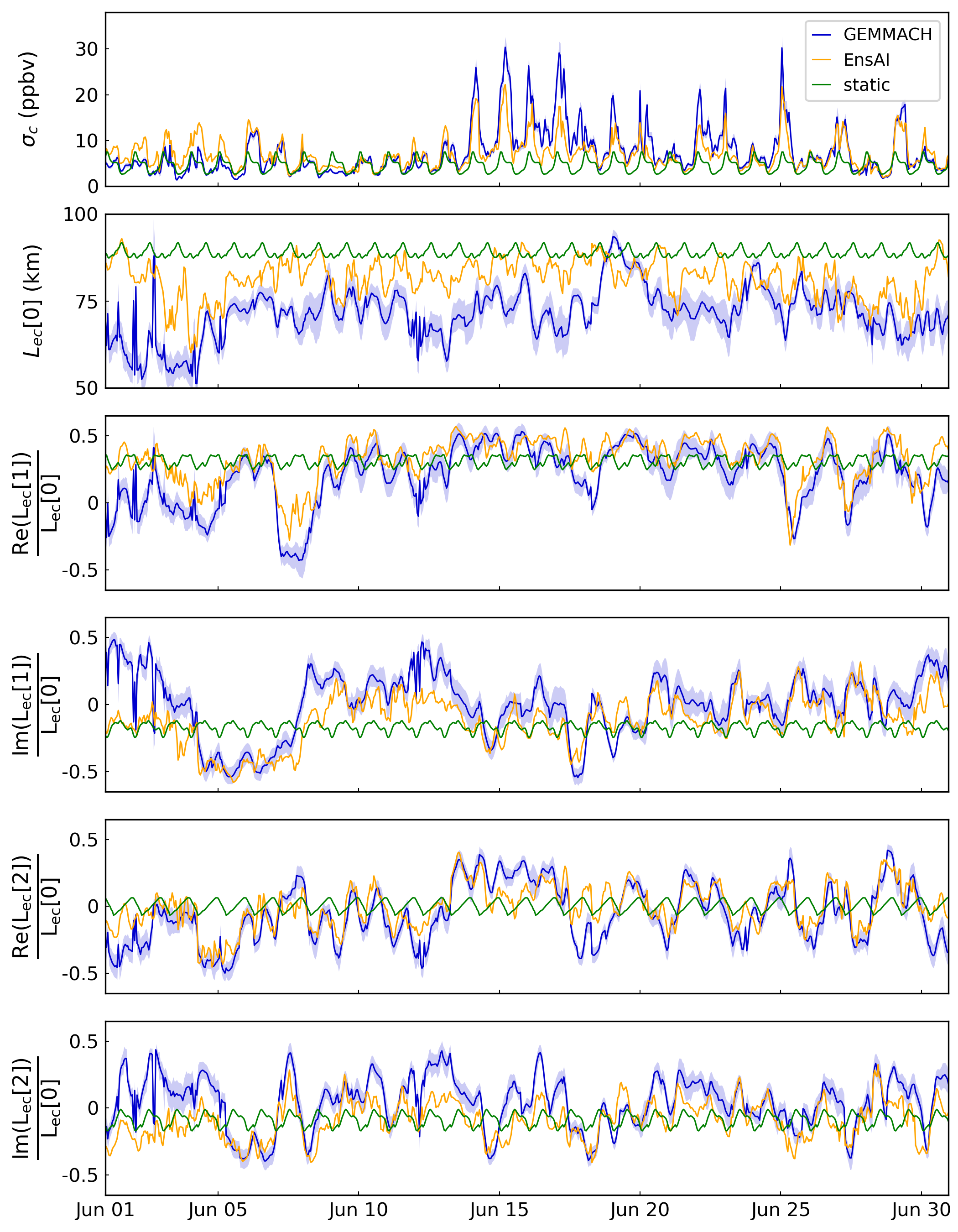}
\end{center}
\caption{Time series of components of the error covariances for the location in North Carolina (exact location shown in Fig. \ref{fig:mean_emissions}) for the GEM-MACH ensemble (blue), the EnsAI ensemble (orange), and the static covariance (green) for June 2015. The top plot shows the time series of the standard deviation of surface ammonia $\sigma_c$. Lower plots show time series for the horizontal correlation length modes $L_{ec}[m\le 2]$ for the emissions/surface concentration cross-correlation.}
\label{fig:timeseries_north_carolina}
\end{figure}

\begin{figure}
\begin{center}
\noindent\includegraphics[width=0.7\textwidth]{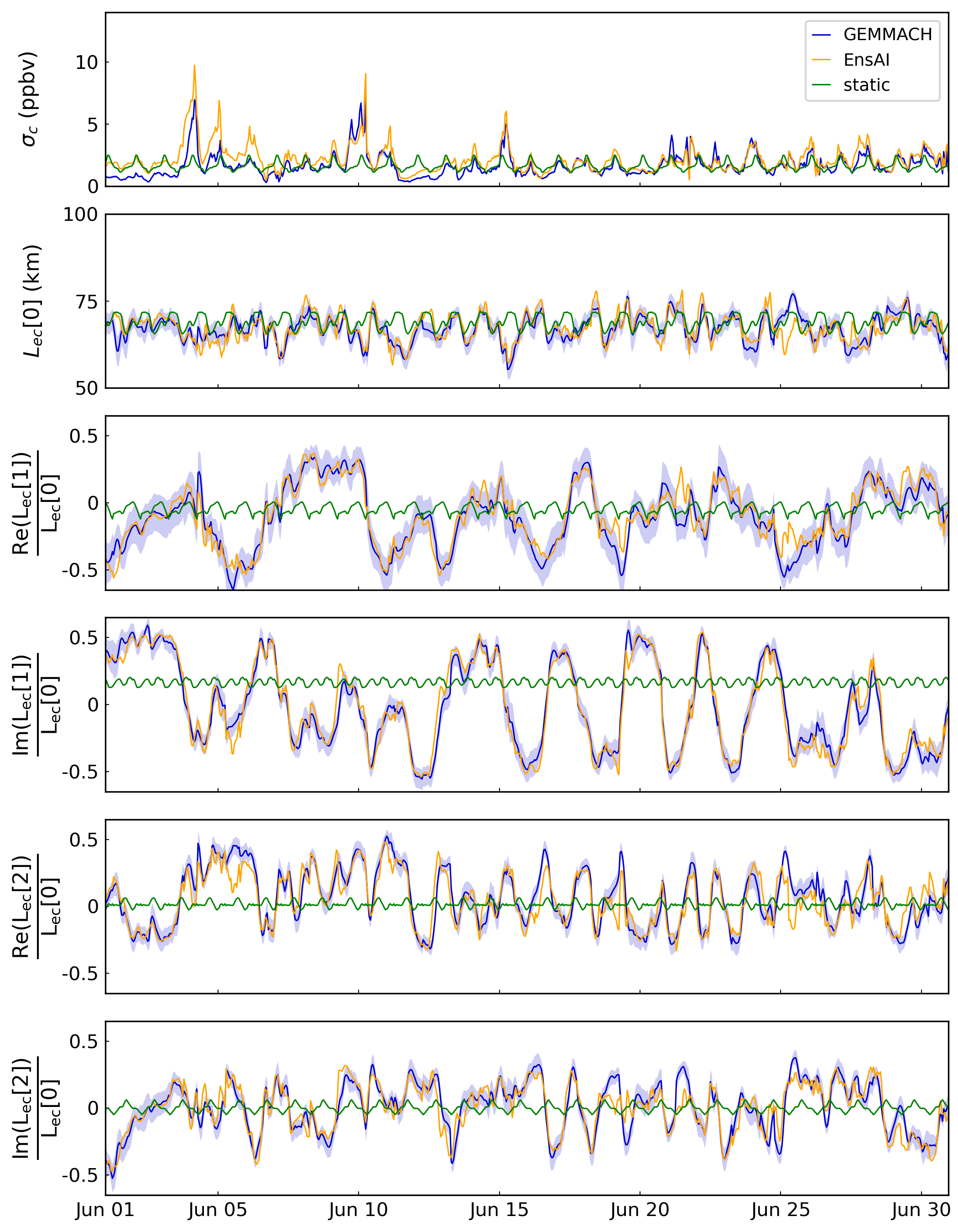}
\end{center}
\caption{Same as Fig. \ref{fig:timeseries_north_carolina}, but for the location in Iowa.}
\label{fig:timeseries_iowa}
\end{figure}

\begin{figure}
\begin{center}
\noindent\includegraphics[width=0.7\textwidth]{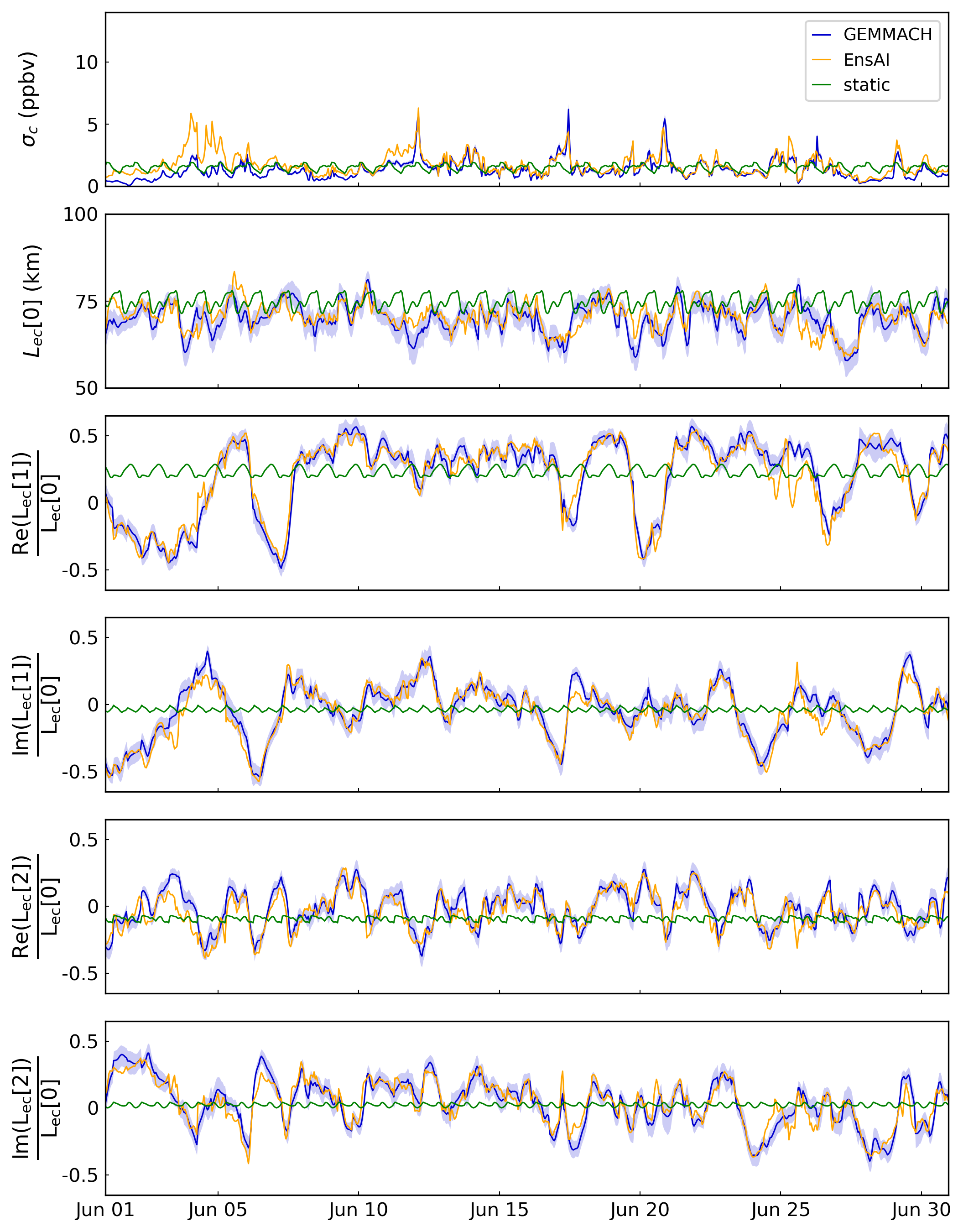}
\end{center}
\caption{Same as Fig. \ref{fig:timeseries_north_carolina}, but for the location in Ohio.}
\label{fig:timeseries_ohio}
\end{figure}

\begin{figure}
\begin{center}
\noindent\includegraphics[width=0.7\textwidth]{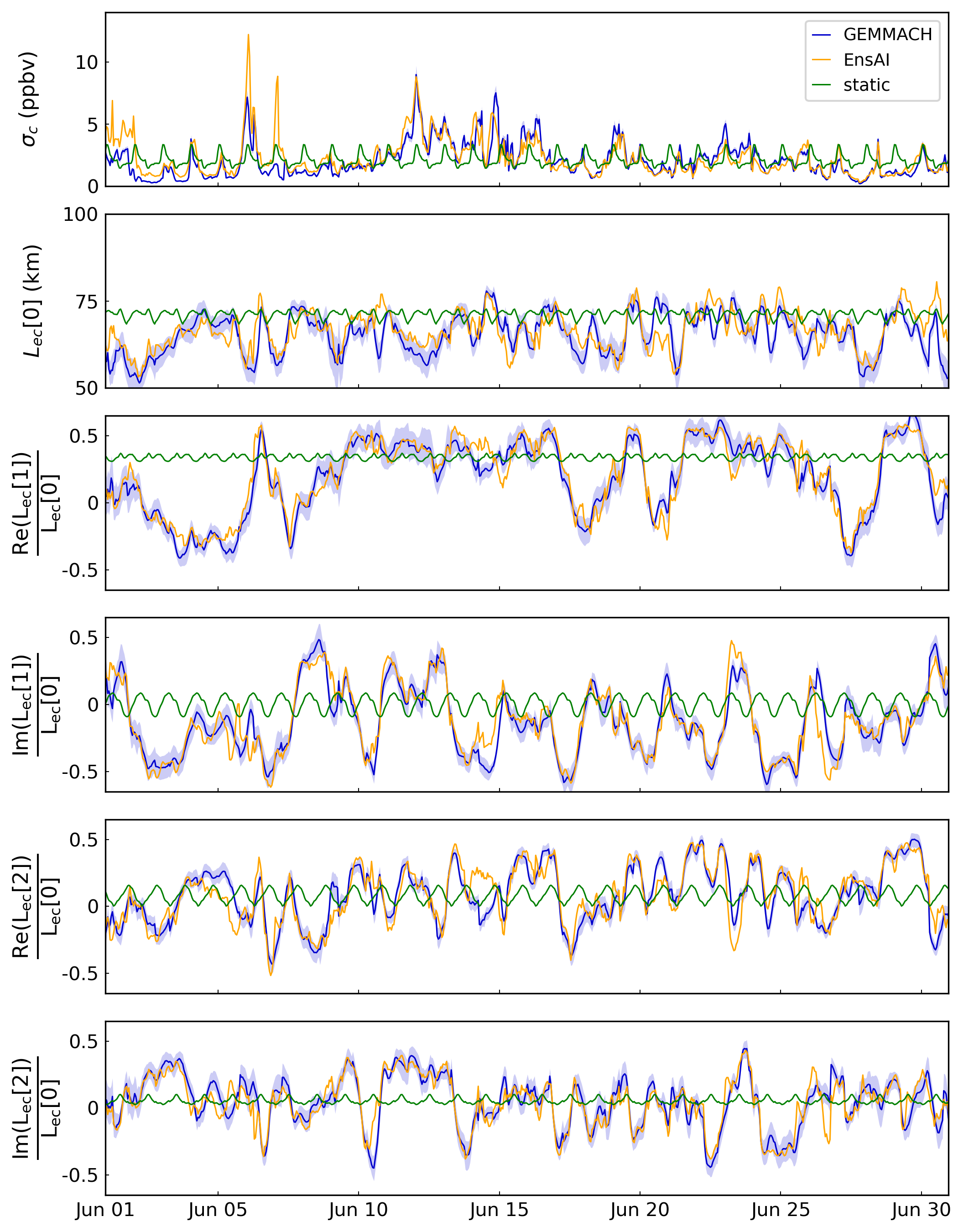}
\end{center}
\caption{Same as Fig. \ref{fig:timeseries_north_carolina}, but for the location in Pennsylvania.}
\label{fig:timeseries_pennsylvania}
\end{figure}

\begin{figure}
\begin{center}
\noindent\includegraphics[width=0.7\textwidth]{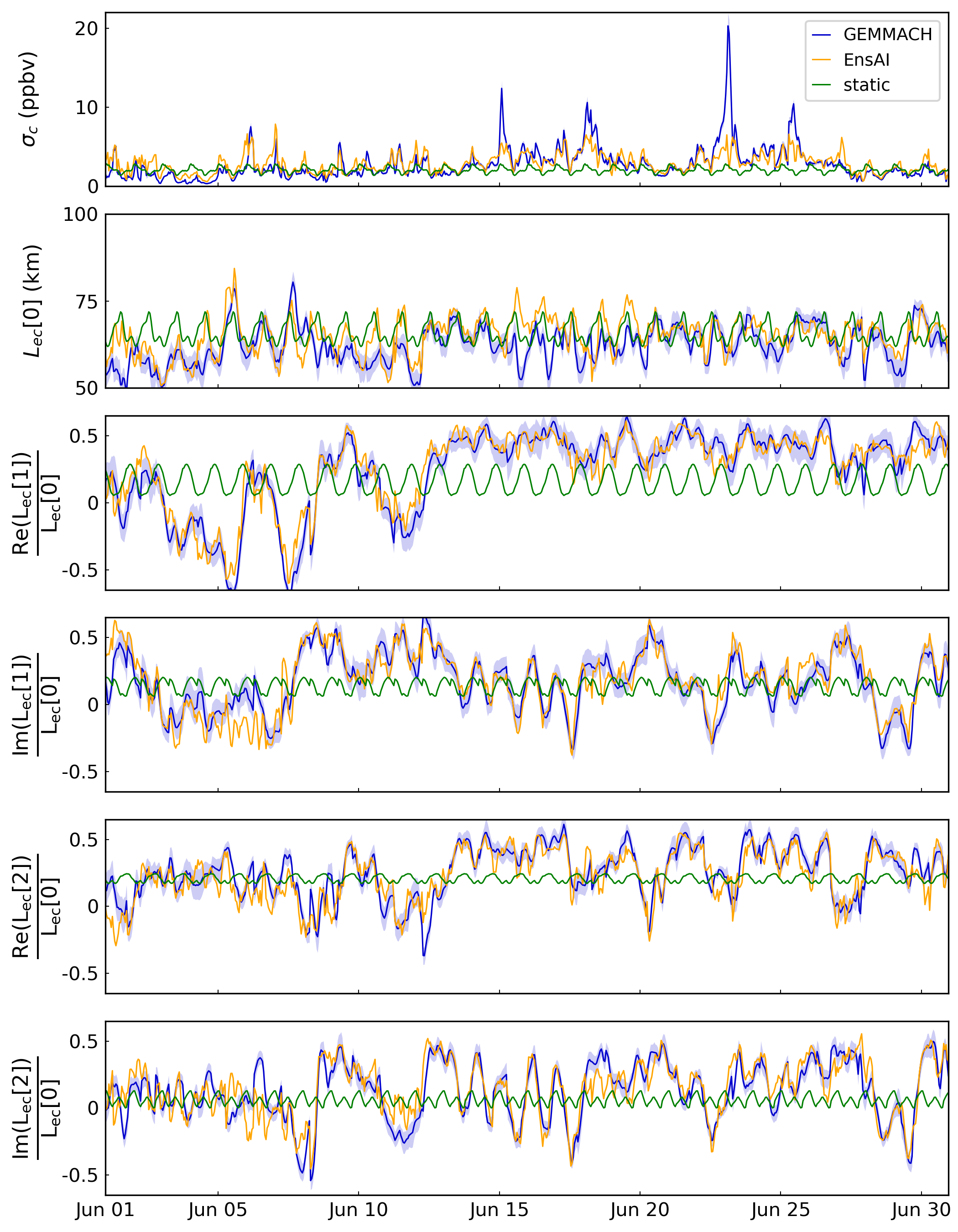}
\end{center}
\caption{Same as Fig. \ref{fig:timeseries_north_carolina}, but for the location in Georgia.}
\label{fig:timeseries_georgia}
\end{figure}

We begin with a comparison of the standard deviations of the ammonia surface concentrations $\sigma_c$. The top panels of Figures \ref{fig:timeseries_north_carolina} to \ref{fig:timeseries_georgia} show time series of $\sigma_c$ for the month of June 2015 at the five locations shown in Fig. \ref{fig:mean_emissions}. These locations correspond to the places with the highest values in the ammonia emissions inventory and are ordered with decreasing emissions. As seen in these panels, $\sigma_c$ for the original GEM-MACH ensemble (shown with the blue curve) displays sporadic spikes associated with changing weather conditions, many of which can also be seen in $\sigma_c$ from the EnsAI ensemble (shown with the orange curve) but not for the static error covariance (shown with the green curve). In these time series, we can see that EnsAI captures a considerable amount of the temporal variation of the GEM-MACH ensemble variances. There are some notable differences, such as EnsAI's over-prediction of variance at the location in Ohio on June 4-5 (Fig. \ref{fig:timeseries_iowa}) and in Pennsylvania on June 6 (Fig. \ref{fig:timeseries_pennsylvania}) and under-prediction in Georgia on June 23, but for the most part EnsAI produces values for $\sigma_c$ that are closer to the GEM-MACH ensemble value than that derived from the static error covariances.

\begin{figure}
\begin{center}
\noindent\includegraphics[width=0.8\textwidth]{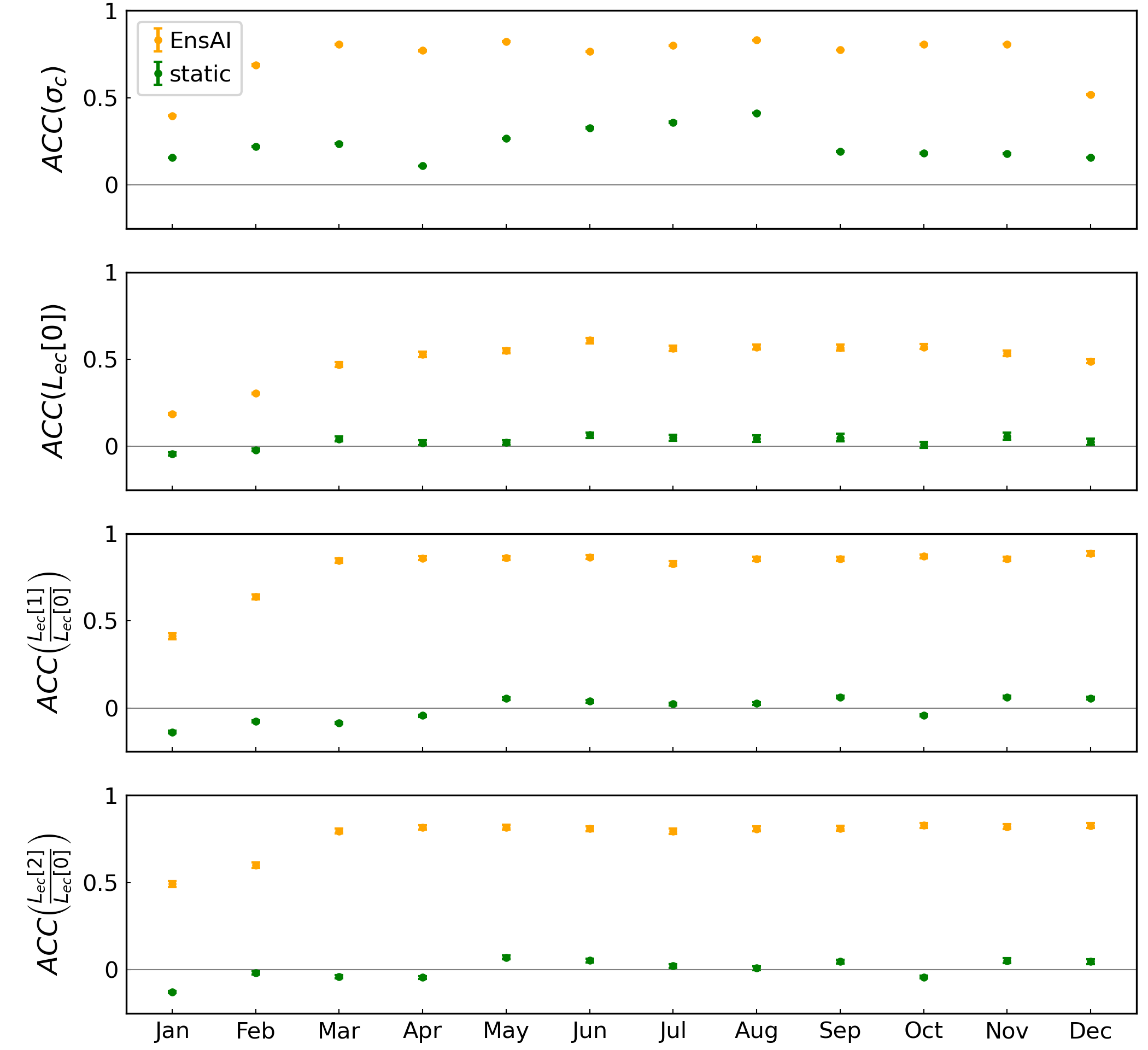}
\end{center}
\caption{$ACC$ values for $\sigma_c$, $L_{ec}[0]$, $L_{ec}[1]/L_{ec}[0]$, and $L_{ec}[2]/L_{ec}[0]$ over the model domain for each month in 2015.}
\label{fig:acc_overall}
\end{figure}

The top panel of Figure \ref{fig:acc_overall}, which shows $ACC(\sigma_c)$ throughout the test period, shows dramatic differences between EnsAI and the static covariances, with EnsAI having larger $ACC$ values for all months. EnsAI has $ACC(\sigma_c)$ values around 0.8 for February to November, and values around 0.4 and 0.5 for January and December, respectively. In contrast, the static covariances have values between 0.2 and 0.4 for February to November and values not far from zero for December and January, indicating that the EnsAI ensemble was much better at capturing the time variation of $\sigma_c$ in the GEM-MACH ensemble.

\begin{figure}
\begin{center}
\noindent\includegraphics[width=0.8\textwidth]{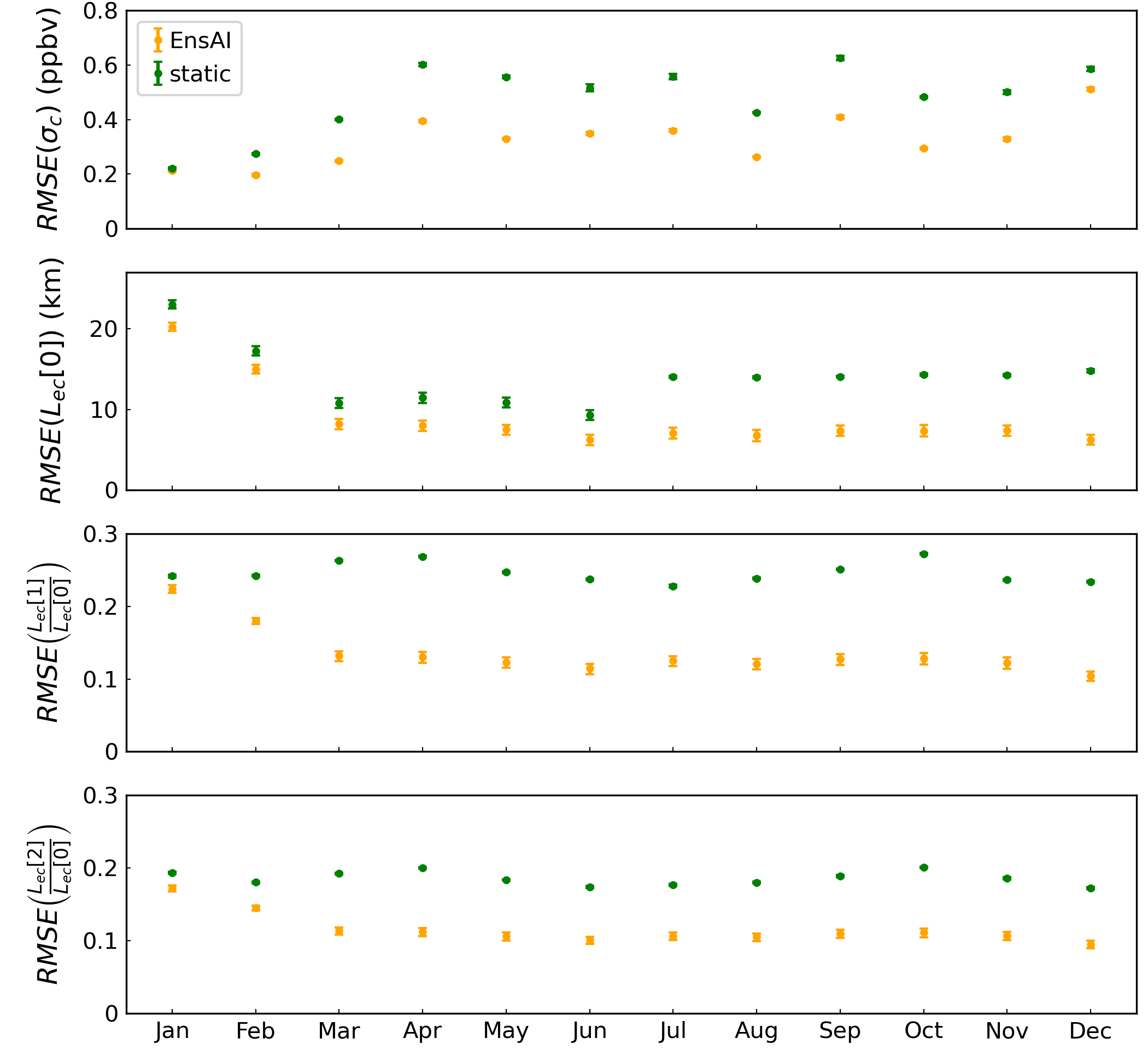}
\end{center}
\caption{$RMSE$ values for $\sigma_c$, $L_{ec}[0]$, $L_{ec}[1]/L_{ec}[0]$, and $L_{ec}[2]/L_{ec}[0]$ over the model domain for each month in 2015.}
\label{fig:rmse_overall}
\end{figure}

The $RMSE$ values for $\sigma_c$ over the model domain for each month in 2015 are shown in the top panel of Figure \ref{fig:rmse_overall}. The $RMSE(\sigma_c)$ values for EnsAI are lower than for the static covariance values for all months, although the difference is negligible for January. While the differences in $RMSE$ values between EnsAI and the static covariances taken over the whole model domain are less than a ppbv (see Figure S3 of the Supplement for a comparison to the mean values of $\sigma_c$), the $RMSE(\sigma_c)$ values near locations with high emissions can vary significantly. For instance, in the summer, the $RMSE(\sigma_c)$ for the static covariances are close to 4 ppbv higher at the location in North Carolina than for EnsAI (see Figure S4 of the Supplement). In addition, the $RMSE(\sigma_c)$  values for the static covariances are higher than that for EnsAI at nearly every point in the domain.

\subsection{Comparison Results for Spatial Correlations}
\label{sec:results_correlations}

\begin{figure}
\begin{center}
\noindent\includegraphics[width=0.8\textwidth]{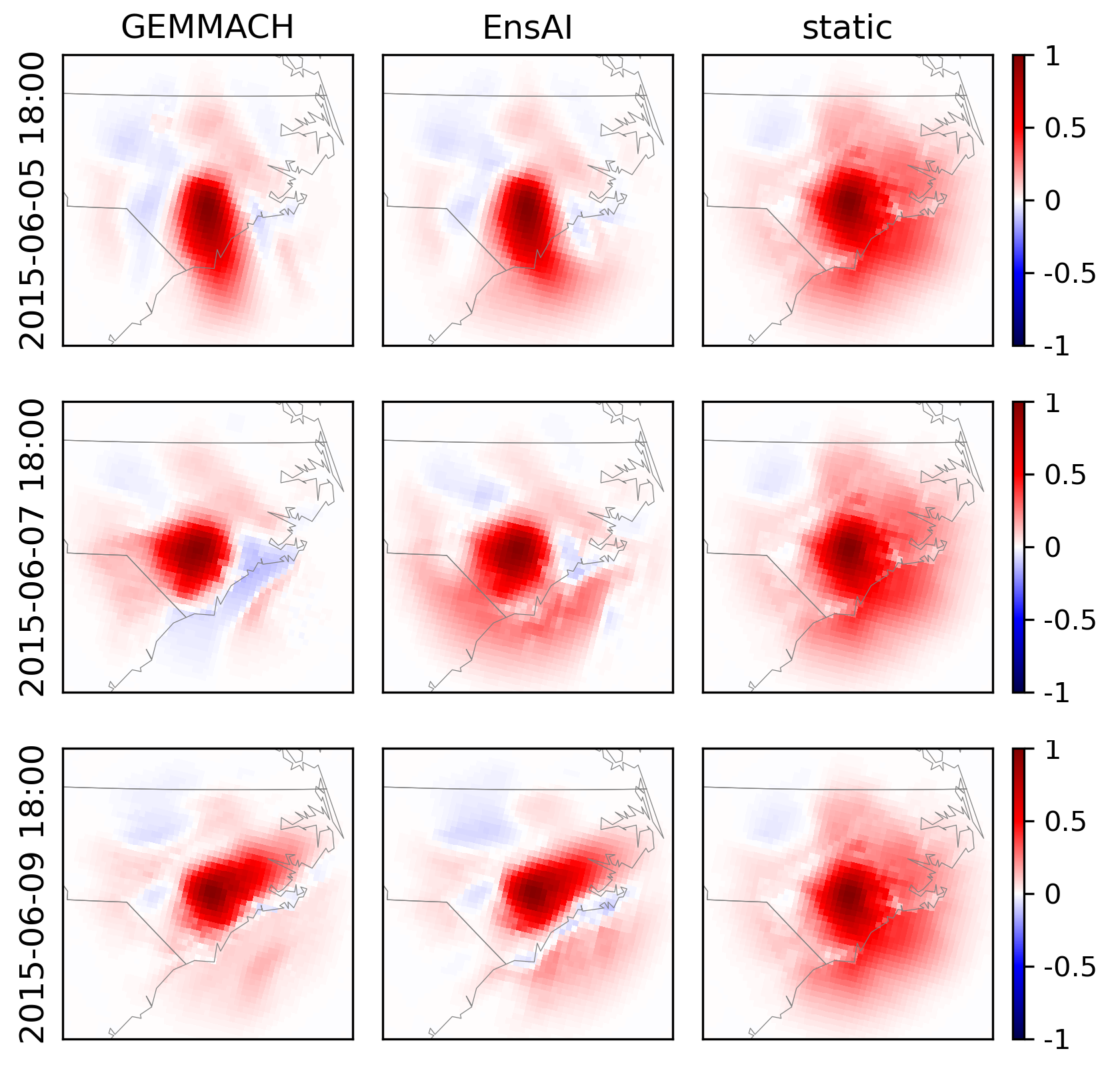}
\end{center}
\caption{Emissions/surface concentration spatial correlations $\rho_{ec}$ for the GEM-MACH ensemble (left column), the EnsAI ensemble (middle column), and the static covariance (right column) at the location in North Carolina at 2015-06-05 18:00 (top row), 2015-06-07 18:00 (middle row), and 2015-06-09 18:00 (bottom row).}
\label{fig:map_corr_north_carolina}
\end{figure}

As an illustration of the time-dependency of the spatial correlations, Figure \ref{fig:map_corr_north_carolina} shows the emissions/surface concentration spatial correlation $\rho_{ec}$ at the location in North Carolina at three different times, separated by two days and all at 18:00 UTC. Looking at the correlations of the GEM-MACH ensemble in the left column of Fig. \ref{fig:map_corr_north_carolina}, we can see that the correlations are often significantly anisotropic and spread out in different directions for different days, reflecting the different meteorological conditions at each time. Comparing this to the correlation in the EnsAI ensemble in the middle column of this figure, it is evident that EnsAI can reasonably reproduce this time-dependence, as the shape of the spatial correlations between the two ensembles looks similar. In contrast, the static correlations displayed in the right column of the figure are much closer to being isotropic (at all times).

\begin{figure}
\begin{center}
\noindent\includegraphics{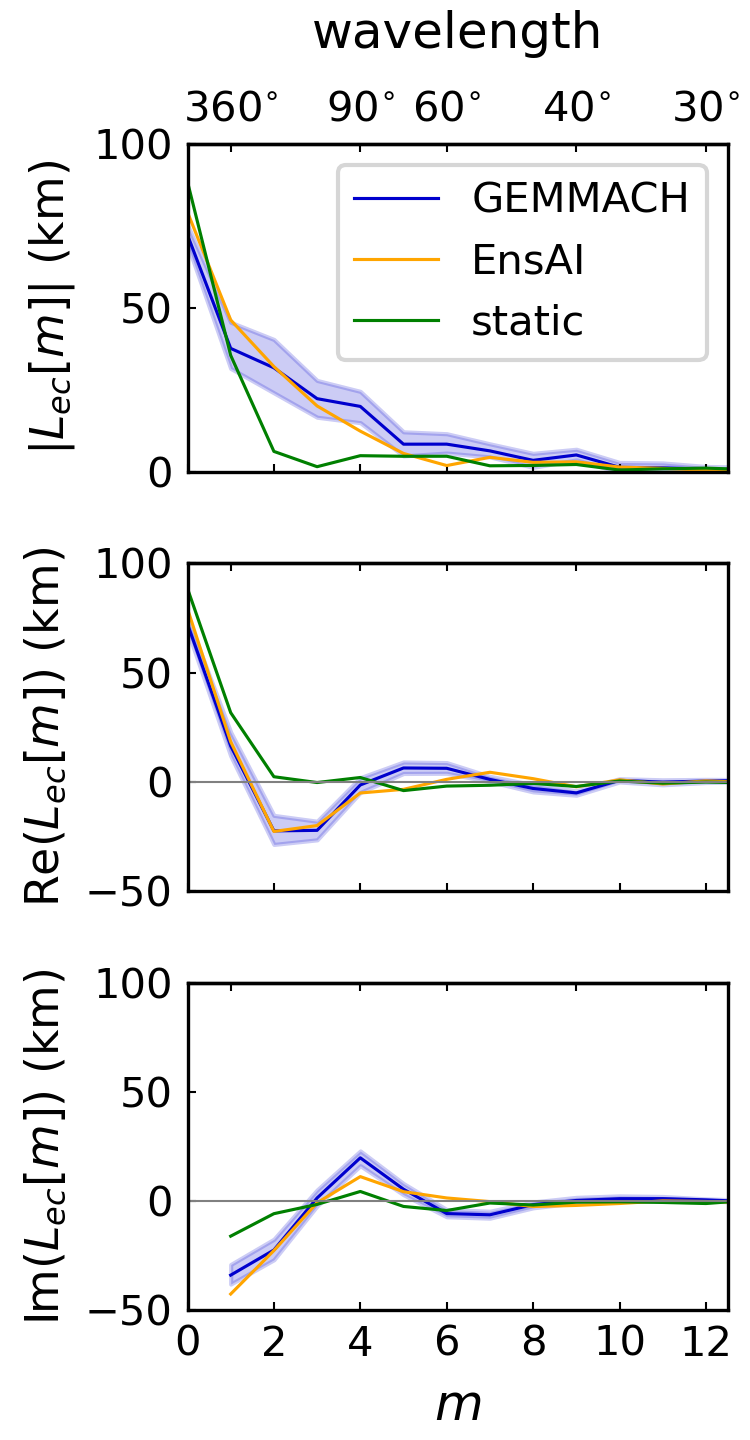}
\end{center}
\caption{Horizontal correlation length modes $L_{ec}[m]$ as a function of wavenumber $m$ for the emissions/surface concentration cross-correlation at the location in North Carolina at 2015-06-05 18:00. The top, middle, and bottom panels show the absolute, real, and imaginary parts, respectively.}
\label{fig:corrlen_example}
\end{figure}

As mentioned in Section \ref{sec:cov_param}, to make the examination of the correlation matrix more tractable, we examine the different modes $m$ of the correlation length $L_{ec}[m]$ instead of dealing with the full correlation matrix directly. To illustrate the expansion into modes, Figure \ref{fig:corrlen_example} shows the correlation length modes $L_{ec}[m]$ as a function of wavenumbers $m$ for the correlations in the top row of Fig. \ref{fig:map_corr_north_carolina}. In this example, the isotropic correlation lengths $L_{ec}[0]$ for the different covariances are between 65 and 90 km, while the anisotropic correlation length modes decrease rapidly with increasing wavenumber. Looking at the absolute value of the correlation length modes (top panel of Fig. \ref{fig:corrlen_example}), we can see that the non-isotropic modes for the static covariance are much smaller than that for the GEM-MACH and EnsAI ensembles, indicating less anisotropy in the static correlation. The EnsAI values for $L_{ec}[m]$ are within the 1--$\sigma$ confidence interval of the GEM-MACH value for many modes, and is generally much closer to the GEM-MACH values than the static covariance is. As $L_{ec}[m]$ decreases rapidly with increasing wavenumber, for the remainder of this paper we concentrate mainly on results for modes with $m \le 2$ for brevity.

Time series of $L_{ec}[m]$ for $m\le 2$ for the five locations identified in Fig. \ref{fig:mean_emissions} are shown in the bottom four panels of Figs. \ref{fig:timeseries_north_carolina} to \ref{fig:timeseries_georgia}. In all of these plots, $L_{ec}[m]$ for the EnsAI ensemble are much closer to the GEM-MACH ensemble value than the static covariance is and captures most of its time variation. While a diurnal pattern is evident in the GEM-MACH ensemble, there are significant non-diurnal time variations that are dependent on meteorological conditions. In Fig. \ref{fig:timeseries_north_carolina}, we can see that the EnsAI correlation length modes generally follow the pattern of the GEM-MACH ensemble, but the results are even more striking for the other four locations (Figs. \ref{fig:timeseries_iowa} to \ref{fig:timeseries_georgia}), where the differences between the EnsAI and GEM-MACH ensembles are often smaller (or not much larger) than the sample variance in the GEM-MACH ensemble, or in other words, the differences between these values are statistically insignificant throughout much of the evaluation time period.

The three lower panels of Fig. \ref{fig:acc_overall} show the $ACC$ values for $L_{ec}[m \le 2]$. For all months, EnsAI has much higher $ACC$ values than the static covariances. The $ACC(L_{ec}[m])$ values with $m \le 2$ for the static covariances are close to zero for all cases, while the EnsAI values are above 0.5 for all cases with the exception of the $m=0$ mode in January and February and the $m=1$ mode in January, and are at or above 0.8 for $m=1,2$ for March to December. From this we can conclude that the EnsAI ensemble can accurately reproduce the time variations in the horizontal correlations (for modes with $m \le 2$) that are not captured by the diurnal pattern of the static covariances. Similar results are seen for the $ACC$ values for $L_{cc}[m \le 2]$ (see Figure S5 of the Supplement), although the $ACC(L_{cc}[1]/L_{cc}[0])$ values are closer to 0.6 for most months, in contrast to $ACC(L_{ec}[1]/L_{ec}[0])$ which is closer to 0.8.

EnsAI had lower $RMSE$ values than the static covariances for all months, as seen in the three lower panels of Fig. \ref{fig:rmse_overall}. For July to December, $RMSE(L_{ec}[0])$ for EnsAI is about 7 km less than the $RMSE$ value for the static covariances, which corresponds to about a 10\% larger $RMSE$ value for the static covariances (see Figure S2 of the Supplement for the mean values of $L_{ec}[0]$). In some areas of the model domain during the summer and fall, $RMSE(L_{ec}[0])$ for the static covariances are over 25 km larger than the values for EnsAI (see Figure S6 of the Supplement), which correspond to static covariance $RMSE$ values which are between 50\% to 490\% larger than for EnsAI. The $RMSE$ values for $L_{ec}[1]/L_{ec}[0]$ taken over the whole domain are 0.02 to 0.14 less for EnsAI than for the static covariance, and between 0.02 and 0.09 lower for EnsAI for $L_{ec}[2]/L_{ec}[0]$, as displayed in the two lowest panels of Fig. \ref{fig:rmse_overall}. Additionally, there are many locations within the model domain where $RMSE(L_{ec}[1]/L_{ec}[0])$ and $RMSE(L_{ec}[2]/L_{ec}[0])$ for the static covariances are over 0.3 and 0.2 higher, respectively, as compared to the $RMSE$ value for EnsAI (see Figures S7 and S8 of the Supplement). Lastly, we note that the EnsAI $RMSE$ values for the surface concentration univariate correlation length modes $L_{cc}[m]$ are uniformly less than those for the static covariances for $m \le 2$ (see Figure S9 of the Supplement).

\begin{figure}
\begin{center}
\noindent\includegraphics{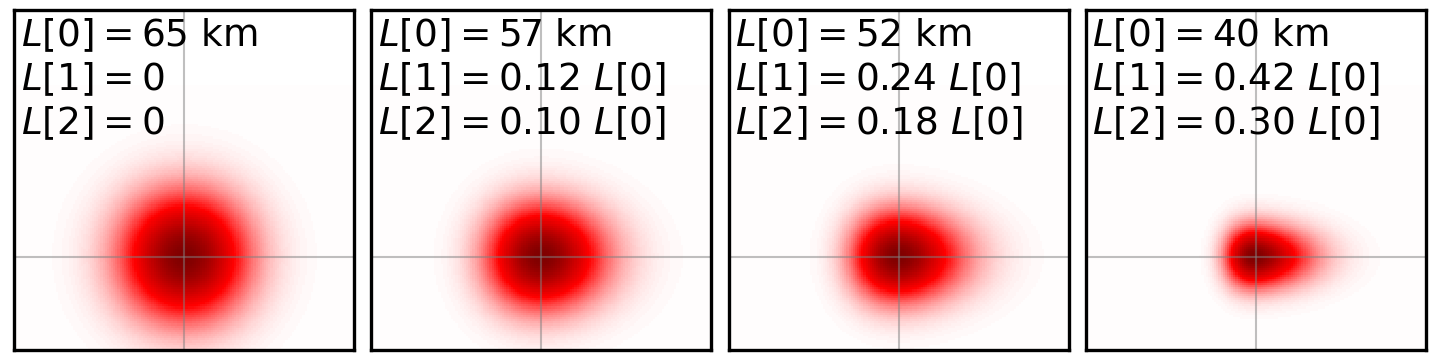}
\end{center}
\caption{Example correlations with their corresponding $L[0]$, $L[1]$, and $L[2]$ values displayed at the top of each panel.}
\label{fig:corr_ref}
\end{figure}

To get an intuitive understanding of what these differences in correlation length modes correspond to in terms of spatial patterns, Figure \ref{fig:corr_ref} shows another set of example correlations that illustrate the deviations in correlation length mode values mentioned in the previous paragraph. The leftmost panel of Fig. \ref{fig:corr_ref} shows an isotropic correlation with $L[0]=65$ km. For illustrative purposes, we will think of this isotropic correlation as a target correlation (i.e. from the GEM-MACH ensemble) and compare it to correlations with different $L[m]$ values to see the difference in the spatial patterns. The correlation shown in the panel second from the left has $L[m \leq 2]$ values that deviate from the isotropic correlation in the leftmost panel by amounts representative of the $RMSE$ values found in Fig. \ref{fig:rmse_overall} for EnsAI. This correlation is only slightly anisotropic and slightly smaller than the correlation in the leftmost panel. The correlation in the panel second from the right has values that deviate from the isotropic correlation close to the $RMSE$ values for the static covariance in Fig. \ref{fig:rmse_overall} and is noticeably more anisotropic than the two correlations to the left. The correlation in the rightmost panel has values that deviate from the isotropic correlation by amounts equal to differences between the GEM-MACH and static covariance values that are commonly found in the model domain (see Figures S7 and S8 of the Supplement), which produces a correlation with a drastically different spatial pattern than the correlation in the leftmost panel.

\begin{figure}
\begin{center}
\noindent\includegraphics{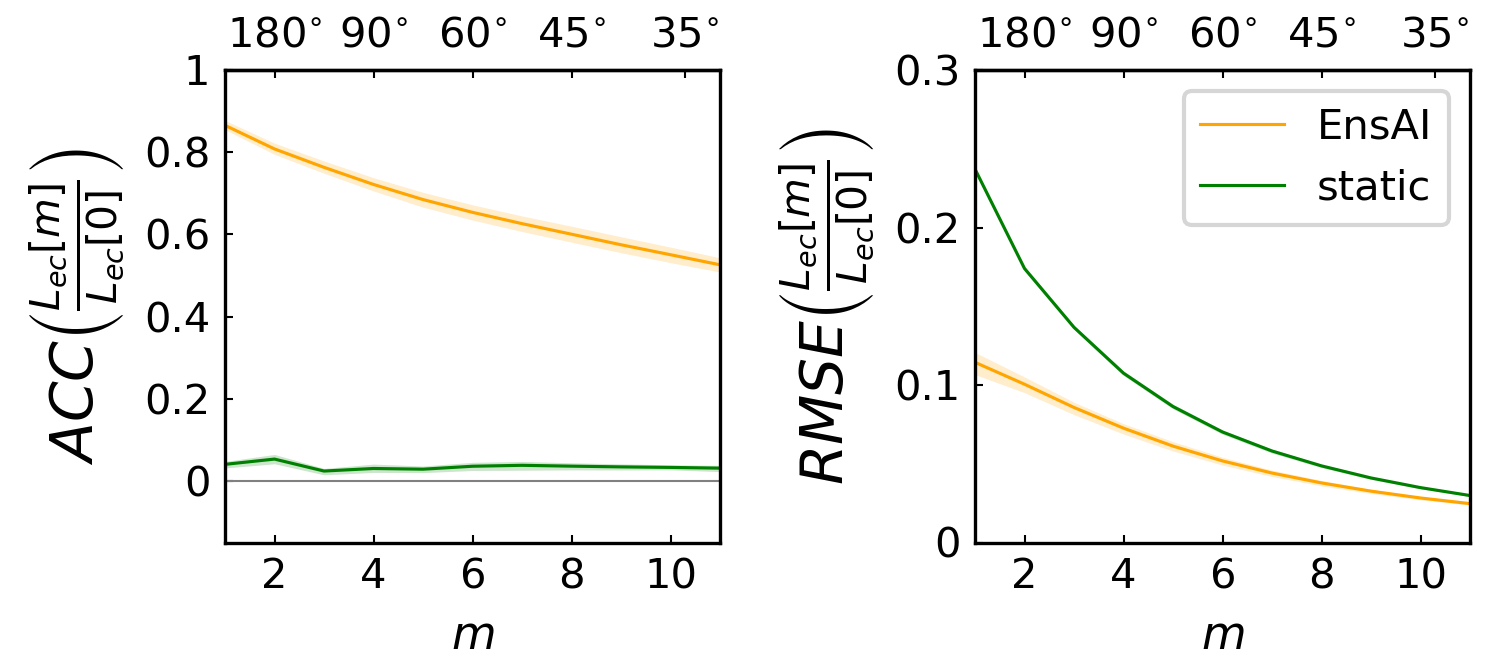}
\end{center}
\caption{$ACC$ (left panel) and $RMSE$ (right panel) values for the horizontal correlation length modes for the emissions/surface concentration correlation $L_{ec}[m]$ as a function of wavenumber $m$ for EnsAI (orange) and the static covariance (green) for June 2015. The wavelength of the mode is displayed on the top axis.}
\label{fig:lcftr_higher_modes}
\end{figure}

To examine the higher correlation length modes with $m>2$, Figure \ref{fig:lcftr_higher_modes} shows the $ACC$ and $RMSE$ values of $L_{ec}[m]/L_{ec}[0]$ for $1 \leq m \leq 11$ for June 2015. As seen in left panel of Fig. \ref{fig:lcftr_higher_modes}, the $ACC$ value for the EnsAI ensemble is above 0.85 for $m=1$ and decreases with $m$ to a value 0.56 by $m=10$. In contrast, the $ACC$ values for the static covariance show little change with wavenumber and is below 0.06 for all wavenumbers. From the right plot, we can see that the $RMSE$ values for both the EnsAI ensemble and static covariance decrease with wavenumber, as all three covariance models typically have less features on smaller scales (the decreasing values of $L_{ec}[m]$ with increasing $m$ that was shown in the example in Fig. \ref{fig:corrlen_example} is a typical feature of the horizontal correlations). The $RMSE$ values for EnsAI are smaller than those for the static covariance for all wavenumbers, but the difference between the two decrease with increasing values of $m$. $ACC$ and $RMSE$ values for the other months of the year look similar to those from June and can be found in Figures S10 and S11 of the Supplement.

Lastly, we mention that the performance of EnsAI is largely translationally invariant, in that only small changes are observed when the input fields are translated to a different location on EnsAI's grid (see Appendix \ref{sec:translational_invariance} for more details).

\section{Inversions}
\label{sec:inversions}

Sections \ref{sec:results_variances} and \ref{sec:results_correlations} showed the results of direct comparisons between the covariances to determine how well EnsAI replicated the GEM-MACH ensemble covariances. For some applications, such as model uncertainty quantification, the determination of the model's error covariance is the end goal. However, for inversions/assimilation, the objective is to produce an analysis (as well as forecasts initialized by the analysis in the case of assimilation). In this section, we compare emissions inversion results made using the different model error covariances, as well as a comparison of the posterior ammonia concentration fields to surface observations.

As the inversions are only sensitive to the model error covariances at times and locations near the available observations, not all differences between the error covariances will show in comparisons of the inversion results and will be dependent on the particular observations used. As such, pairing the comparison of inversion results with the direct comparison between the covariances in the previous section provides a robust examination between the different error covariances.

\subsection{Emissions Inversions}

\begin{figure}
\begin{center}
\noindent\includegraphics{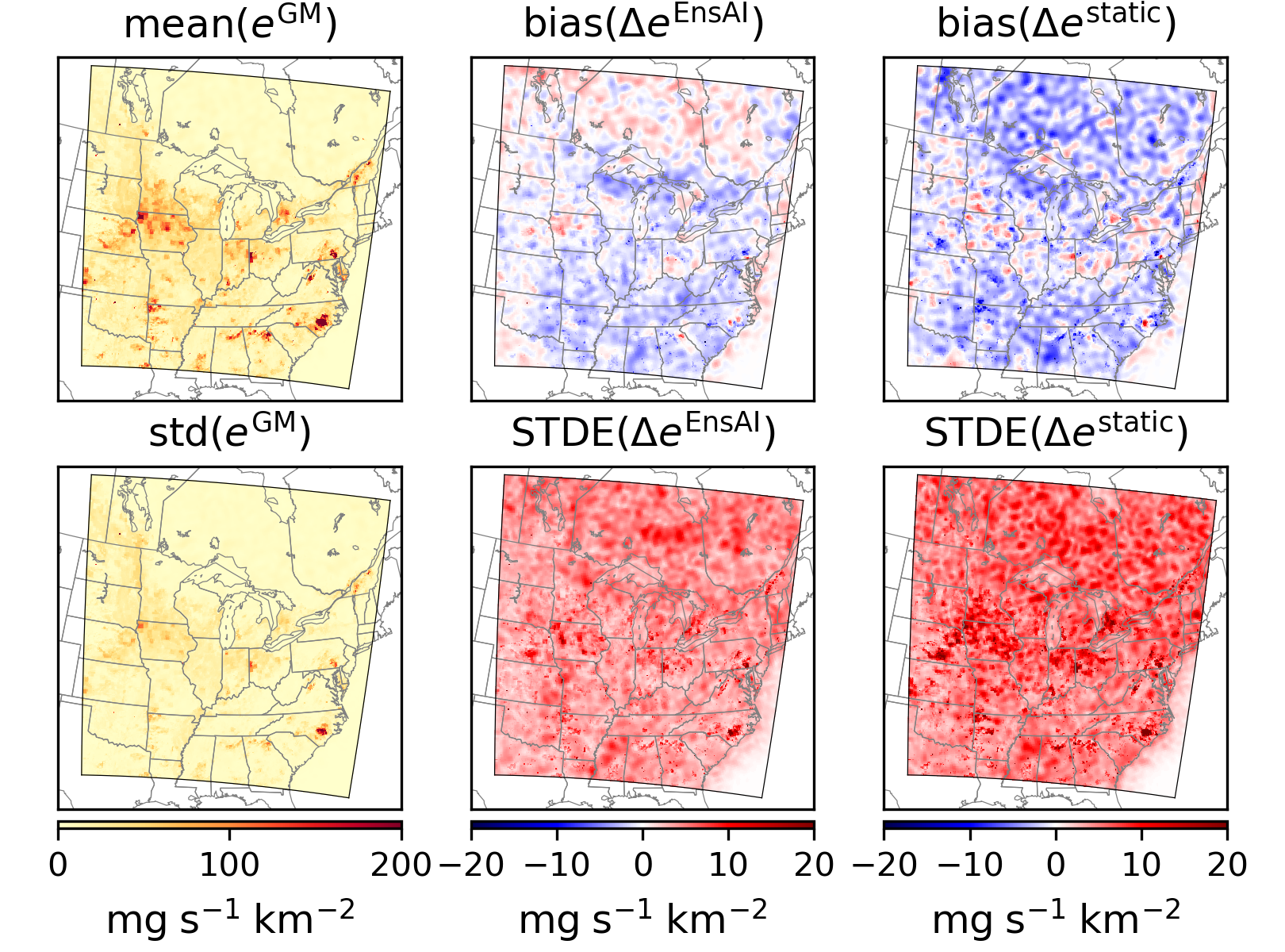}
\end{center}
\caption{Statistics for the ammonia emissions inversions taken over 2015. The left column shows the mean (top panel) and standard deviation (bottom panel) of the  emissions analyses from the inversions that used the GEM-MACH ensemble. The middle column shows the bias (top panel) and $STDE$ (bottom panel) for the emissions increments from the EnsAI inversions. The right column displays the bias and $STDE$ for the increments from the static covariances inversions.}
\label{fig:increment_annual_map}
\end{figure}

The ammonia emissions increments produced using the inversion scheme described in Section \ref{sec:emissions_inversions} using the GEM-MACH ensemble, EnsAI ensemble, and static error covariances (denoted by $\Delta\textbf{e}^{\rm{GM}}$, $\Delta\textbf{e}^{\rm{EnsAI}}$, and $\Delta\textbf{e}^{\rm{static}}$, respectively) can be found in Figures S12, S13, and S14 of the Supplement. Statistics of these increments taken over the year are shown in Figure \ref{fig:increment_annual_map}. The middle and right columns of the figure shows the bias and the standard deviation of errors ($STDE$) of the emissions increments from the EnsAI and static covariances inversions, respectively, which are defined in Section \ref{sec:comparison_statistics} of the Appendix. The left column shows the mean and standard deviation of the emissions analyses from the GEM-MACH ensemble inversions for comparison (for notational simplicity, we will not write the $^{\rm{a}}$ superscript for the emission analysis is this section). From this figure, we can see that the inversions using both EnsAI and the static covariance tend to underestimate the increments in many regions, although the extent and magnitude of this underestimation is larger for the static covariances (also see Figures S15 and S16 of the Supplement for plots of individual months). The $STDE$ values for the emissions increment using the static covariances is also noticeably larger than that for inversions using EnsAI in many regions.

\begin{figure}
\begin{center}
\noindent\includegraphics{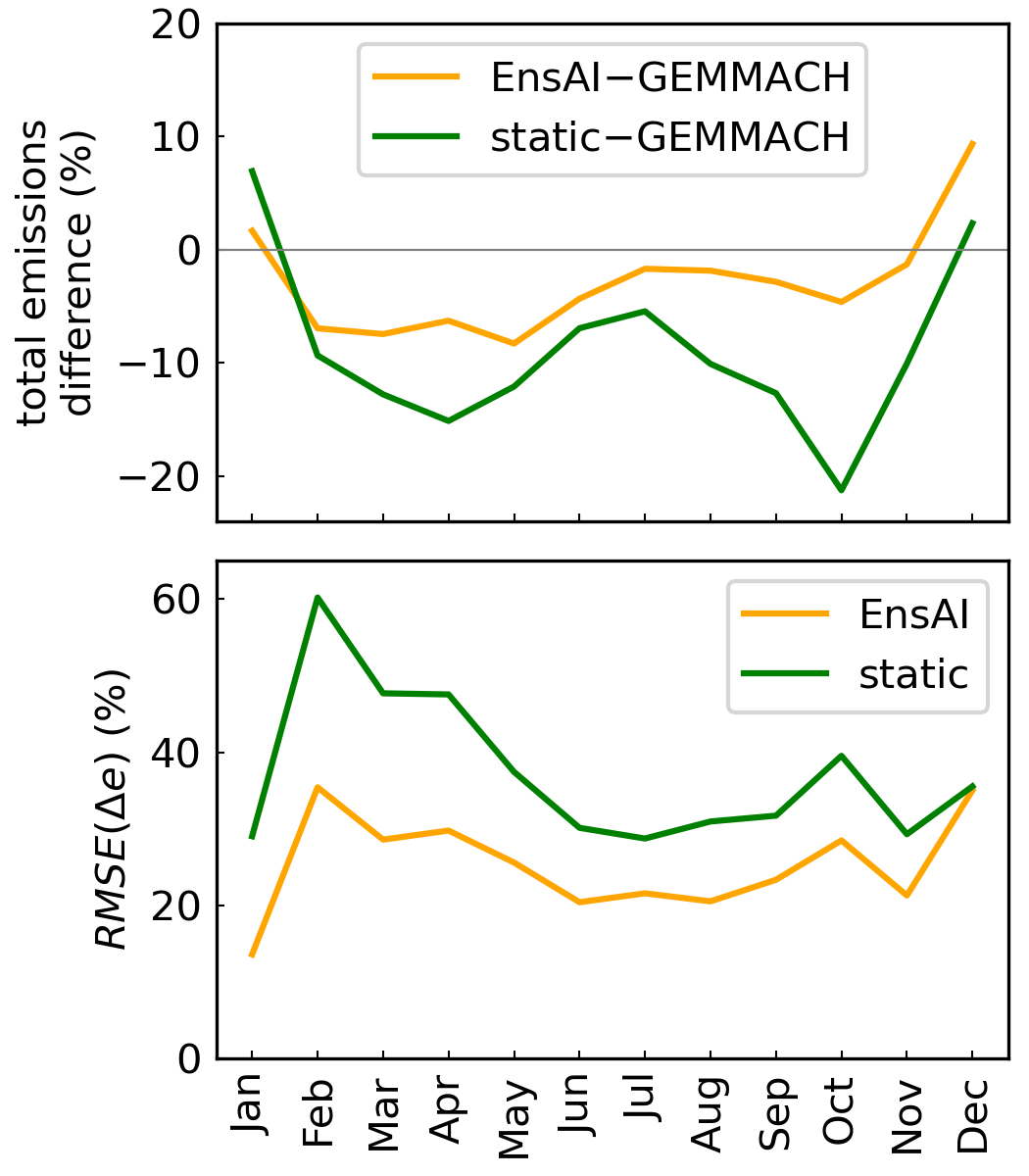}
\end{center}
\caption{Monthly comparison statistics for the ammonia emissions inversions between the GEM-MACH and EnsAI ensembles inversions (orange curves) and between the GEM-MACH ensemble and static covariance inversions (green curve). The top panel shows the differences between the total emissions within the domain. The bottom panel shows the root mean square error of the ammonia emissions increment $RMSE(\Delta \textbf{e})$.}
\label{fig:emissions_curves}
\end{figure}

The differences in the total ammonia emissions within the model domain for each month are shown in the top panel of Figure \ref{fig:emissions_curves}. These curves are expressed as percentages by dividing by the total monthly emissions (averaged over 2015) within the domain as determined by the GEM-MACH ensemble inversions. The monthly total emissions from the EnsAI inversions were within 10\% of those from the GEM-MACH ensemble inversions for all months. The monthly total emissions differences for the static covariance inversions exceeded a magnitude of 10\% for March to May and August to November and had larger differences than the EnsAI inversions for all months except December. The total monthly emissions from both the EnsAI and static covariance inversions were lower than those from the GEM-MACH ensemble inversions for February to November. The total annual emissions from the EnsAI inversions were 2.6\% lower than from the GEM-MACH ensemble inversions, but were 6.9\% lower for the static covariance inversions.

The root mean square error of the ammonia emissions increment $RMSE(\Delta \textbf{e})$ over the domain for each month are shown in the bottom panel of Fig. \ref{fig:emissions_curves} (expressed as percentages relative to the mean annual emissions given by the GEM-MACH ensemble inversions). The $RMSE(\Delta \textbf{e})$ values for EnsAI range from 14\% to 35\%, and between 29\% and 60\% for the static covariances. The $RMSE$ for the increments produced using EnsAI have lower values for all months as compared to the static covariances, although only slightly lower for December. 

\begin{figure}
\begin{center}
\noindent\includegraphics[width=0.7\textwidth]{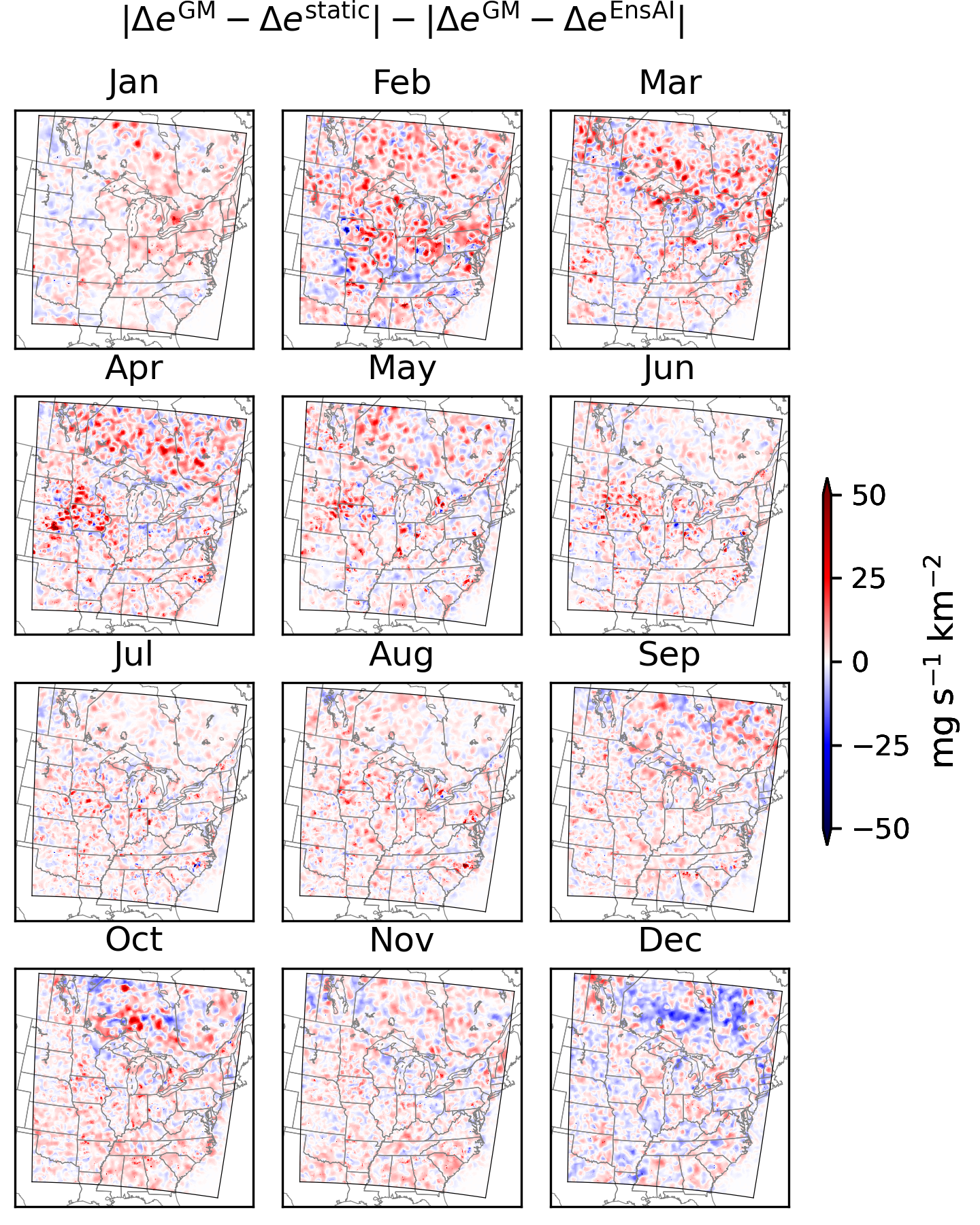}
\end{center}
\caption{Differences between ammonia emissions increments $\Delta\textbf{e}$ from inversions using different background error covariances. Plots show the deviation between the GEM-MACH ensemble and static covariance increments $|\Delta\textbf{e}^{\rm{GM}}-\Delta\textbf{e}^{\rm{static}}|$ minus the deviation between the GEM-MACH and EnsAI ensemble increments $|\Delta\textbf{e}^{\rm{GM}}-\Delta\textbf{e}^{\rm{EnsAI}}|$.}
\label{fig:increment_diff_map}
\end{figure}

To further examine the spatial dependency of the inversion results, Figure \ref{fig:increment_diff_map} shows the deviations between the GEM-MACH ensemble and static covariance increments $|\Delta\textbf{e}^{\rm{GM}}-\Delta\textbf{e}^{\rm{static}}|$ minus the deviation between the GEM-MACH and EnsAI ensemble increments $|\Delta\textbf{e}^{\rm{GM}}-\Delta\textbf{e}^{\rm{EnsAI}}|$ for each month. In this figure, locations where the EnsAI increment is closer to the GEM-MACH ensemble increment are shown in red and locations where the static covariance increment is closer to the GEM-MACH ensemble increment are shown in blue. For most months, there are many locations where the EnsAI increment is closer to the GEM-MACH ensemble increment by between 25 and 50 $\rm{mg}\;\rm{s}^{-1}\;\rm{km}^{-2}$ (see Fig. \ref{fig:mean_emissions} for comparison to the annual mean emissions). With the exception of December, more locations have the GEM-MACH ensemble increment closer to the EnsAI increment than to the static covariance increment. In December, there are larger regions where the static covariance increment is closer to the GEM-MACH ensemble increment, most notably in Ontario and Quebec, although $RMSE(\Delta \textbf{e})$ for EnsAI is still slightly lower than for the static covariance in December.

\subsection{Posterior Surface Concentrations}

GEM-MACH run with the inversion-derived emissions parameters produce the posterior ammonia surface concentrations. The posterior surface concentrations for the GEM-MACH ensemble inversions $\textbf{c}^{\rm{GM}}$ has monthly means that generally ranges from 0 to 12 ppbv, with concentrations highest between April to September (see Figure S17 of the Supplement). The posterior surface concentrations for the EnsAI and static covariance inversions ($\textbf{c}^{\rm{EnsAI}}$ and $\textbf{c}^{\rm{static}}$, respectively) have biases that are generally within 2 ppbv and vary between over- and under-prediction depending on the location (see Figures S18 and S19 of the Supplement). Additionally, a number of areas switch between over- and under-prediction depending on the month for both the EnsAI and static covariance inversions.

\begin{figure}
\begin{center}
\noindent\includegraphics[width=0.7\textwidth]{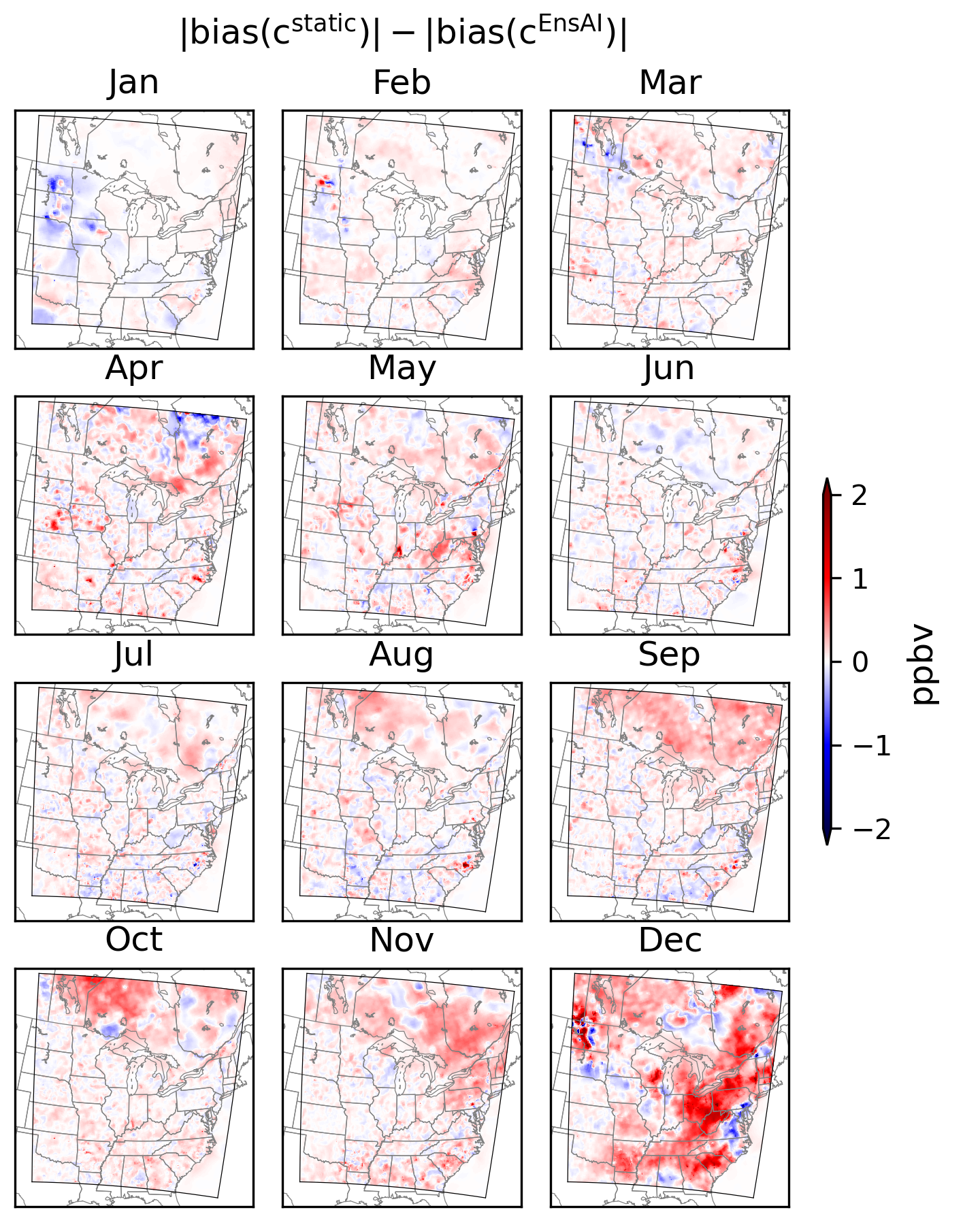}
\end{center}
\caption{Differences in the magnitude of the bias of the posterior ammonia surface concentrations between the EnsAI inversions and the static covariance inversions.}
\label{fig:posterior_concentrations_rmse_diff_map}
\end{figure}

Figure \ref{fig:posterior_concentrations_rmse_diff_map} shows the differences in the magnitudes of the bias of the posterior surface concentration between the EnsAI inversions and the static covariance inversions, where regions in red indicate that the bias for EnsAI was lower than for the static covariances, and vice versa for regions in blue. From this figure, we can see that there are significantly more regions where the bias for EnsAI is lower than for the static covariances, and a number of regions have the EnsAI bias 1-2 ppbv lower than the static covariances bias. Similar figures for the $STDE$ (Figures S20, S21, and S22) show that the EnsAI inversions also produce surface concentrations that have lower $STDE$ values in most areas of the domain.

\subsection{Comparison to Surface Observations}

We now compare the posterior ammonia surface concentrations to observations from the AMoN and NAPS networks. The monthly biases for individual stations from inversions using the GEM-MACH ensemble range from -6 to +8 ppbv, with over-predictions often found near eastern coastal regions and under-predictions more likely to be found closer to the center of the continent (see Figure S23 of the Supplement). The biases for individual stations from the EnsAI inversions range from being 0.5 ppbv lower to 1.3 ppbv higher than the biases from the GEM-MACH ensemble inversions (see Figure S24 of the Supplement). The biases from the static covariance inversions are between 0.9 ppbv lower to 2 ppbv higher than that from the GEM-MACH ensemble inversions (see Figure S25 of the Supplement). For both EnsAI and the static covariances, the biggest differences in bias values occur in December.

\begin{figure}
\begin{center}
\noindent\includegraphics{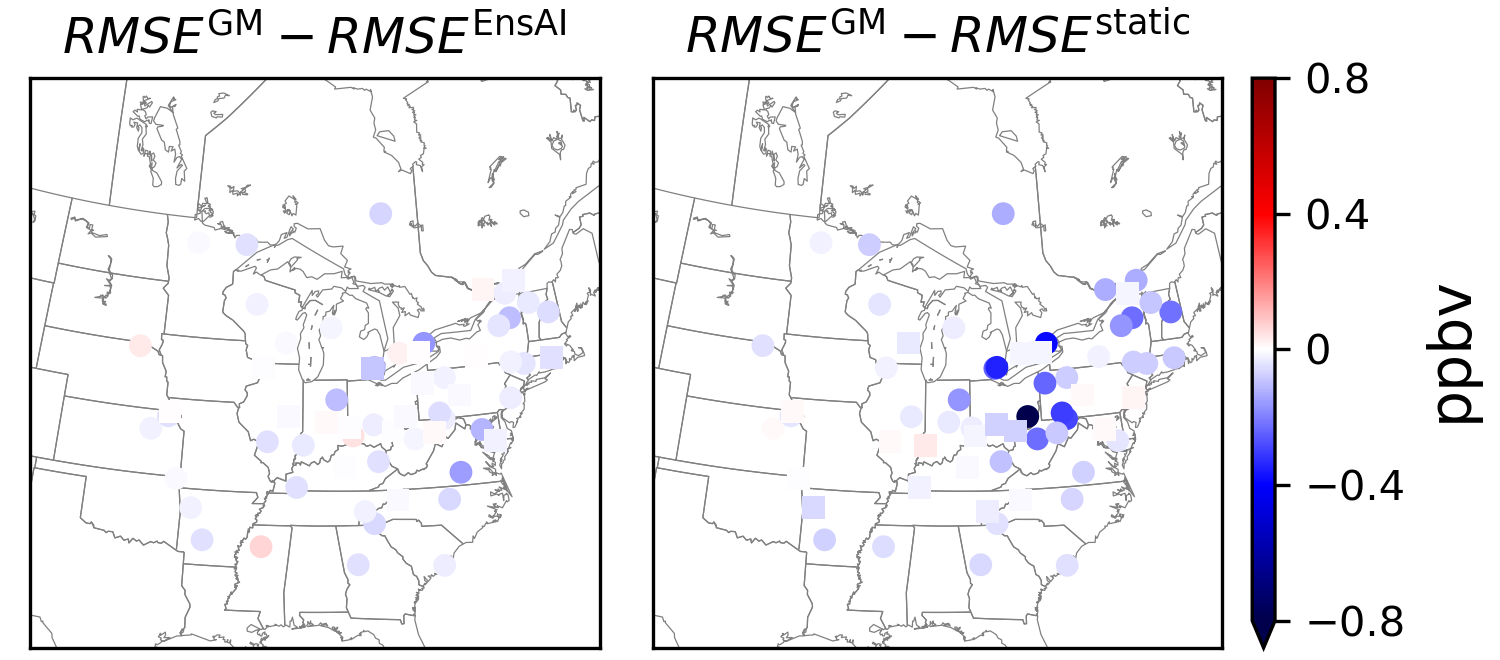}
\end{center}
\caption{Differences in $RMSE$ values for AMoN and NAPS observations for 2015. The left panel shows the differences between $RMSE$ values for the GEM-MACH ensemble ($RMSE^{\rm{GM}}$) and EnsAI ($RMSE^{\rm{EnsAI}}$), while the right panel shows the differences between the GEM-MACH ensemble and the static covariance inversions $RMSE^{\rm{static}}$. Statistically significant/insignificant differences at the 1$\sigma$ confidence level are shown with a circular/square markers.}
\label{fig:surfaceobs_rmse_annual_amon_naps}
\end{figure}

Figure \ref{fig:surfaceobs_rmse_annual_amon_naps} shows differences in the $RMSE$ taken over 2015 for each station, where differences between the GEM-MACH and EnsAI ensembles are shown in the left panel and differences between the GEM-MACH ensemble and the static covariances are shown in the right panel. In this figure, red markers indicate the stations where EnsAI/the static covariances have lower $RMSE$ values than for the GEM-MACH ensemble, and vice versa for the blue markers. While 38 stations had a higher $RMSE$ for EnsAI as compared to the GEM-MACH ensemble, the increases in $RMSE$ values were small, with 0.16 ppbv being the largest increase. EnsAI had lower $RMSE$ values for four stations, where the reductions in $RMSE$ were similarly small, with the largest reduction being 0.07 ppbv (the remaining 19 stations did not have a statistically significant change in the $RMSE$). For the static covariances, 41 stations had an increase in their $RMSE$, but in this case increases were as high as 0.81 ppbv. One station had a lower $RMSE$ for the static covariances, with a decrease of only 0.01 ppbv. For stations that showed statistically significant differences, 25 stations had lower $RMSE$ values for EnsAI than for the static covariances and four stations had lower $RMSE$ values for the static covariances than for EnsAI.

Differences in the surface observations comparison statistics taken over the whole domain are shown in Figure \ref{fig:surfaceobs_rmse_annual_amon_naps}. Each plot shows the change in the magnitude of the bias, $STDE$, or correlation coefficient ($r$, defined in Section \ref{sec:comparison_statistics} of the Appendix) for each month. For each comparison statistic, each difference is composed such that a positive value indicates that EnsAI or the static covariances have improved the statistic relative to the GEM-MACH ensemble inversion, while a negative value indicates a degradation. In these plots, the bias and $STDE$ are expressed as percentages relative to the annual mean observation value taken over all stations.

The monthly biases taken over all stations for the GEM-MACH ensemble are between -6\% and 43\% (see Table S1 of the Supplement for all comparison statistic values). For EnsAI, the change in magnitude of the bias as compared to the GEM-MACH ensemble range from -5\% to +4\% for January to November (top panel of Fig. \ref{fig:surfaceobs_rmse_annual_amon_naps}). During these months, the static covariances changed the bias magnitude by between -6\% and +8\%. For December, the bias is degraded significantly for both EnsAI and the static covariances, although the degradation is much more severe for the static covariances (-53\% as compared to -19\%). For February, May, September, and November, the bias from EnsAI is lower than from the static covariances by 2\% to 5\%. Both EnsAI and the static covariances improve the bias in October, with a decrease of 4\% for EnsAI and a decrease of 8\% for the static covariances, and is the only month where the static covariances significantly improve the bias as compared to EnsAI.

The monthly $STDE$ values for the GEM-MACH ensemble are between 43\% and 182\%. EnsAI noticeably improves the $STDE$ in January, July, and October as compared to the GEM-MACH ensemble, but degrades the $STDE$ April, August, November, and December (middle panel of Fig. \ref{fig:surfaceobs_rmse_annual_amon_naps}). The static covariances improve the $STDE$ in October, but degrade it in January, April, August, September, November, and December. The $STDE$ values from EnsAI are lower than for the static covariances in January, April, July, September, and December, and the converse is true for August and October.

Lastly, only small changes are seen in the correlation coefficient $r$ (bottom panel of Fig. \ref{fig:surfaceobs_rmse_annual_amon_naps}). Nevertheless, EnsAI does have slightly higher correlation coefficients values than the static covariances for all months except February (although the statistical significance of this difference is low for a number of these months).

\begin{figure}
\begin{center}
\noindent\includegraphics{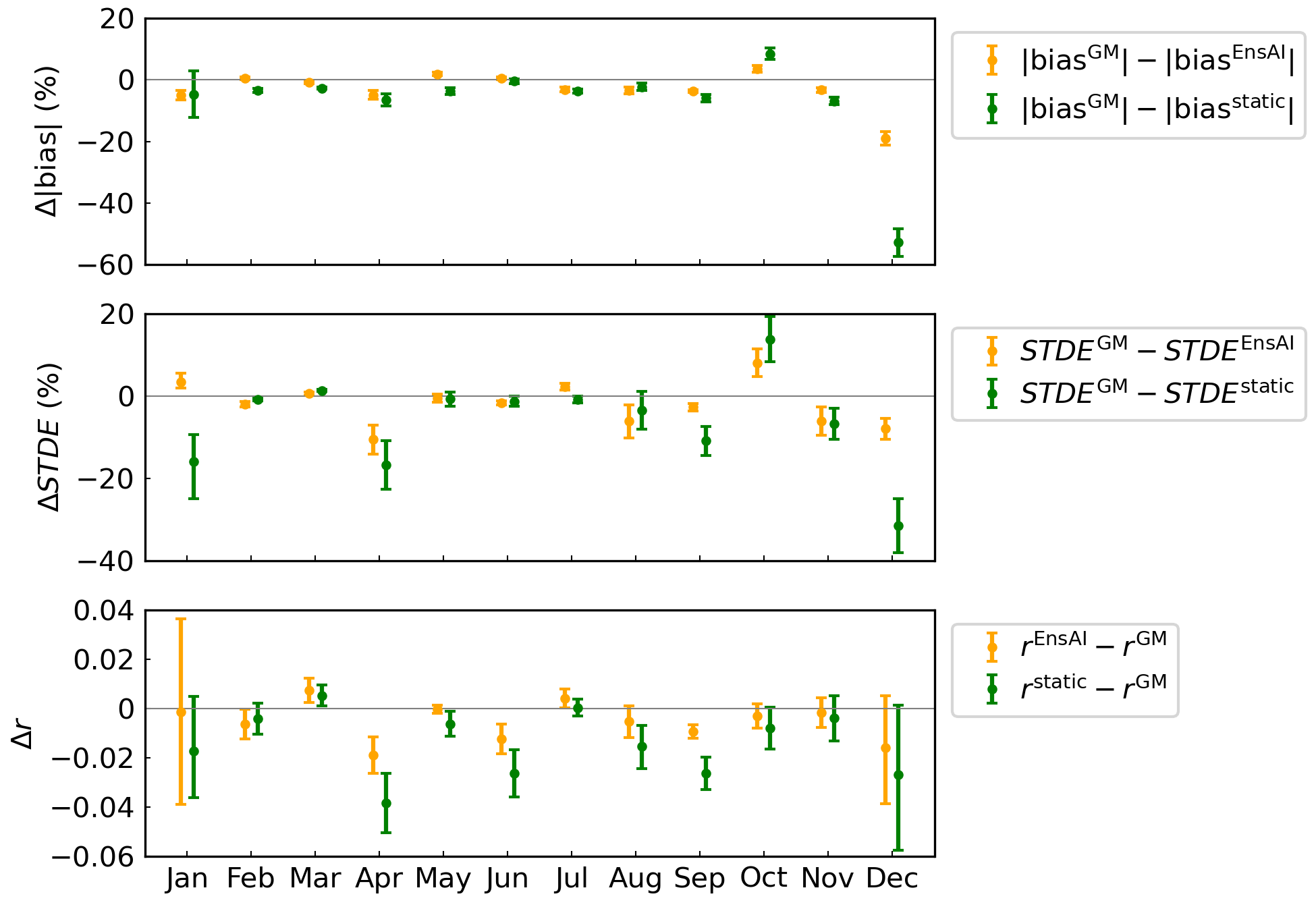}
\end{center}
\caption{Changes in surface observation comparison statistics for the combined set of AMoN and NAPS observations for 2015.}
\label{fig:surfaceobs_statdiff_amon_naps}
\end{figure}

\section{Conclusions}

Obtaining flow-dependent model error covariances is an important part of modern data assimilation and inversion systems. Ensemble-based methods have become a popular choice for adding flow-dependence to the error covariances as it does not require the use of a model adjoint that can be time consuming to create and maintain. The downside however is the need to repeatedly run the model to generate the ensemble, which for atmospheric models can be a significant burden on computational resources.

In this work we introduced EnsAI, an AI-based ensemble generation model for atmospheric surface ammonia that can run many orders of magnitude faster than traditional physics-based models. When trained on an ensemble generated using the GEM-MACH air quality forecasting model, EnsAI can accurately reproduce many features of the original ensemble. By comparing the original GEM-MACH and EnsAI ensembles to a static error covariance, it was shown that including flow-dependence in the error covariances has a considerable effect on the covariances and that EnsAI can encode much of the meteorology-dependent features of the original ensemble. Furthermore, it was demonstrated that including only surface information in EnsAI was sufficient to reproduce most of the time-dependent features of the ensemble, thereby allowing EnsAI to produce accurate ensemble information for ammonia with much smaller memory requirements as compared to systems that require multiple vertical levels.

In emissions inversion systems, accurate emissions/concentration spatial correlations are necessary for proper spatial source attribution of observations. As such, capturing the time-dependency of these correlations is important for successful inversions. It was demonstrated that EnsAI-based emissions inversions produced very similar results to inversions performed using the GEM-MACH ensemble. Being able to generate ensembles in a fraction of the time required for physics-based models significantly eases the computational cost of running such inversion systems, thereby making it much more feasible to run frequent inversions.

EnsAI was trained on a year's worth of data from the GEM-MACH regional model, which contained data only from North America. Although this exposed EnsAI to regions with a variety of different land-use categories, agricultural types, and meteorological conditions during training, the training data may have not included drastically different meteorological conditions or chemical regimes than those seen in North America during the year used for training. As such, until further tests can be conducted demonstrating its performance in different regions, EnsAI in its present form should be considered to be restricted to its current spatial domain. Additionally, if future meteorological or chemical conditions change significantly from those in the training period, continued training (or fine-tuning) on data from future years might be necessary as this data becomes available to maintain EnsAI's accuracy.

While EnsAI was successfully implemented in an emissions inversion system for ammonia, a species with a short atmospheric lifetime, expanding EnsAI to work with longer lived species will require expanding the input capacity to the network. Most significantly, longer lived species will require EnsAI to model a prognostic relationship, which would add fields from previous time steps to the input. Input and output fields for the vertical level above the surface would also be required for long lived species.

\section*{Code and Data Availability}

The source code for EnsAI, as well as the code used for generating the evaluation statistics, is available from \url{https://doi.org/10.5281/zenodo.15342768} under the Creative Commons Attribution-NonCommercial-NoDerivatives 4.0 International license. GEM-MACH version 3.1.0 is available from \url{https://doi.org/10.5281/zenodo.8346780}, under the GNU Lesser General Public License. The GEM-MACH training data is available at \url{https://doi.org/10.5281/zenodo.18843196} and \url{https://doi.org/10.5281/zenodo.18843210}, the validation data is available at \url{https://doi.org/10.5281/zenodo.18820049}, and the test data is available at \url{https://doi.org/10.5281/zenodo.18825231} and \url{https://doi.org/10.5281/zenodo.18839599}, all under the Creative Commons Attribution 4.0 International license. The CrIS ammonia retrieval data used for evaluation is available at \url{https://doi.org/10.5281/zenodo.18865369}, under the Creative Commons Attribution 4.0 International license. AMoN and NAPS observations are available at \url{https://nadp.slh.wisc.edu/networks/ammonia-monitoring-network} and \url{https://data-donnees.az.ec.gc.ca/data/air/monitor/national-air-pollution-surveillance-naps-program}, respectively.

\section*{Author Contributions}

MS wrote the manuscript, developed the EnsAI model, and performed all analysis.

\section*{Competing Interests}

The author declares that he has no conflict of interest.

\appendix

\section*{Acknowledgments}

The author would like to thank Colin Lee for discussions on AI and GPU computing, Mark Shephard for providing advice on utilizing the CrIS ammonia retrievals, and the GEM-MACH development team for the creation and maintenance of the GEM-MACH model. The author would also like to thank Colin Lee, Dominique Brunet, Vikram Khade, and Mark Buehner for providing feedback on a draft of this manuscript.

\section*{Appendix}

\section{Ensemble-Variational Emissions Inversion}
\label{sec:inversion_details}

Emissions inversions were performed using a variational algorithm, in which a cost function is minimized to find the optimal combination between a priori (background) and observational information. The cost function is minimized with respect to the increment $\Delta\textbf{x}$, which when added to the background state $\textbf{x}^{\rm{b}}$ yields the analysis $\textbf{x}^{\rm{a}} = \textbf{x}^{\rm{b}} + \Delta\textbf{x}$. In our case, the cost function $J$ is given by
\begin{equation}
J = \frac{1}{2}\Delta\textbf{x}^{\mathrm{T}}\textbf{B}^{-1}\Delta\textbf{x} + \frac{1}{2}(\textbf{d}-\textbf{H}\Delta\textbf{x})^{\mathrm{T}}\textbf{R}^{-1}(\textbf{d}-\textbf{H}\Delta\textbf{x}) ,
\label{eq:costfnc}
\end{equation}
where $\textbf{d}=\textbf{y}-H(\textbf{x}^{\mathrm{b}})$ is the difference between the observations $\textbf{y}$ and the background in observation space. $H$ is the (nonlinear) observation operator that maps the model state into observation space, while $\textbf{H}$ denotes its linearization. The first term in the right hand side of Equation (\ref{eq:costfnc}) quantifies the deviation from the background state and is weighted by the inverse of the background error covariance $\textbf{B}$. Similarly, the second term in the right hand side of Eq. (\ref{eq:costfnc}) quantifies the deviation from the observations and is weighted by the inverse of the observation error covariance $\textbf{R}$.

In the case of an emissions inversion where the unobserved emissions $\textbf{e}$ are constrained using observations of the atmospheric concentration $\textbf{c}$, using Eq. (\ref{eq:bmatrix}), the cost function becomes
\begin{equation}
J = \frac{1}{2}\Delta\textbf{e}^{\mathrm{T}}\textbf{B}_{ee}^{-1}\Delta\textbf{e} + \frac{1}{2}(\textbf{d}-\textbf{H}\Delta\textbf{c})^{\mathrm{T}}\textbf{R}^{-1}(\textbf{d}-\textbf{H}\Delta\textbf{c}) ,
\end{equation}
where $\Delta\textbf{e}$ and $\Delta\textbf{c}$ are the emissions and atmospheric concentration increments, respectively. Since we assume that uncertainties in the atmospheric concentrations are due solely to the uncertainties in emissions, the emissions and concentration increments are related to each other by
\begin{equation}
\Delta\textbf{e} = \textbf{B}_{ec} \textbf{B}_{cc}^{-1} \Delta\textbf{c} .
\end{equation}
It is useful to rewrite the cost function in terms of a control vector $\boldsymbol\chi$, which defines the control vector transforms
\begin{subequations}
\begin{equation}
\Delta\textbf{c}=\textbf{B}_{cc}^{1/2}\boldsymbol\chi ,
\end{equation}
\begin{equation}
\Delta\textbf{e}=\textbf{B}_{ee}^{1/2}\boldsymbol\chi ,
\end{equation}
\label{eq:cvt}
\end{subequations}
where $\textbf{B}_{cc}^{1/2}$ and $\textbf{B}_{ee}^{1/2}$ are the square roots of $\textbf{B}_{cc}$ and $\textbf{B}_{ee}$, respectively. The cost function can then be expressed in terms of the control vector as
\begin{equation}
J(\boldsymbol\chi) = \frac{1}{2}\boldsymbol\chi^{\mathrm{T}}\boldsymbol\chi + \frac{1}{2}(\textbf{d}-\textbf{H}\textbf{B}_{cc}^{1/2}\boldsymbol\chi)^{\mathrm{T}}\textbf{R}^{-1}(\textbf{d}-\textbf{H}\textbf{B}_{cc}^{1/2}\boldsymbol\chi) .
\end{equation}
The emissions inversion then consists of minimizing this cost function with respect to the control vector $\boldsymbol\chi$, followed by the transformation of the optimal control vector to the emissions (or concentration) increment given by Equation (\ref{eq:cvt}).

Ensemble-variational methods are a class of variational algorithms that model the background error covariance using an ensemble and is the inversion method used in this study. As mentioned in Section \ref{sec:inversion_methods}, covariances that are formed solely from ensemble perturbations will in general have spurious correlations and low rank due to the limited size of the ensemble. As such, localization of the ensemble covariances is included, as shown in Eq. (\ref{eq:bens}). With this, $\textbf{B}_{cc}^{1/2}$ and $\textbf{B}_{ee}^{1/2}$ can then be expressed as
\begin{subequations}
\begin{equation}
\textbf{B}_{cc}^{1/2} = \frac{1}{\sqrt{N-1}} \begin{bmatrix} \mathrm{diag}(\delta\textbf{c}_1)\mathcal{L}^{1/2} & \dots & \mathrm{diag}(\delta\textbf{c}_{N})\mathcal{L}^{1/2} \end{bmatrix} ,
\end{equation}
\begin{equation}
\textbf{B}_{ee}^{1/2} = \frac{1}{\sqrt{N-1}} \begin{bmatrix} \mathrm{diag}(\delta\textbf{e}_1)\mathcal{L}^{1/2} & \dots & \mathrm{diag}(\delta\textbf{e}_{N})\mathcal{L}^{1/2} \end{bmatrix} ,
\end{equation}
\end{subequations}
where $\{ \delta\textbf{e}_i \}$ and $\{ \delta\textbf{c}_i \}$ are the emissions and concentration perturbation ensembles, respectively, $\mathrm{diag}(\cdot)$ is the diagonalization operator, and $\mathcal{L}^{1/2}$ is the square root of the localization matrix. Here, we have chosen the same localization matrix for both the emissions and concentrations, which was chosen to be isotropic with a half-width at half-maximum of 120 km with the shape prescribed by the fifth-order function of \citet{gaspari1999}.

\section{Network Architecture, Training, and Inference}

\subsection{Components of the U-Net}
\label{sec:unet_details}

Here we describe the different components that comprise the U-Net used in EnsAI. All references to graphical representations of the U-Net components refer to Fig. \ref{fig:unet}.

Residual blocks \citep{he2016} (represented by the blue right arrows) are used throughout the network and consist of two sequential convolutional layers with a residual connection. Each convolutional layer is comprised by (in sequential order) a $3\times 3$ two-dimensional convolution, a group normalization \citep{wu2018}, and a Sigmoid Linear Unit (SiLU) \citep{hendrycks2016} activation function.

In the encoding branch, the residual blocks are used to increase the number of channels, while a $2\times 2$ max pooling (represented by the orange downward arrows) is used to half each spatial dimension. In the decoding branch, upscaling blocks (denoted by the green upward arrows), consisting of a nearest neighbor upscaling followed by a $3\times 3$ two-dimensional convolution, are used to double each spatial dimension and half the number of channels. Skip connections (represented by the grey dashed arrows), comprised of a copy and concatenation, are made across the encoding and decoding branches. A final two-dimensional convolution is made at the end of the decoding branch (denoted by the purple right arrow) to reduce the array to a single channel that represents the surface ammonia concentration perturbation.

\subsection{Network Training}
\label{sec:training_details}

Training was done on a single NVIDIA A100 GPU. The network was trained using the Adam optimizer with a batch size of 32 and learning rate of $10^{-4}$. The learning curve for the training is plotted in Figure S1 of the Supplement. The weights that produced the minimum validation loss were selected for the evaluation, which occurred at the 9th epoch. Selecting the weights that yielded the minimum validation loss reduces the risk of overfitting that might have occurred in the later epochs of the training.

\subsection{Inference}
\label{sec:inference_details}

As EnsAI describes a diagnostic relationship, inference can be done in parallel and any number of different time steps could in principle be processed simultaneously. For the 7 second run time to produce a week's worth of simulated data per ensemble member that was quoted in Section \ref{sec:runtime}, inferences were performed in batches of 16. 

\section{Details of Correlation Length Decomposition}
\label{sec:corrlen_details}

With the decomposition of the correlation length defined in Equation (\ref{eq:lcftr_def}), the first three modes are given by
\begin{equation}
L[0] = \sqrt{\frac{2}{\pi}} \int l^2 \varrho(l,\phi) dl d\phi ,
\label{eq:L0}
\end{equation}
\begin{equation}
\begin{split}
L[1] & = \sqrt{\frac{2}{\pi}} \int (\cos(\phi) + i\sin(\phi)) l^2 \varrho(l,\phi) dl d\phi \\
& = \sqrt{\frac{2}{\pi}} \int  (x + iy) \varrho(x,y) dx dy ,
\end{split}
\label{eq:L1}
\end{equation}
\begin{equation}
\begin{split}
L[2] & = \sqrt{\frac{2}{\pi}} \int (\cos(2\phi) + i\sin(2\phi)) l^2 \varrho(l,\phi) dl d\phi \\
& = \sqrt{\frac{2}{\pi}} \int (\cos^2(\phi) - \sin^2(\phi) + i\sin(\phi)\cos(\phi)) l^2 \varrho(l,\phi) dl d\phi \\
& = \sqrt{\frac{2}{\pi}} \int (x^2 - y^2 + ixy) \frac{\varrho(x,y)}{\sqrt{x^2 + y^2}} dx dy ,
\end{split}
\label{eq:L2}
\end{equation}
where in these equations the correlations are computed between the origin and the point $(x, y)$, which when expressed in polar coordinates is a distance $l$ from the origin at an angle $\phi$ from the $x$-axis. 

Examining the integral in Equation (\ref{eq:L0}), at each angle $\phi$, the integral over $l$ is equal to the (one-sided) variance of $\varrho$ in that particular direction. The integral over $\phi$ then takes the angle-averaged value of this one-sided spatial variance. As an illustration of this quantity, consider the spatial correlation function of a 2D isotropic Gaussian $\rho_G$, given by
\begin{equation}
\rho_G(x,y) = \exp \left( -\frac{x^2 + y^2}{2 L_G^2} \right) ,
\end{equation}
where $L_G$ is the scale parameter of the Gaussian function, so that its normalized spatial correlation $\varrho_G$ is
\begin{equation}
\varrho_G(x,y) = \frac{1}{2 \pi L_G^2} \exp \left( -\frac{x^2 + y^2}{2 L_G^2} \right) .
\end{equation}
In this case, $L[0]$ is given by
\begin{equation}
\begin{split}
L[0] & =  \frac{1}{\sqrt{2}\pi^{3/2} L_G^2} \int_0^{\infty} l^2 \exp \left( -\frac{l^2}{2 L_G^2} \right) dl \int_0^{2\pi} d\phi \\
& = \frac{\sqrt{2}}{\sqrt{\pi} L_G^2} \int_0^{\infty} l^2 \exp \left( -\frac{l^2}{2 L_G^2} \right) dl \\
& = L_G ,
\end{split}
\end{equation}
which shows that $L[0]$ is equal to the Gaussian scale parameter $L_G$ in this example.

From the second line of Equation (\ref{eq:L1}), it is apparent that the real and imaginary parts of the $L[1]$ are proportional to the dipole moments of $\varrho$ in the $x$ and $y$ directions, respectively. Functions that are symmetric about the x-axis (y-axis) will have vanishing real (imaginary) parts of $L[1]$.

The third line of Equation (\ref{eq:L2}) shows that $L[2]$ is related to the quadrupole moments of the function $\varrho(x,y)/\sqrt{x^2+y^2}$, where the real part is proportional to the difference between the $x,x$ and $y,y$ quadrupole components and the imaginary part is proportional to the $x,y$ component.

\section{Comparison Statistics}
\label{sec:comparison_statistics}

For a parameter $\theta$, which could represent a covariance parametrization variable or a field produced by the inversions, the bias, the standard deviation of errors $STDE$, the root mean square error $RMSE$, the anomaly correlation coefficient $ACC$, and Pearson correlation coefficient $r$, are defined by
\begin{equation}
\rm{bias} = \bar{\theta}^{\rm{mod}} - \bar{\theta}^{\rm{ref}} ,
\end{equation}
\begin{equation}
STDE = \sqrt{ \frac{1}{N-1} \sum_i^N (\theta_i^{\rm{mod}} - \bar{\theta}^{\rm{mod}} - \theta^{\rm{ref}}_i  + \bar{\theta}^{\rm{ref}})^2 } ,
\end{equation}
\begin{equation}
RMSE = \sqrt{ \frac{1}{N} \sum_i^N | \theta_i^{\rm{mod}} - \theta_i^{\rm{ref}} |^2 } ,
\end{equation}
\begin{equation}
ACC = \frac{\sum_i^N \left( \theta_i^{\rm{ref}} - \theta_i^{\rm{c}} - \bar{\theta}^{\rm{ref}} + \bar{\theta}^{\rm{c}} \right) \left( \theta_i^{\rm{mod}} - \theta_i^{\rm{c}} - \bar{\theta}^{\rm{mod}} + \bar{\theta}^{\rm{c}} \right)}{\sqrt{\sum_i^N ( \theta_i^{\rm{ref}} - \theta_i^{\rm{c}} - \bar{\theta}^{\rm{ref}} + \bar{\theta}^{\rm{c}} )^2} \sqrt{\sum_i^N ( \theta_i^{\rm{mod}} - \theta_i^{\rm{c}} - \bar{\theta}^{\rm{mod}} + \bar{\theta}^{\rm{c}} )^2 }} ,
\label{eq:acc}
\end{equation}
\begin{equation}
r = \frac{\sum_i^N (\theta_i^{\rm{mod}} - \bar{\theta}^{\rm{mod}})(\theta_i^{\rm{ref}} - \bar{\theta}^{\rm{ref}})}{\sqrt{\sum_i^N (\theta_i^{\rm{mod}}-\bar{\theta}^{\rm{mod}})^2}\sqrt{\sum_i^N (\theta_i^{\rm{ref}}-\bar{\theta}^{\rm{ref}})^2}} ,
\end{equation}
where $i$ indexes each quantity (in space and/or time) with a total of $N$ elements, and overbars denote the mean of that quantity. In the equations above, the superscripts `mod' and `ref' stand for `model' and `reference', respectively. For comparisons between covariance parameters or inversion fields, `mod' will refer to either EnsAI or the static covariances, while `ref' refers to the GEM-MACH ensemble. For comparisons to surface observations, `mod' will stand for any one of the covariance models (EnsAI ensemble, GEM-MACH ensemble, or static covariances) and `ref' refers to the surface observations. In Eq. (\ref{eq:acc}), the superscript `c' refers to a climatological (or other reference) value, which for this work we take as the two-month mean value derived from the GEM-MACH ensemble for 2016. 

\section{Translational Invariance of the Network}
\label{sec:translational_invariance}

While convolutional neural networks are in principle translationally invariant, it has been shown that in some situations, particular choices of network architecture can allow the network to deduce positional information \citep{kayhan2020}. As a test of EnsAI's translational invariance, the input fields from the leftmost 90 grid points were translated to the 90 rightmost grid points and input into EnsAI. In this translation, a point in Nebraska is translated to the point `(d)' in Fig. \ref{fig:mean_emissions} in Pennsylvania. Only small differences were observed in the covariance parameters after this translation (see Figure S26 of the Supplement), suggesting that the network is largely translationally invariant.

\bibliographystyle{plainnat}
\bibliography{biblio}

\pagebreak
\begin{center}
\textbf{\large Supplement to `EnsAI: An Emulator for Atmospheric Ammonia Ensembles'}
\end{center}
\setcounter{section}{0}
\setcounter{equation}{0}
\setcounter{figure}{0}
\setcounter{table}{0}
\setcounter{page}{1}
\makeatletter
\renewcommand \thesection{S\@arabic\c@section}
\renewcommand{\theequation}{S\arabic{equation}}
\renewcommand{\thefigure}{S\arabic{figure}}
\renewcommand{\bibnumfmt}[1]{[S#1]}
\renewcommand{\citenumfont}[1]{S#1}

\begin{figure}[h]
\begin{center}
\noindent\includegraphics{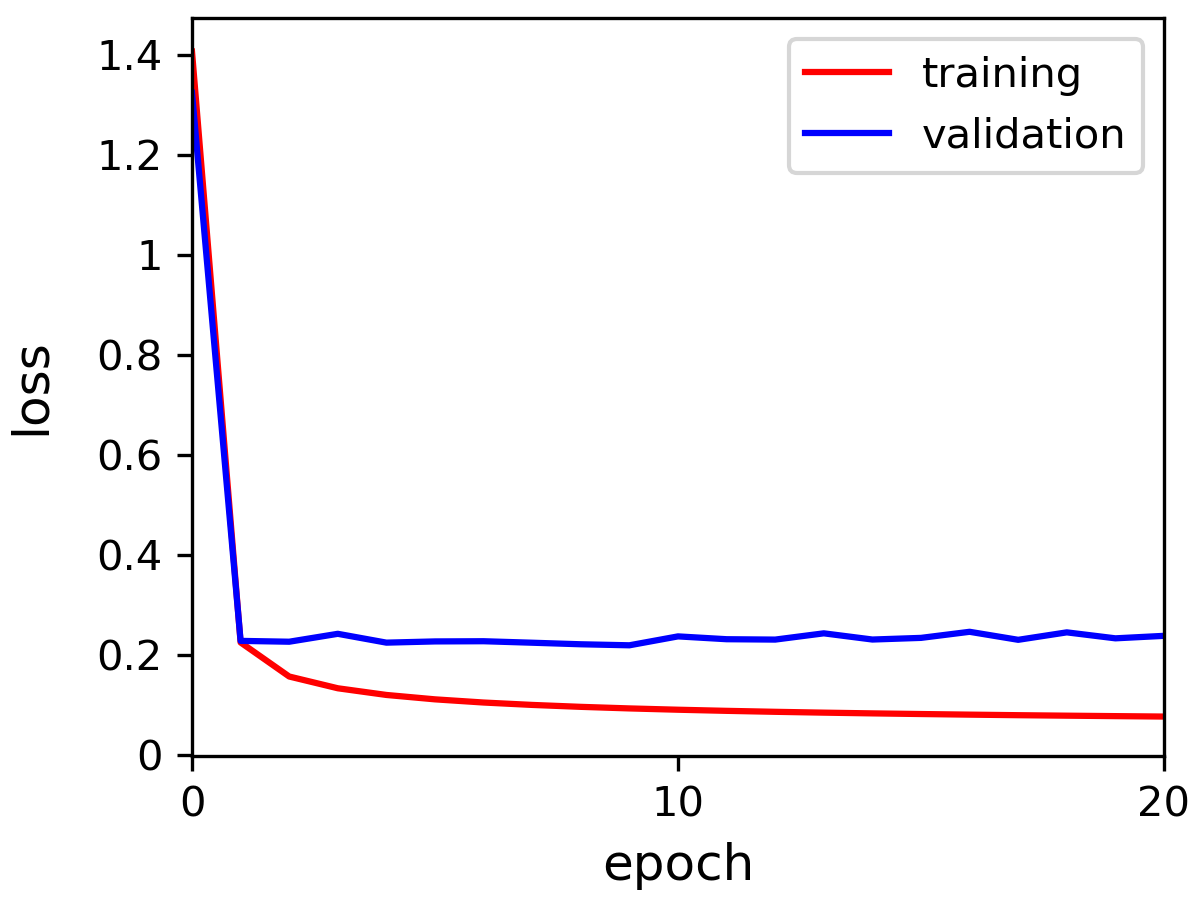}
\end{center}
\caption{The learning curve produced during the training of the EnsAI model.}
\end{figure}

\begin{figure}
\begin{center}
\noindent\includegraphics{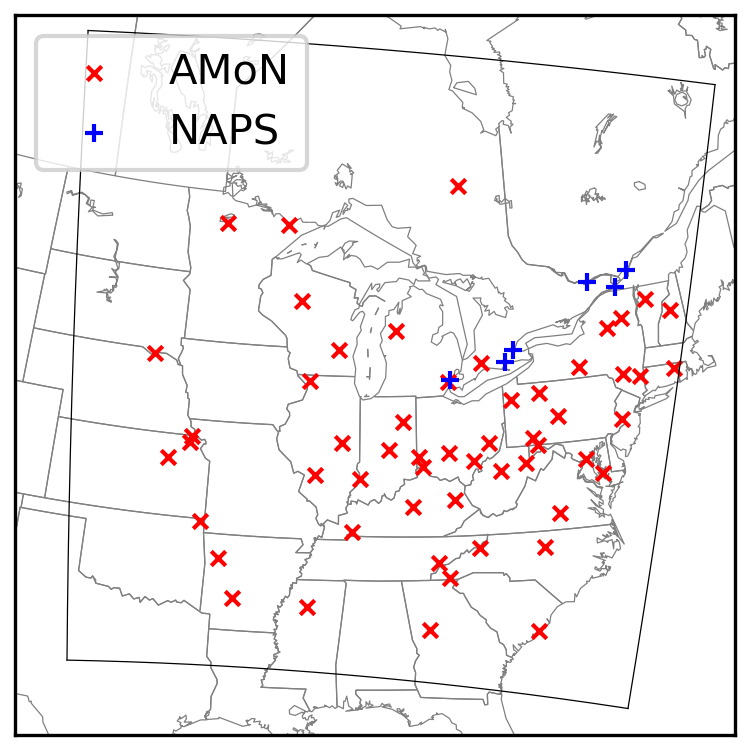}
\end{center}
\caption{AMoN and NAPS station locations within the EnsAI model domain that have valid observations for the test period (2015).}
\end{figure}

\begin{figure}
\begin{center}
\noindent\includegraphics{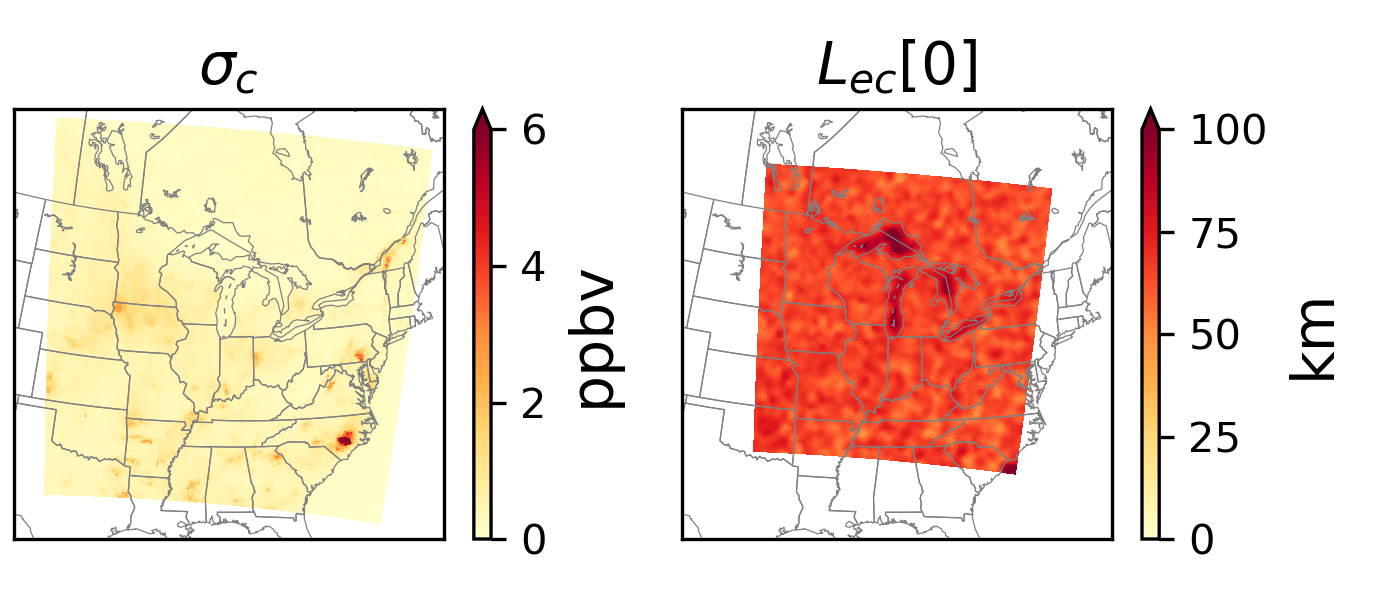}
\end{center}
\caption{Time means of the surface ammonia concentration error standard deviation $\sigma_c$ and emissions/surface concentration isotropic correlation length $L_{ec}[0]$ of the GEM-MACH ensemble ammonia error covariance taken over 2015.}
\label{fig:stats_ref_map}
\end{figure}

\begin{figure}
\begin{center}
\noindent\includegraphics{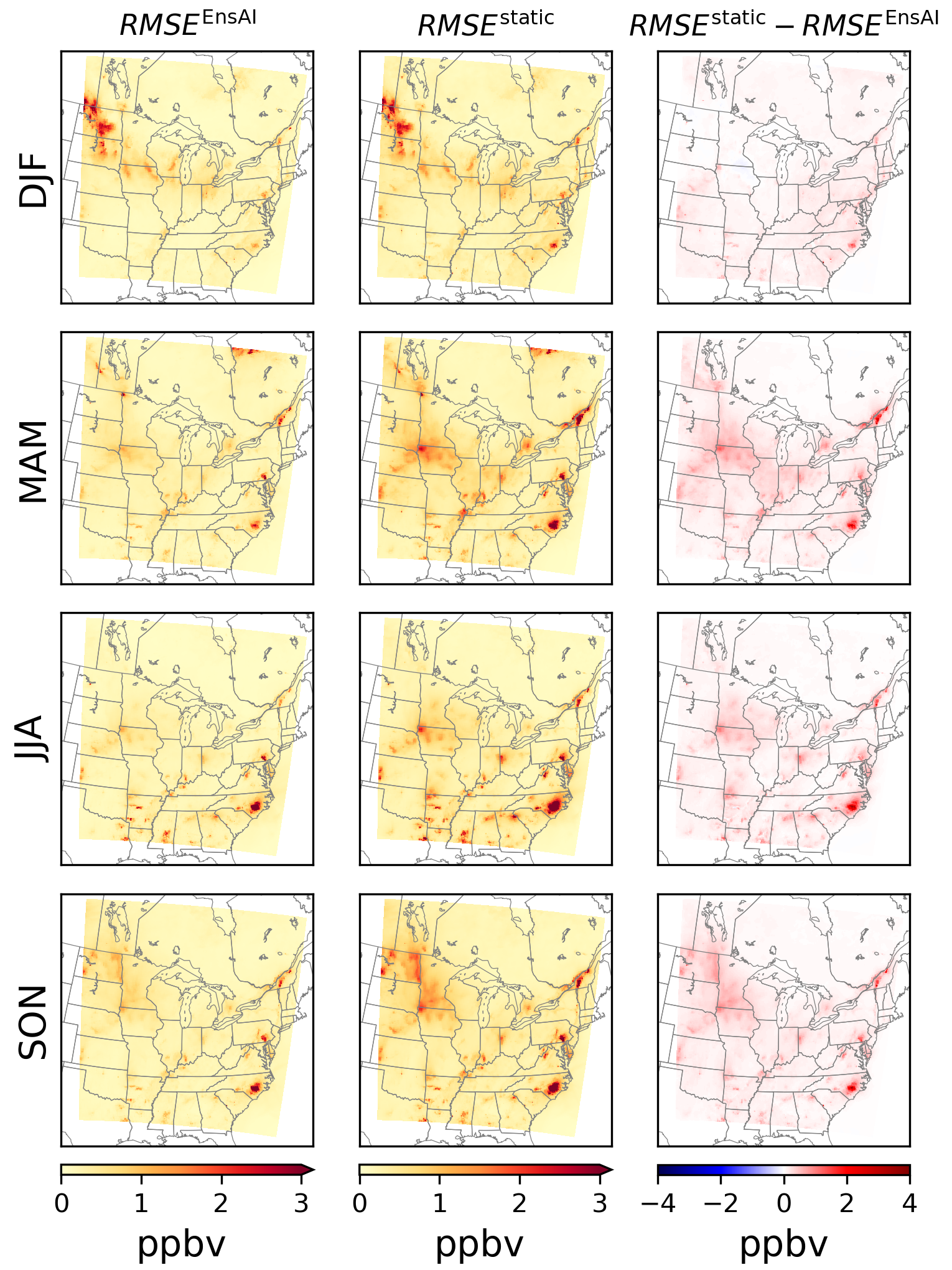}
\end{center}
\caption{$RMSE$ values for the surface ammonia concentration error standard deviation $\sigma_c$ for EnsAI (left column), the static error covariances (middle column), and their difference (right column) for December-January-February (DJF), March-April-May (MAM), June-July-August (JJA), and September-October-November (SON) of 2015.}
\label{fig:map_std_rmse}
\end{figure}

\begin{figure}
\begin{center}
\noindent\includegraphics[width=0.8\textwidth]{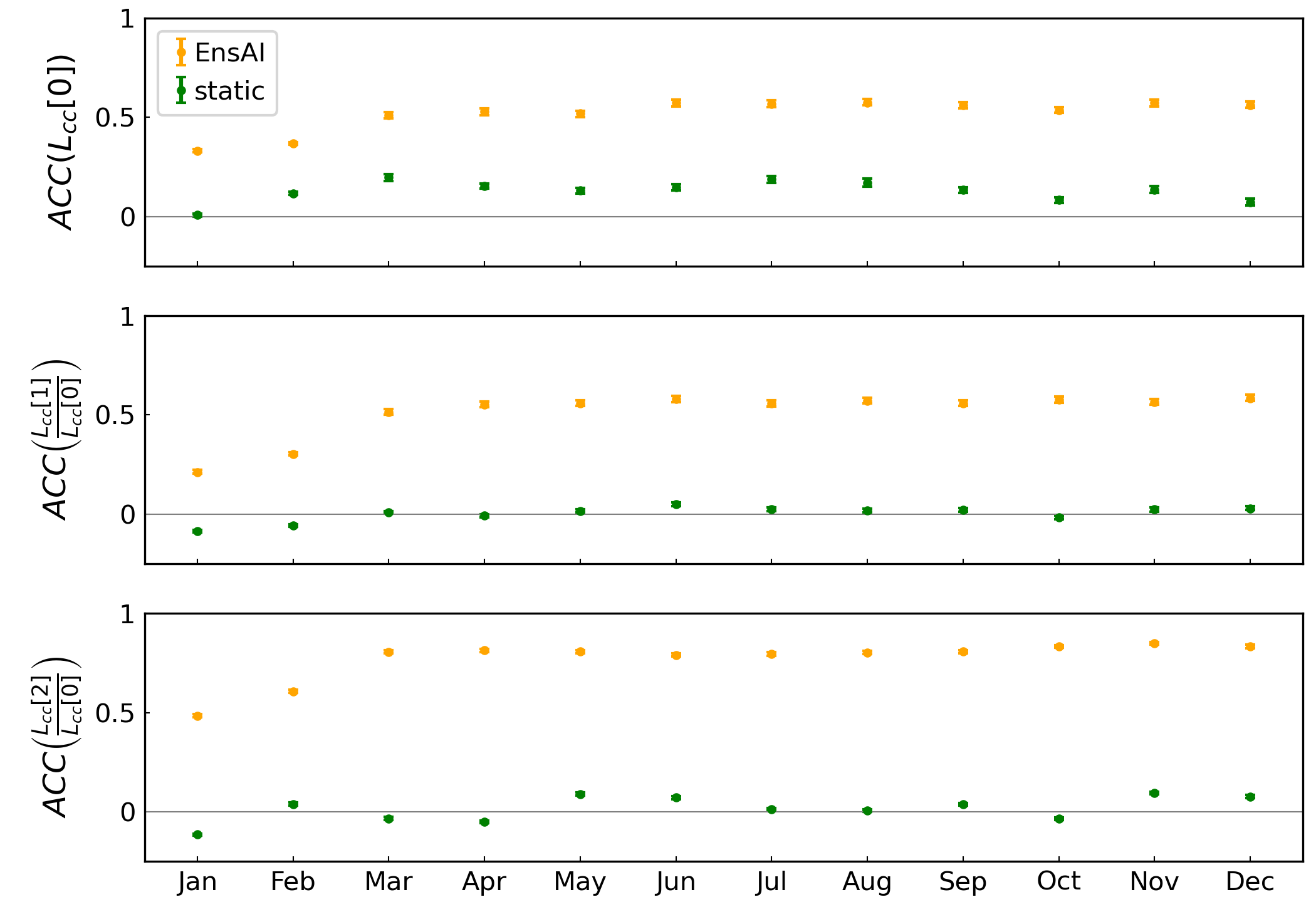}
\end{center}
\caption{$ACC$ values for $L_{cc}[0]$, $L_{cc}[1]/L_{cc}[0]$, and $L_{cc}[2]/L_{cc}[0]$ over the model domain for each month in 2015.}
\end{figure}

\begin{figure}
\begin{center}
\noindent\includegraphics{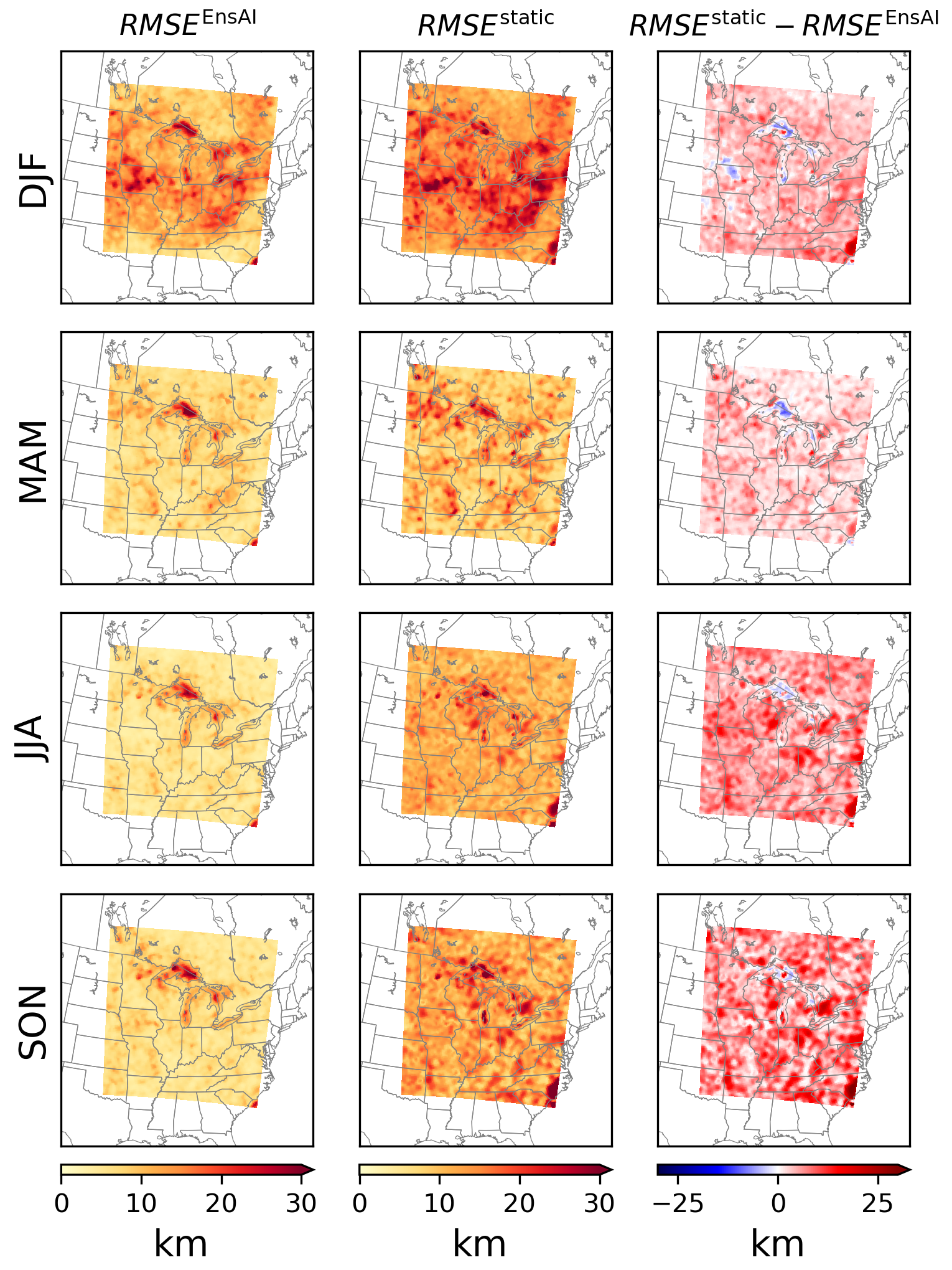}
\end{center}
\caption{Same as Figure \ref{fig:map_std_rmse} but for $RMSE(L_{ec}[0])$.}
\label{fig:map_Lec_m0_rmse}
\end{figure}

\begin{figure}
\begin{center}
\noindent\includegraphics{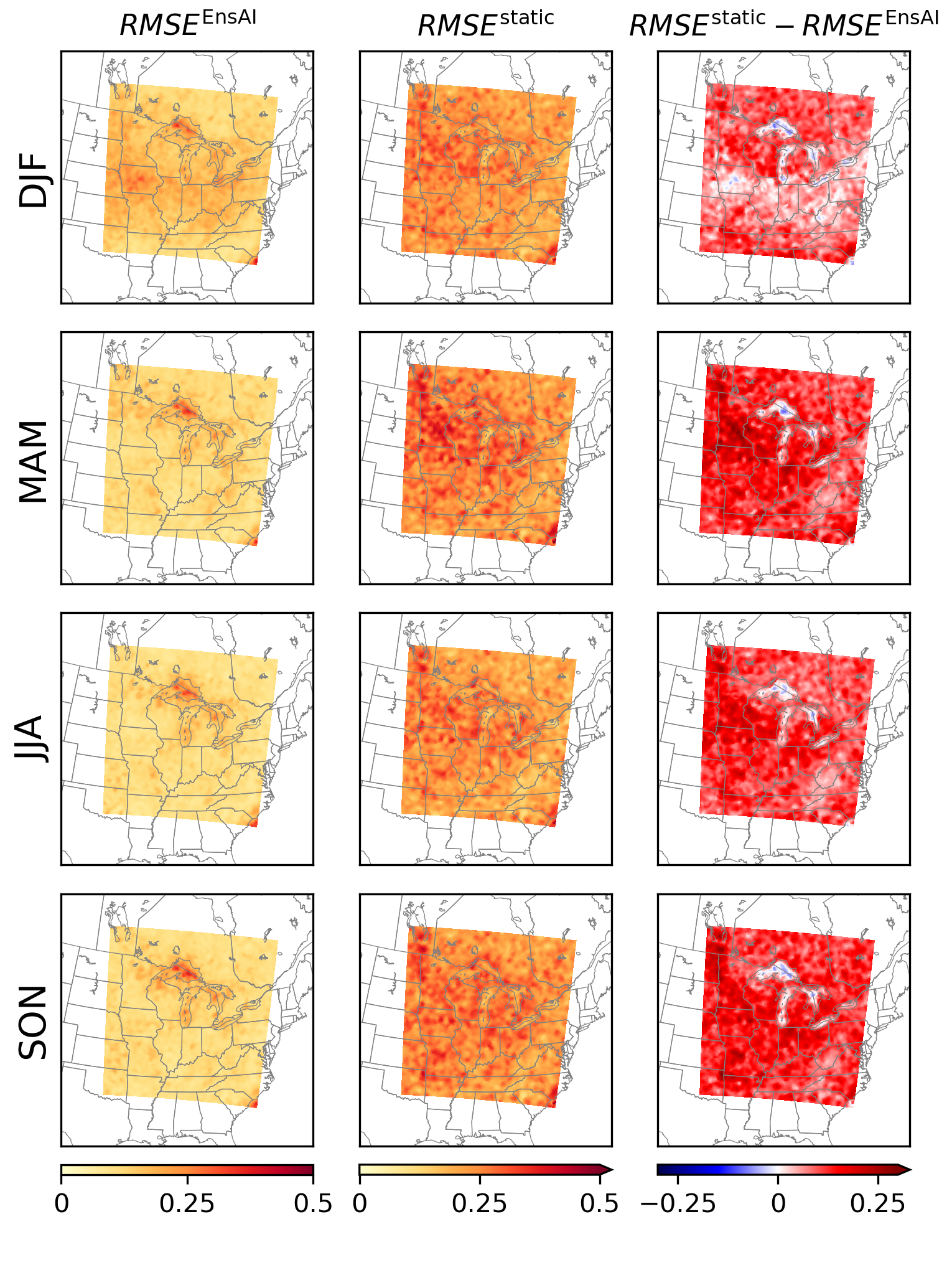}
\end{center}
\caption{Same as Figure \ref{fig:map_std_rmse} but for $RMSE(L_{ec}[1]/L_{ec}[0])$.}
\label{fig:map_Lec_m1_rmse}
\end{figure}

\begin{figure}
\begin{center}
\noindent\includegraphics{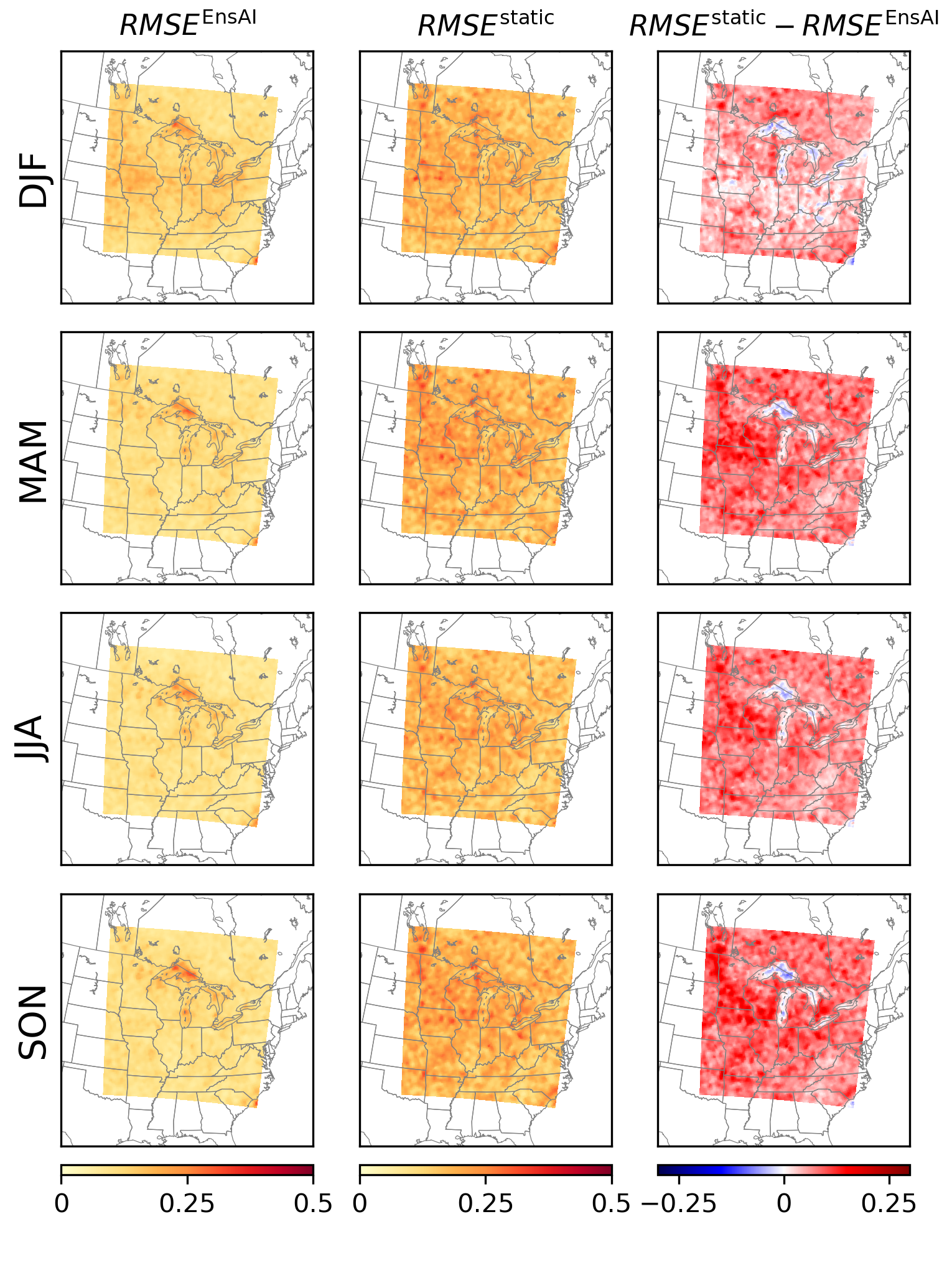}
\end{center}
\caption{Same as Figure \ref{fig:map_std_rmse} but for $RMSE(L_{ec}[2]/L_{ec}[0])$.}
\label{fig:map_Lec_m2_rmse}
\end{figure}

\begin{figure}
\begin{center}
\noindent\includegraphics[width=0.8\textwidth]{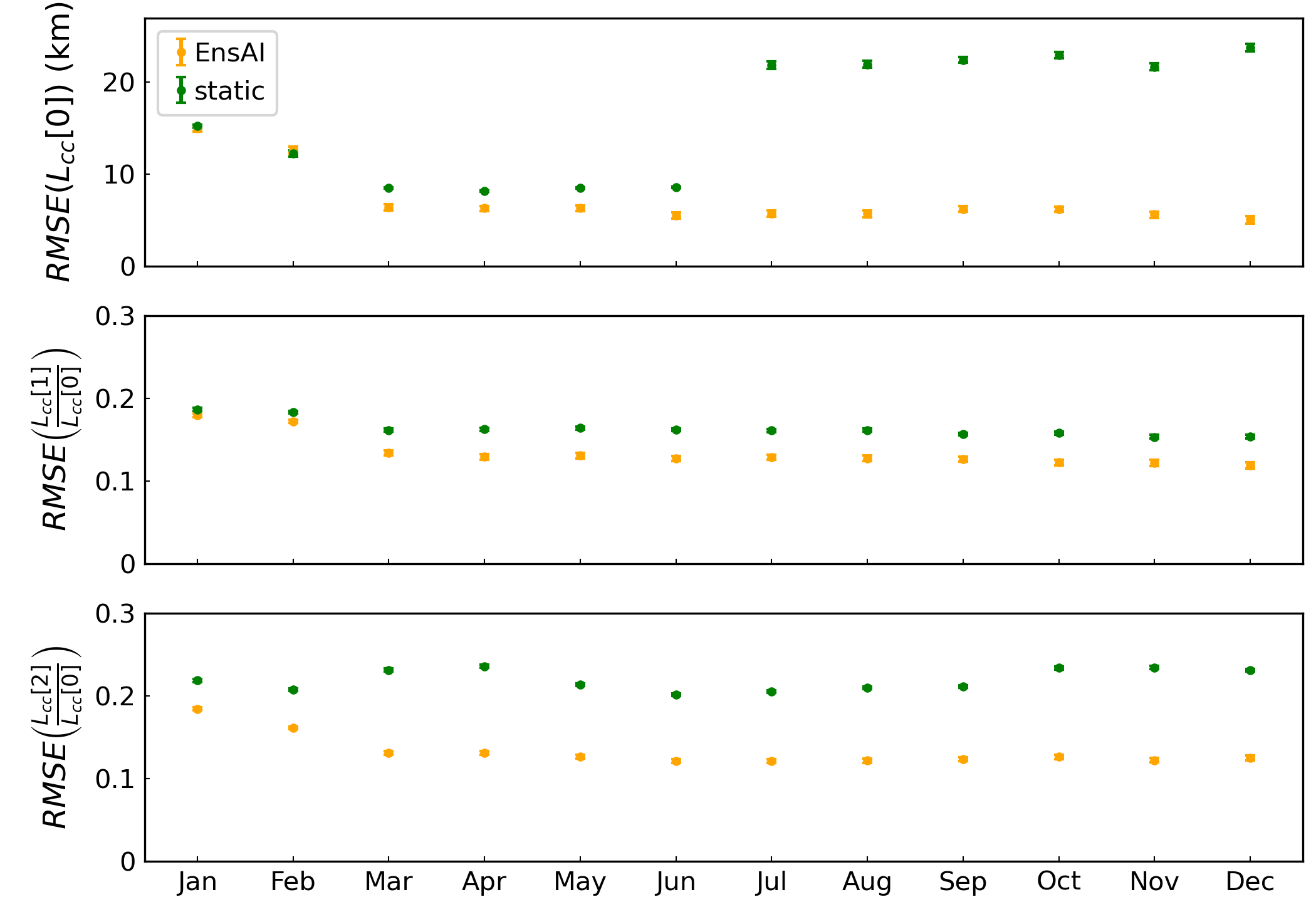}
\end{center}
\caption{$RMSE$ values for $L_{cc}[0]$, $L_{cc}[1]/L_{cc}[0]$, and $L_{cc}[2]/L_{cc}[0]$ over the model domain for each month in 2015.}
\end{figure}

\begin{figure}
\begin{center}
\noindent\includegraphics[scale=0.75]{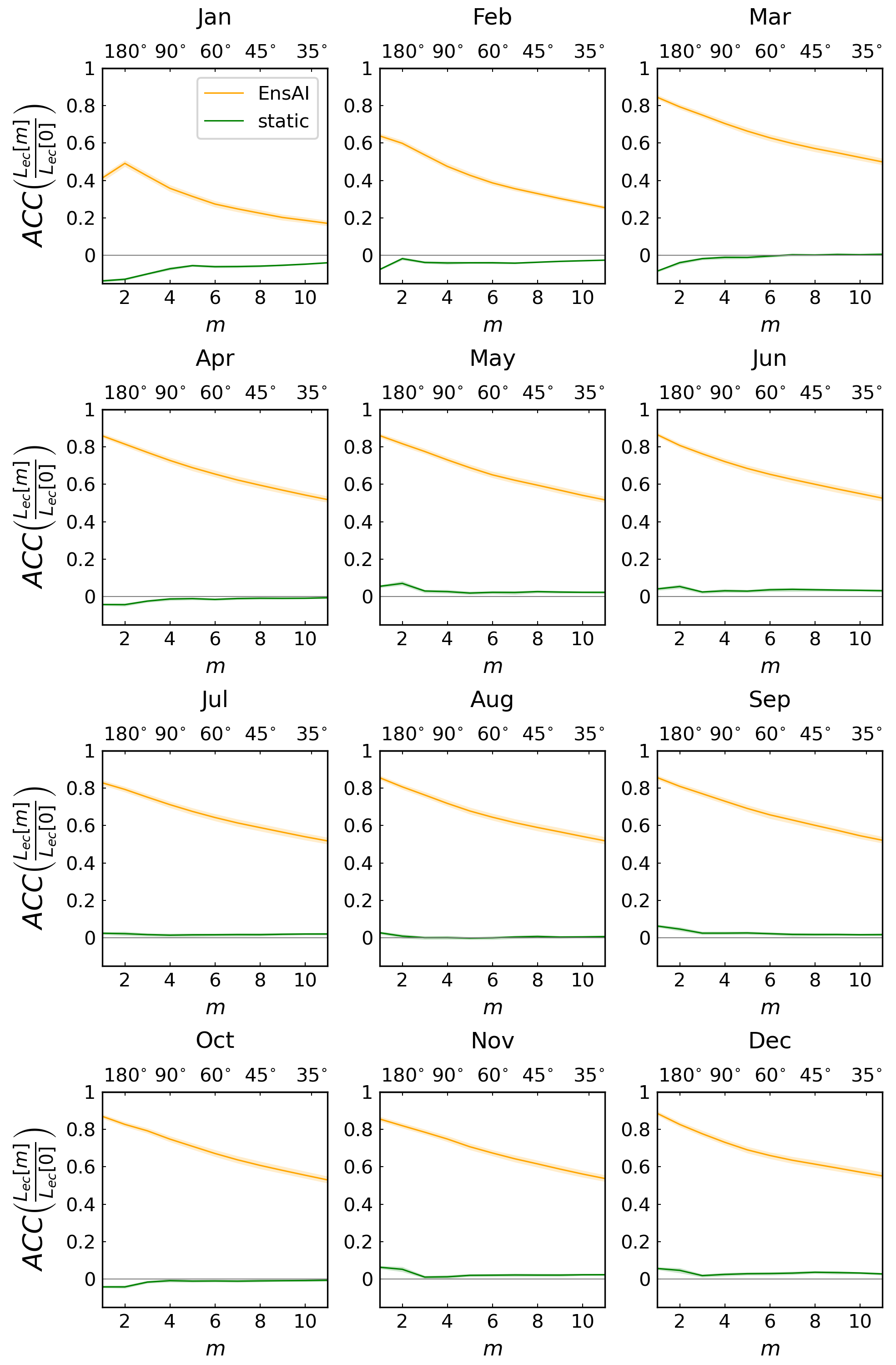}
\end{center}
\caption{$ACC$ values for the horizontal correlation length modes for the emissions/surface concentration correlation $L_{ec}[m]$ as a function of wavenumber $m$ for EnsAI (orange) and the static covariance (green). The corresponding wavelength of each mode is shown on the upper axis.}
\label{fig:lcftr_higher_modes_acc}
\end{figure}

\begin{figure}
\begin{center}
\noindent\includegraphics[scale=0.75]{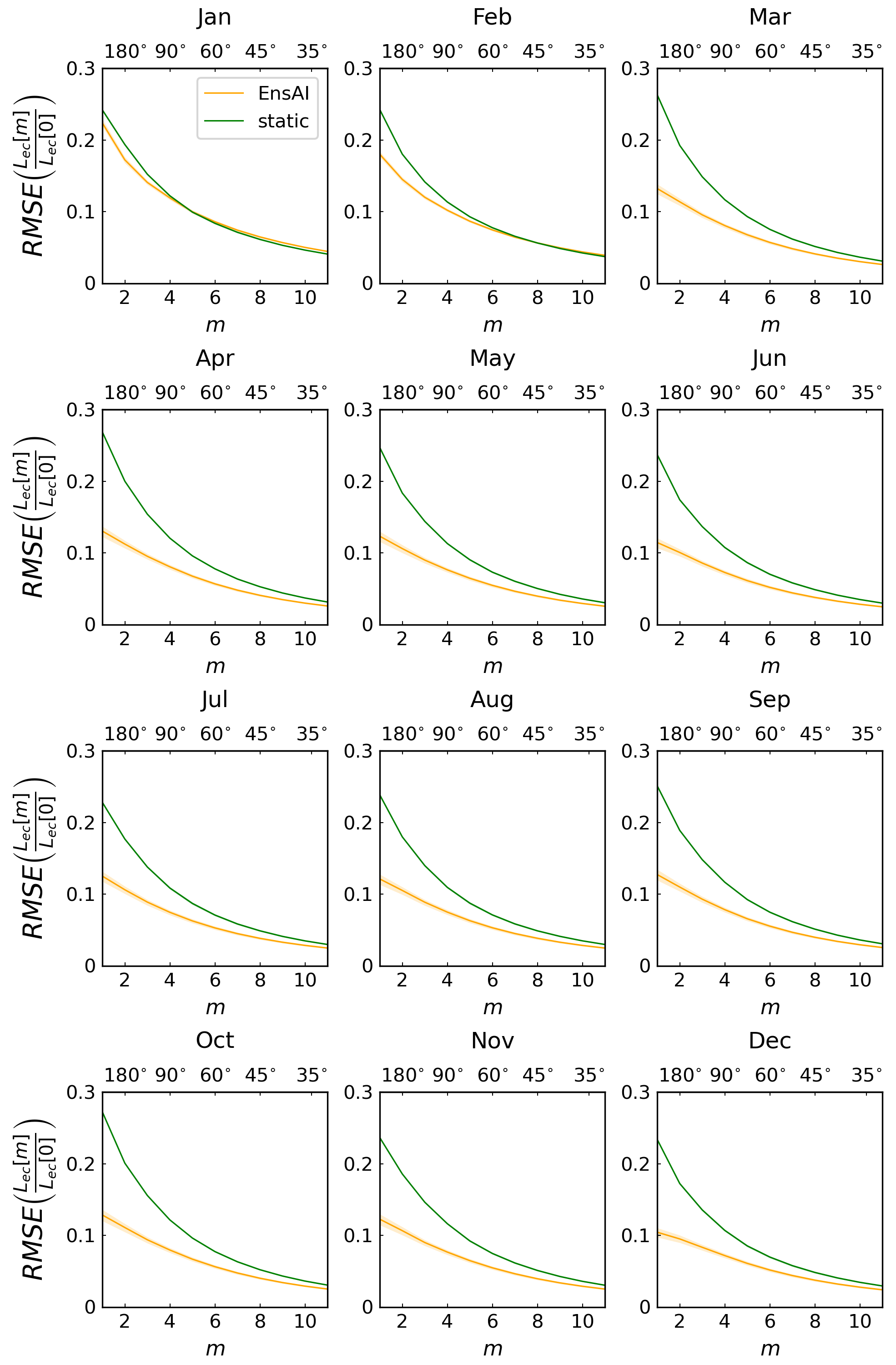}
\end{center}
\caption{Same as Figure \ref{fig:lcftr_higher_modes_acc} but for the $RMSE$.}
\label{fig:lcftr_higher_modes_rmse}
\end{figure}

\begin{figure}
\begin{center}
\noindent\includegraphics[width=0.7\textwidth]{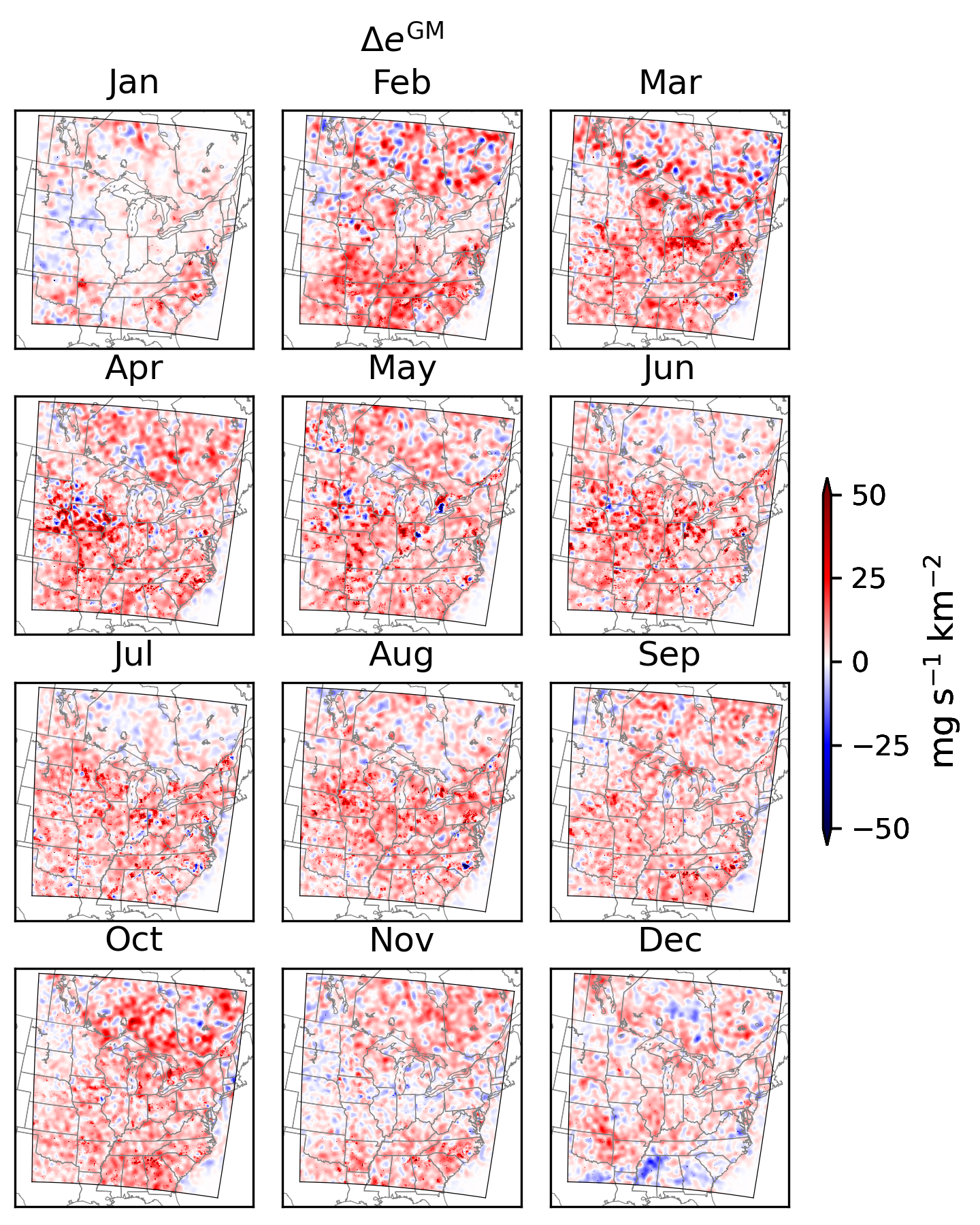}
\end{center}
\caption{Ammonia emissions increments $\Delta\textbf{e}^{\rm{GM}}$ from inversions using the GEM-MACH ensemble to form the background error covariances.}
\end{figure}

\begin{figure}
\begin{center}
\noindent\includegraphics[width=0.7\textwidth]{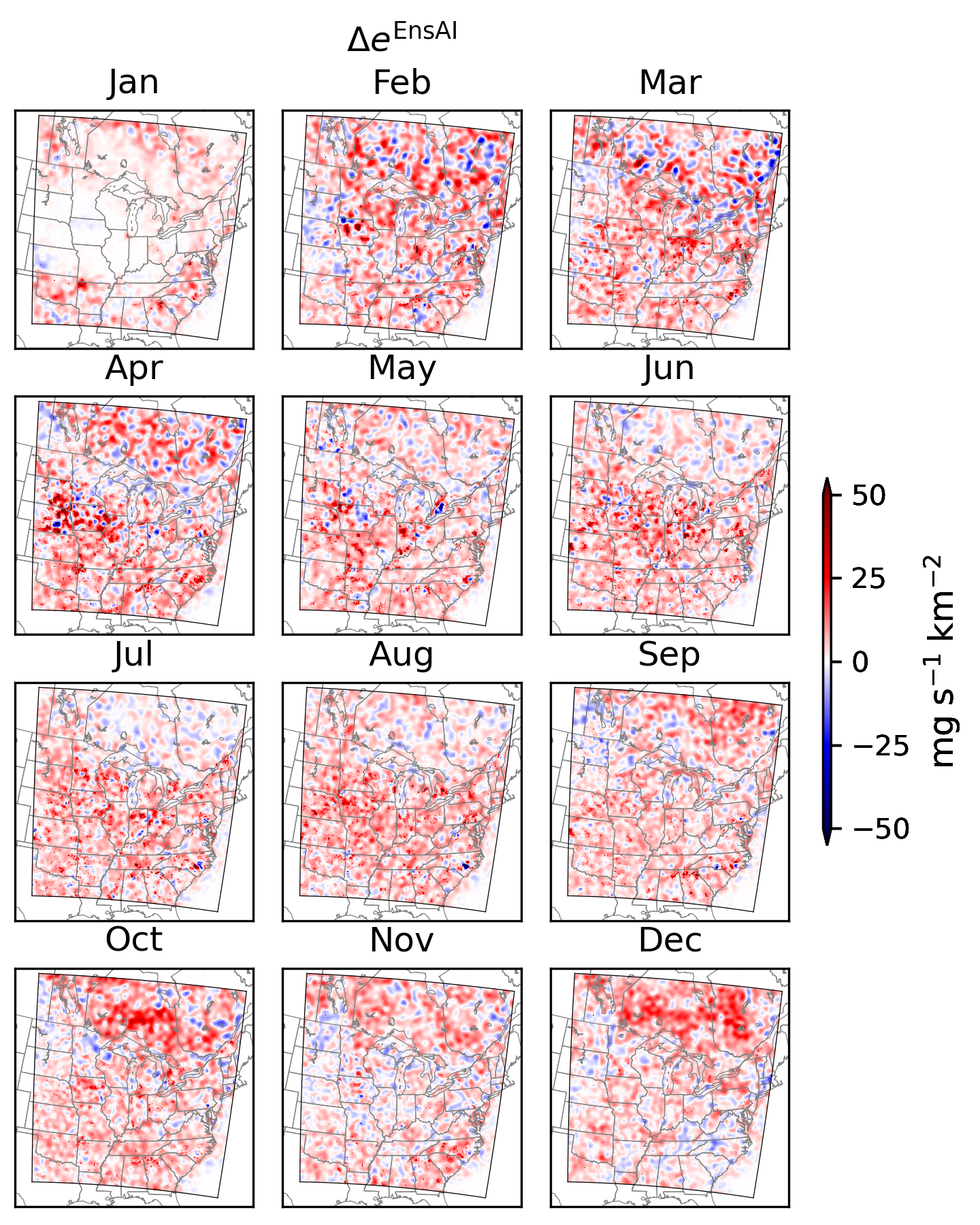}
\end{center}
\caption{Ammonia emissions increments $\Delta\textbf{e}^{\rm{EnsAI}}$ from inversions using the EnsAI ensemble to form the background error covariances.}
\end{figure}

\begin{figure}
\begin{center}
\noindent\includegraphics[width=0.7\textwidth]{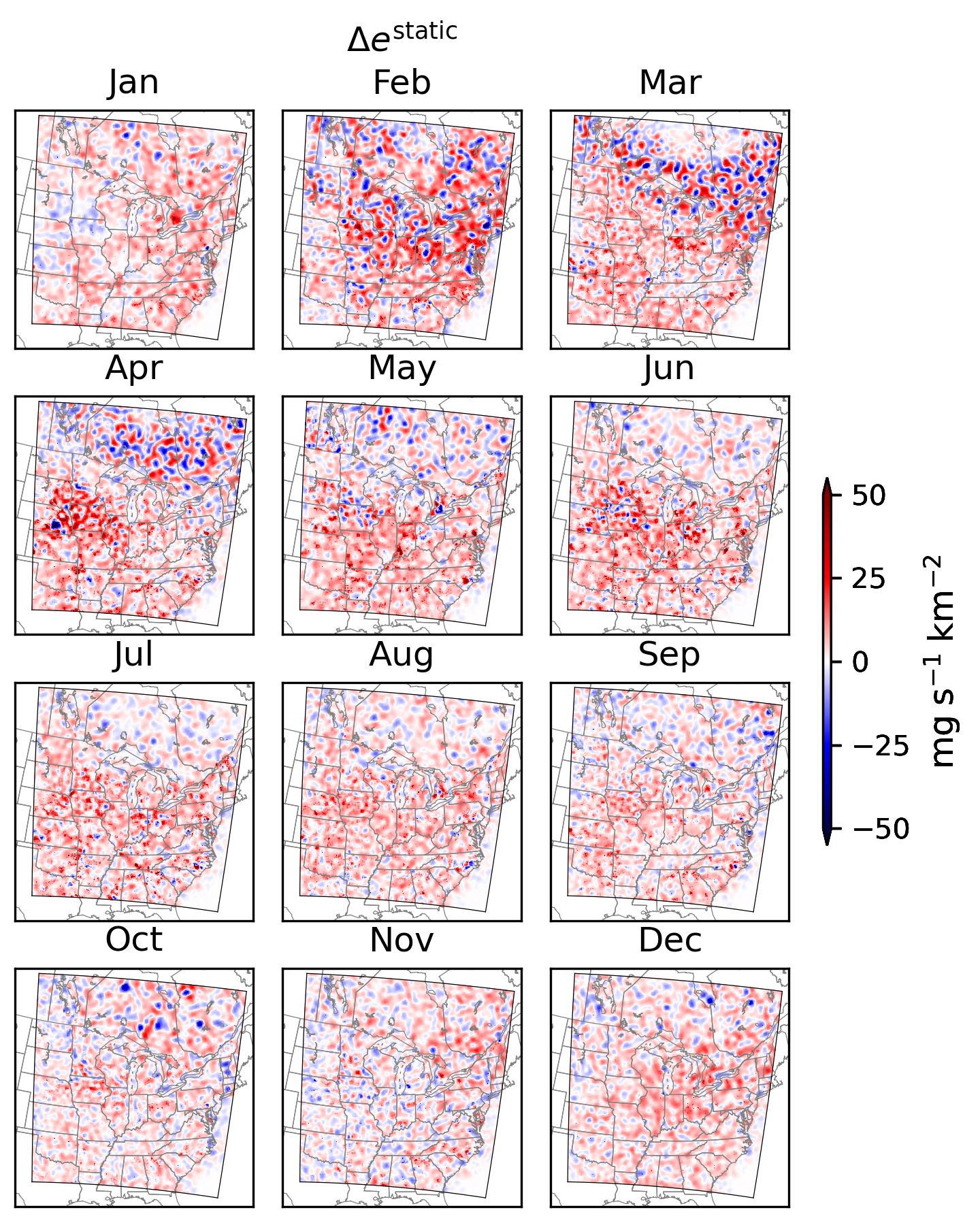}
\end{center}
\caption{Ammonia emissions increments $\Delta\textbf{e}^{\rm{static}}$ from inversions using the static covariances as the background error covariances.}
\end{figure}

\begin{figure}
\begin{center}
\noindent\includegraphics[width=0.7\textwidth]{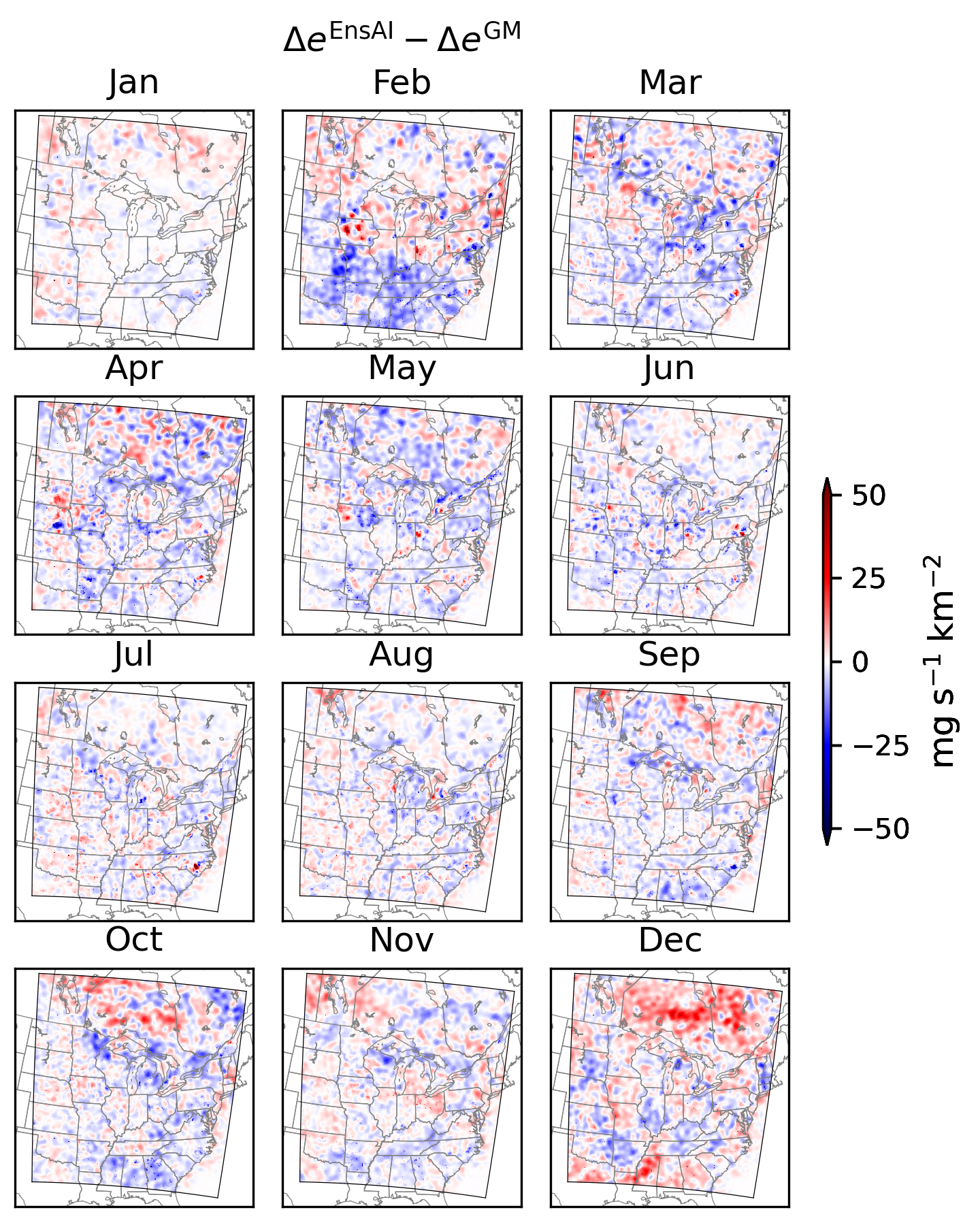}
\end{center}
\caption{Differences between the ammonia emissions increments from inversions using the EnsAI ensemble ($\Delta\textbf{e}^{\rm{EnsAI}}$) and using the GEM-MACH ensemble ($\Delta\textbf{e}^{\rm{GM}}$).}
\end{figure}

\begin{figure}
\begin{center}
\noindent\includegraphics[width=0.7\textwidth]{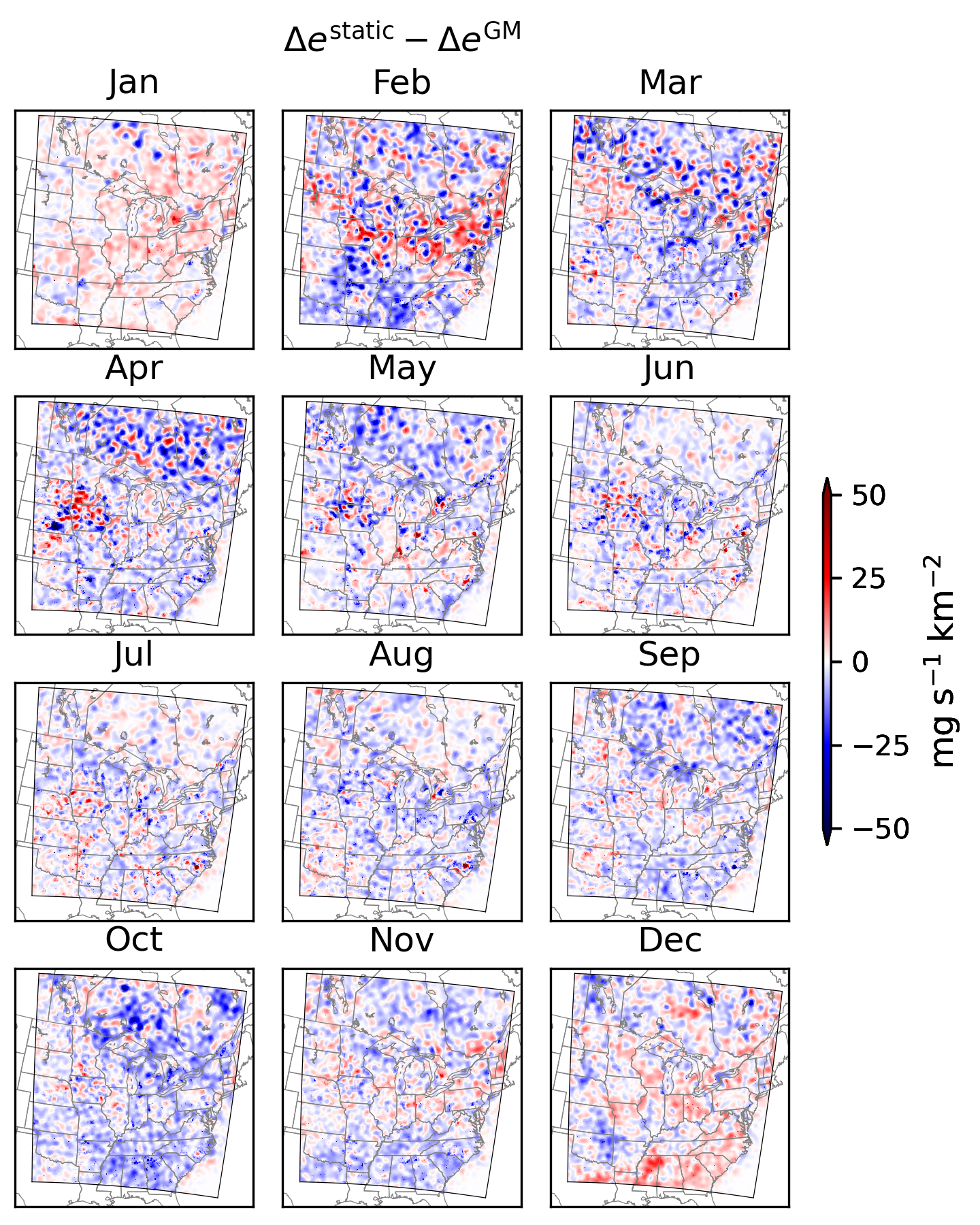}
\end{center}
\caption{Differences between the ammonia emissions increments from inversions using the static covariances ($\Delta\textbf{e}^{\rm{static}}$) and using the GEM-MACH ensemble ($\Delta\textbf{e}^{\rm{GM}}$).}
\end{figure}

\begin{figure}
\begin{center}
\noindent\includegraphics[width=0.7\textwidth]{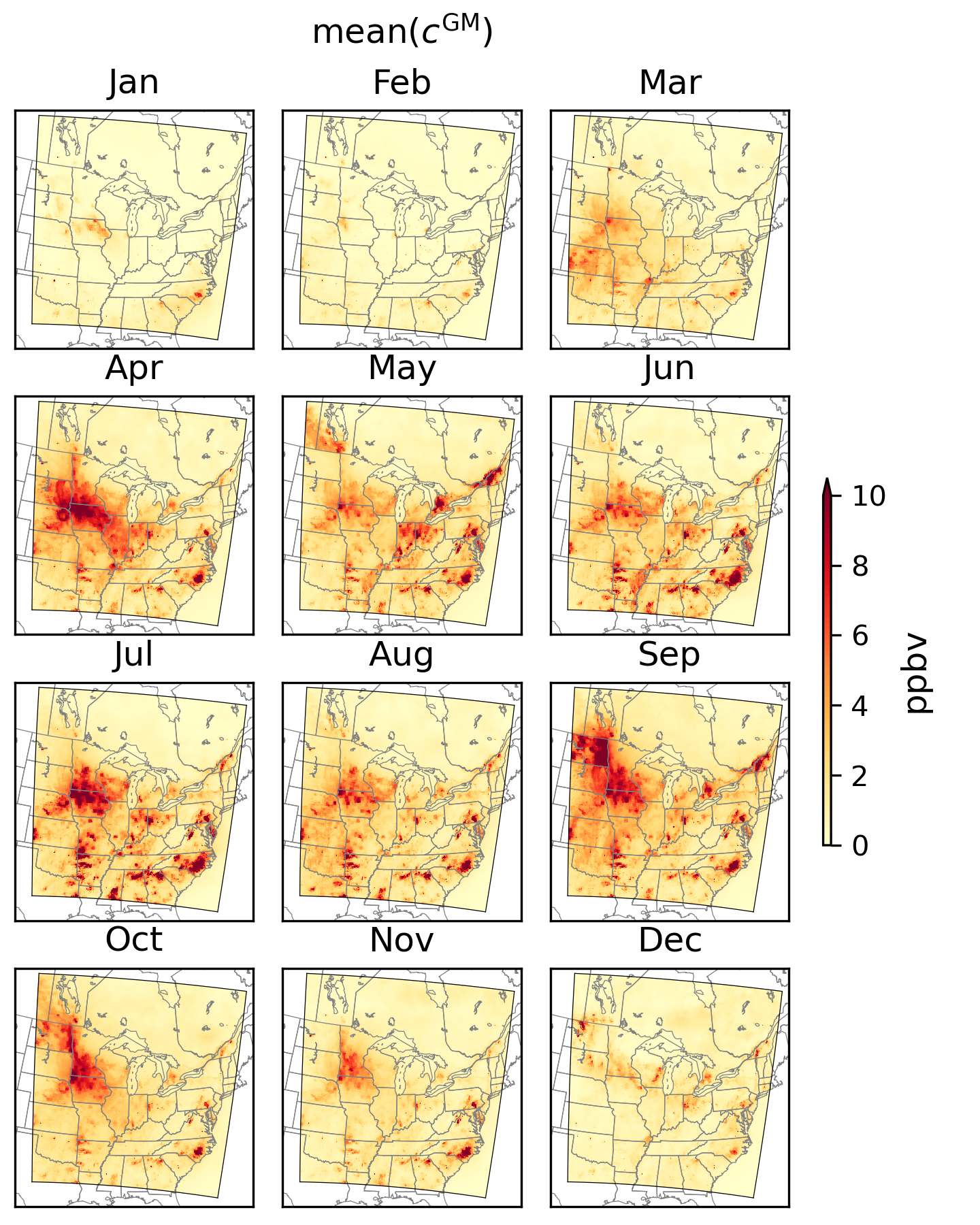}
\end{center}
\caption{Mean of the posterior ammonia surface concentrations from inversions using the GEM-MACH ensemble.}
\end{figure}

\begin{figure}
\begin{center}
\noindent\includegraphics[width=0.7\textwidth]{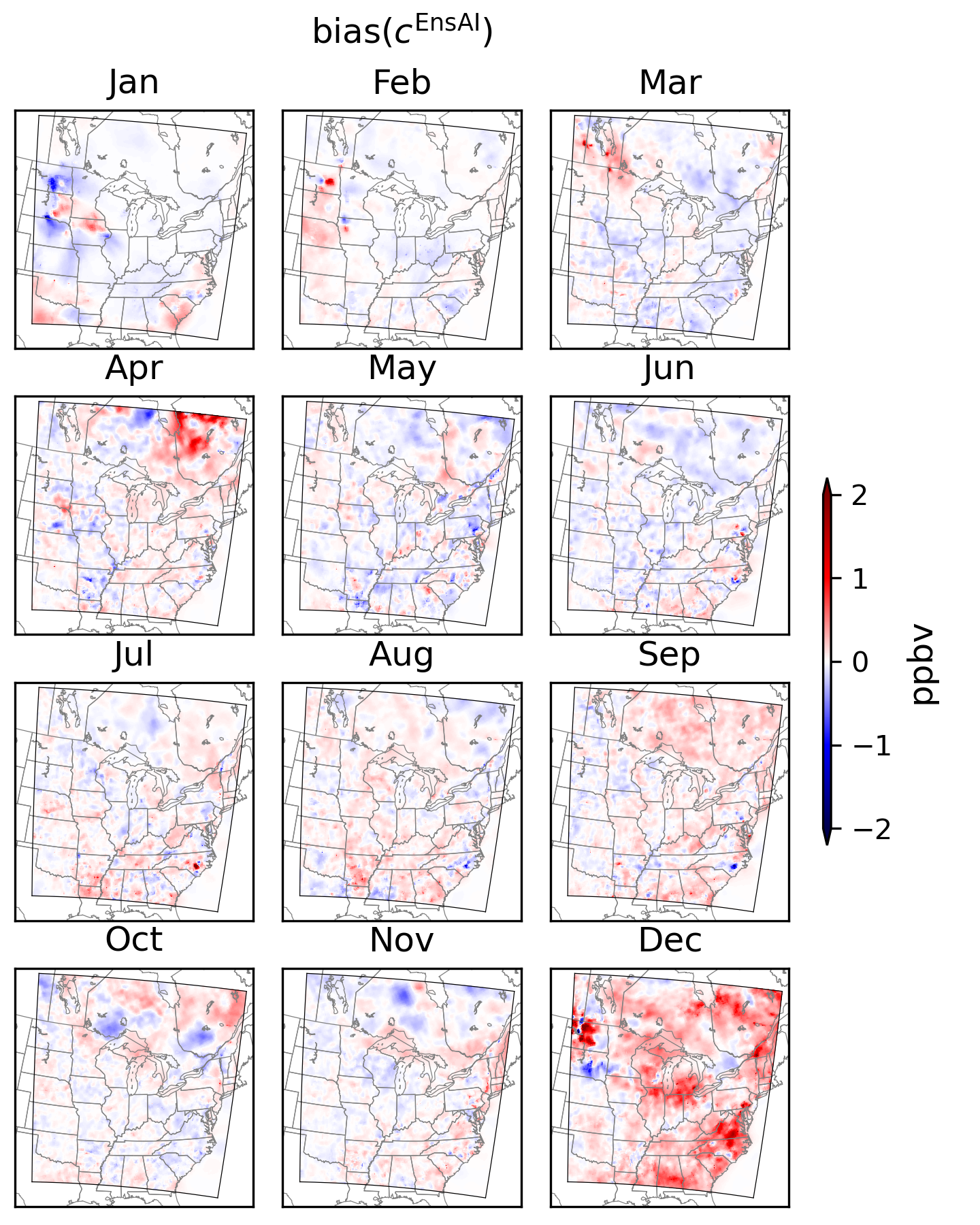}
\end{center}
\caption{Mean differences between the posterior ammonia surface concentrations from inversions using the EnsAI ensemble and using the GEM-MACH ensemble.}
\end{figure}

\begin{figure}
\begin{center}
\noindent\includegraphics[width=0.7\textwidth]{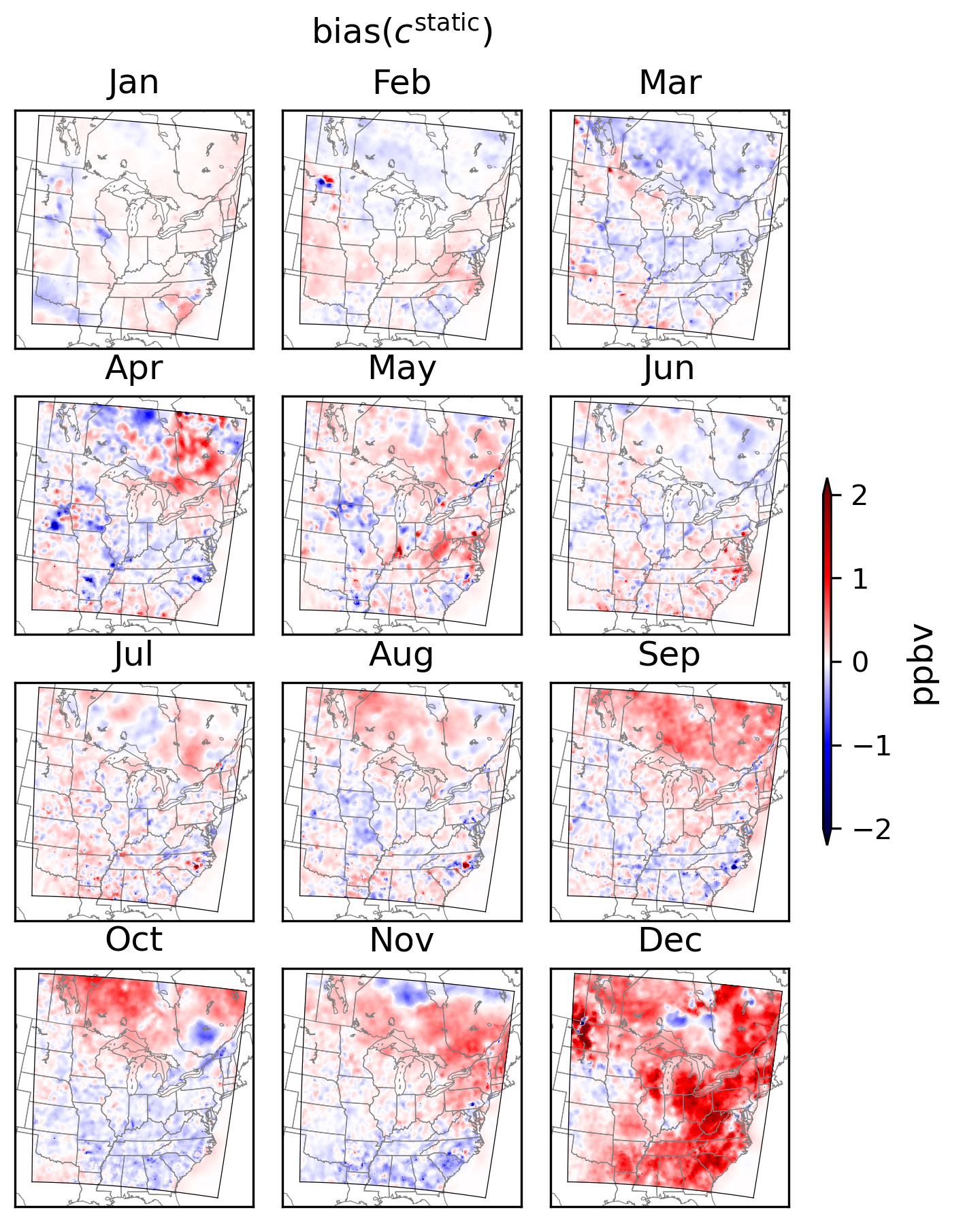}
\end{center}
\caption{Mean differences between the posterior ammonia surface concentrations from inversions using the static covariances and using the GEM-MACH ensemble.}
\end{figure}

\begin{figure}
\begin{center}
\noindent\includegraphics[width=0.7\textwidth]{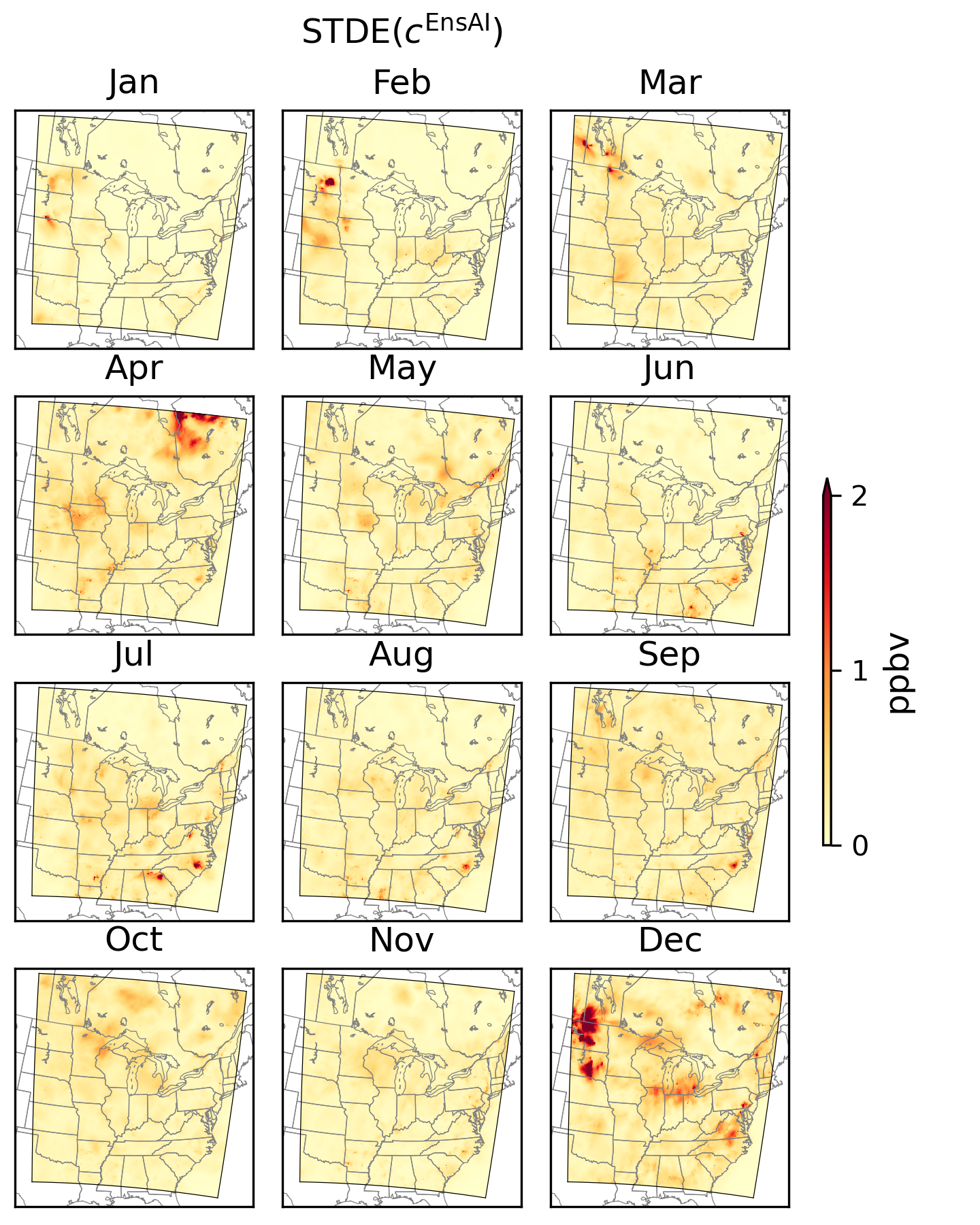}
\end{center}
\caption{Standard deviation of the differences between the posterior ammonia surface concentrations from inversions using the EnsAI ensemble and using the GEM-MACH ensemble.}
\end{figure}

\begin{figure}
\begin{center}
\noindent\includegraphics[width=0.7\textwidth]{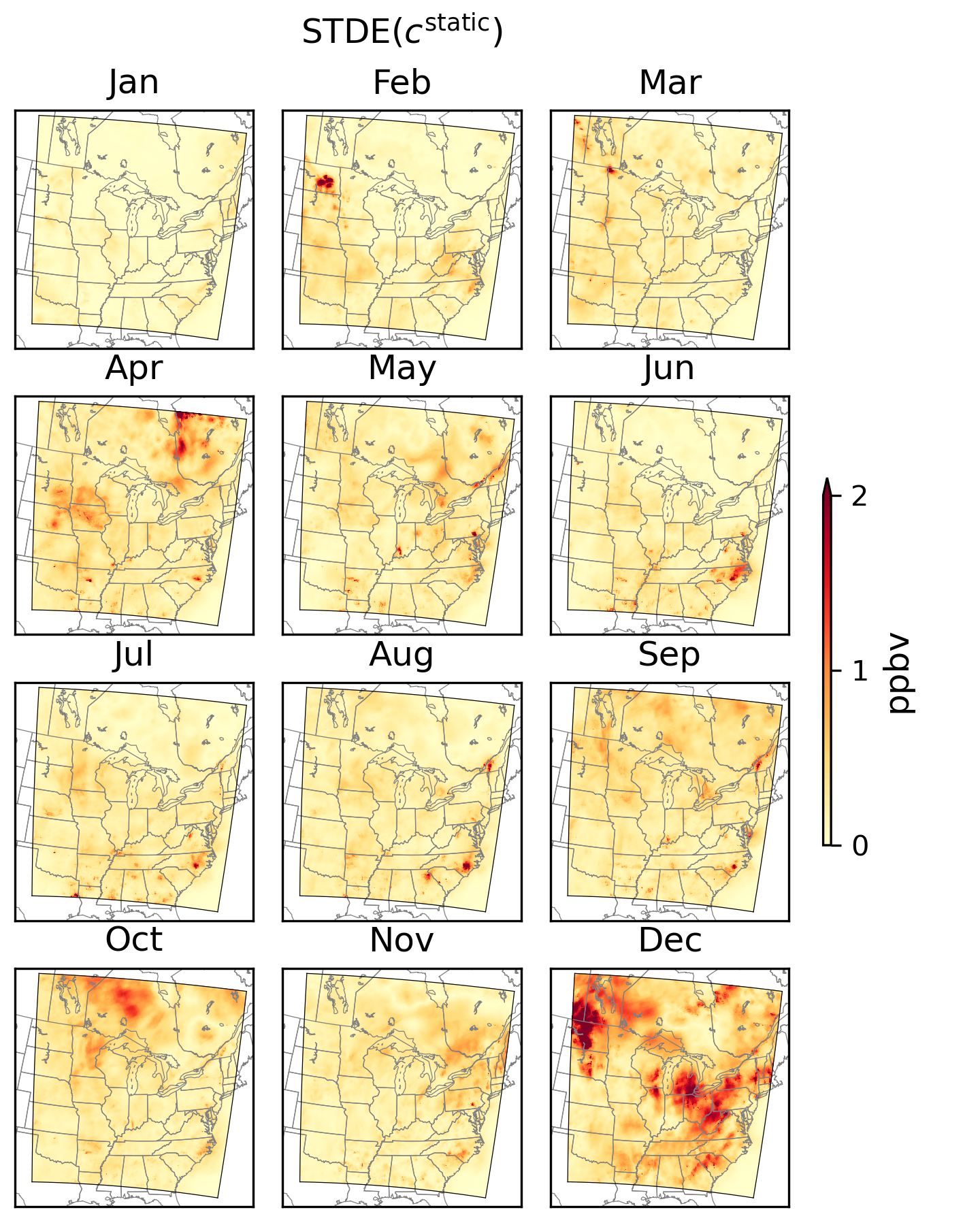}
\end{center}
\caption{Standard deviation of the differences between the posterior ammonia surface concentrations from inversions using the static covariances and using the GEM-MACH ensemble.}
\end{figure}

\begin{figure}
\begin{center}
\noindent\includegraphics[width=0.7\textwidth]{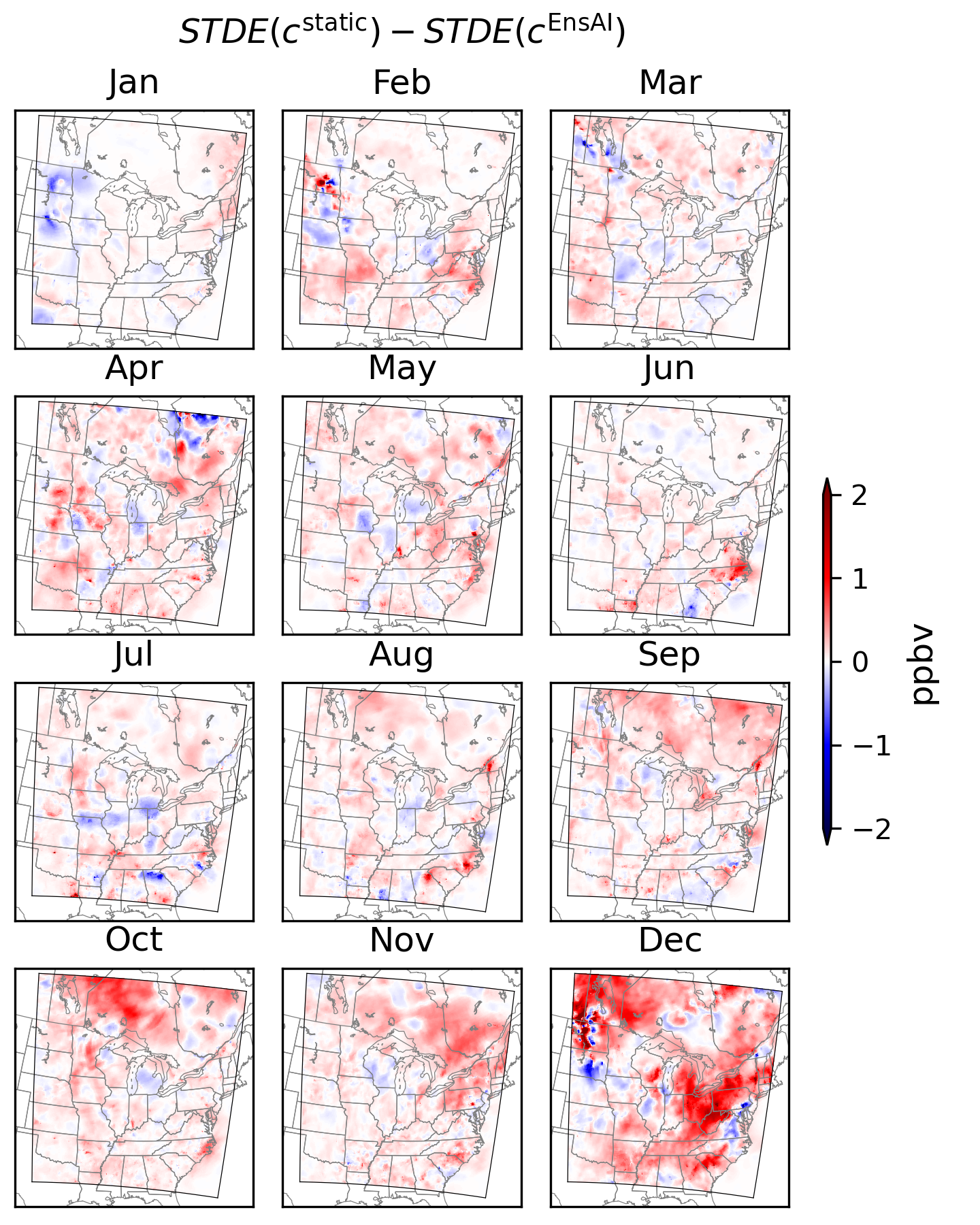}
\end{center}
\caption{Differences in the $STDE$ of the posterior ammonia surface concentrations between the EnsAI inversions ($STDE(\textbf{c}^{\rm{EnsAI}})$) and the static covariance inversions ($STDE(\textbf{c}^{\rm{static}})$).}
\end{figure}

\begin{figure}
\begin{center}
\noindent\includegraphics[width=0.7\textwidth]{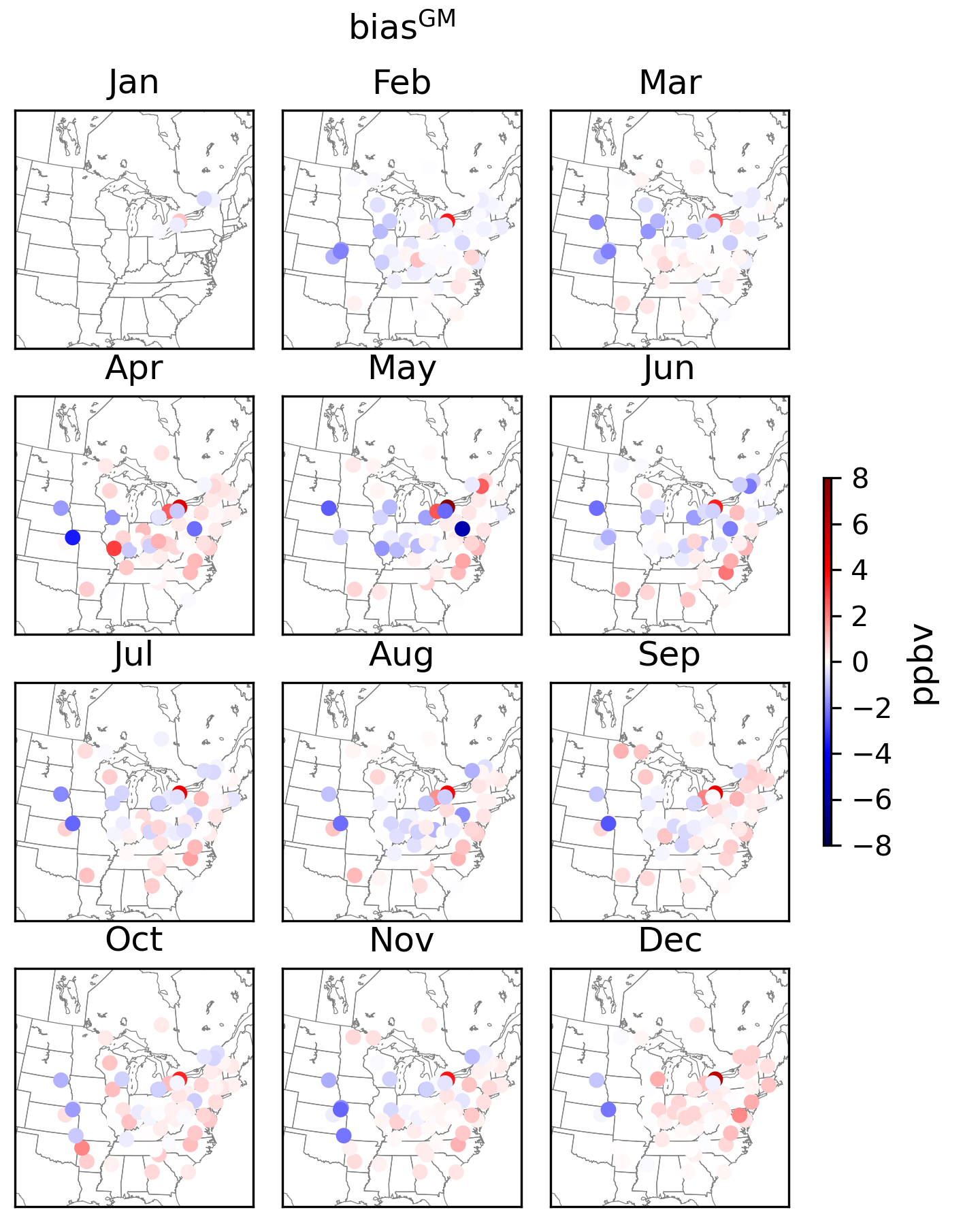}
\end{center}
\caption{Biases of the posterior ammonia surface concentrations from inversions using the GEM-MACH ensemble ($\rm{bias}^{\rm{GM}}$) in comparison to AMoN and NAPS observations for 2015.}
\end{figure}

\begin{figure}
\begin{center}
\noindent\includegraphics[width=0.7\textwidth]{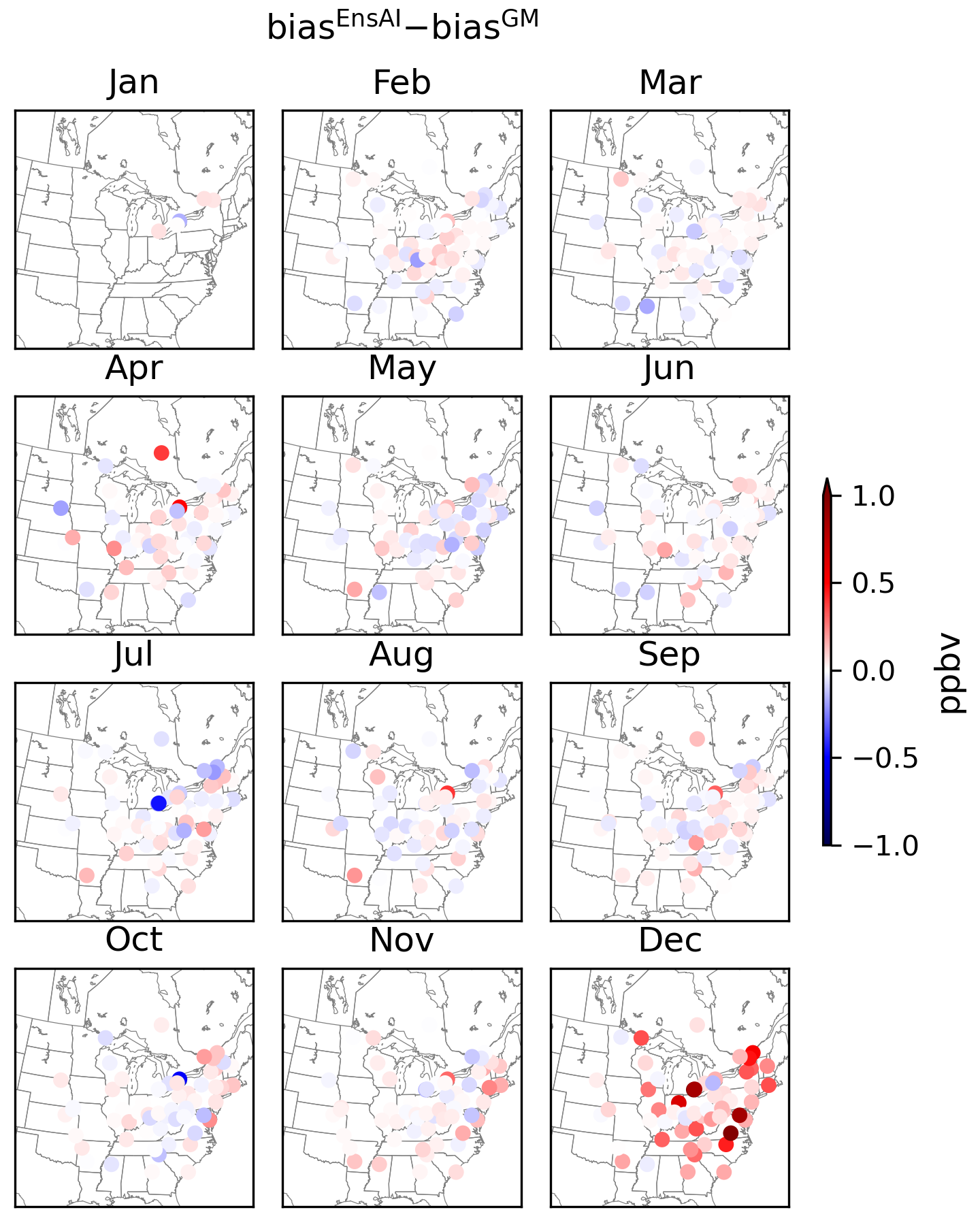}
\end{center}
\caption{Differences in biases of the posterior ammonia surface concentrations between the EnsAI ($\rm{bias}^{\rm{EnsAI}}$) and GEM-MACH ensemble ($\rm{bias}^{\rm{GM}}$) inversions for AMoN and NAPS observations for 2015.}
\end{figure}

\begin{figure}
\begin{center}
\noindent\includegraphics[width=0.7\textwidth]{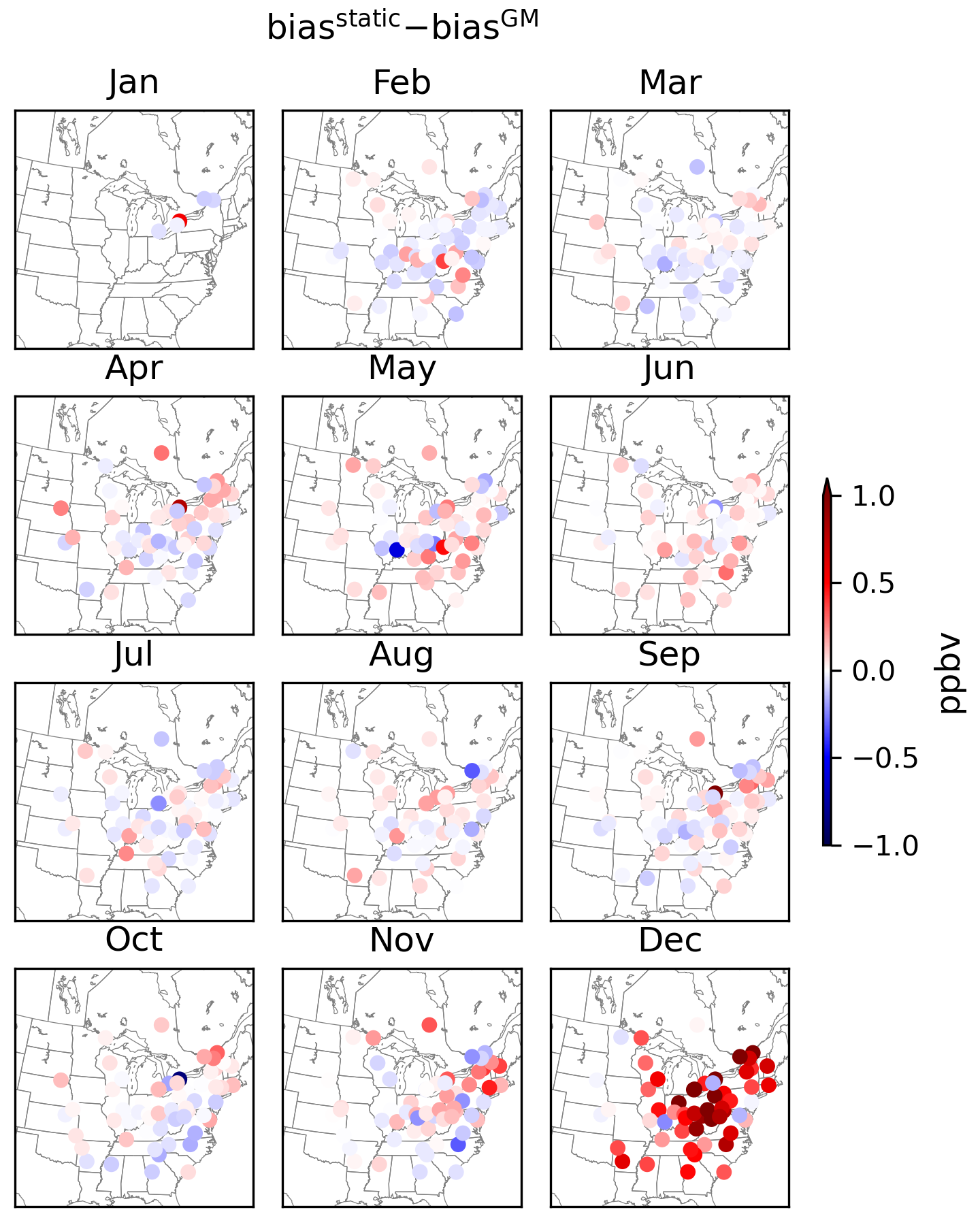}
\end{center}
\caption{Differences in biases of the posterior ammonia surface concentrations between the static covariances ($\rm{bias}^{\rm{static}}$) and GEM-MACH ensemble ($\rm{bias}^{\rm{GM}}$) inversions for AMoN and NAPS observations for 2015.}
\end{figure}

\begin{figure}
\begin{center}
\noindent\includegraphics[width=0.7\textwidth]{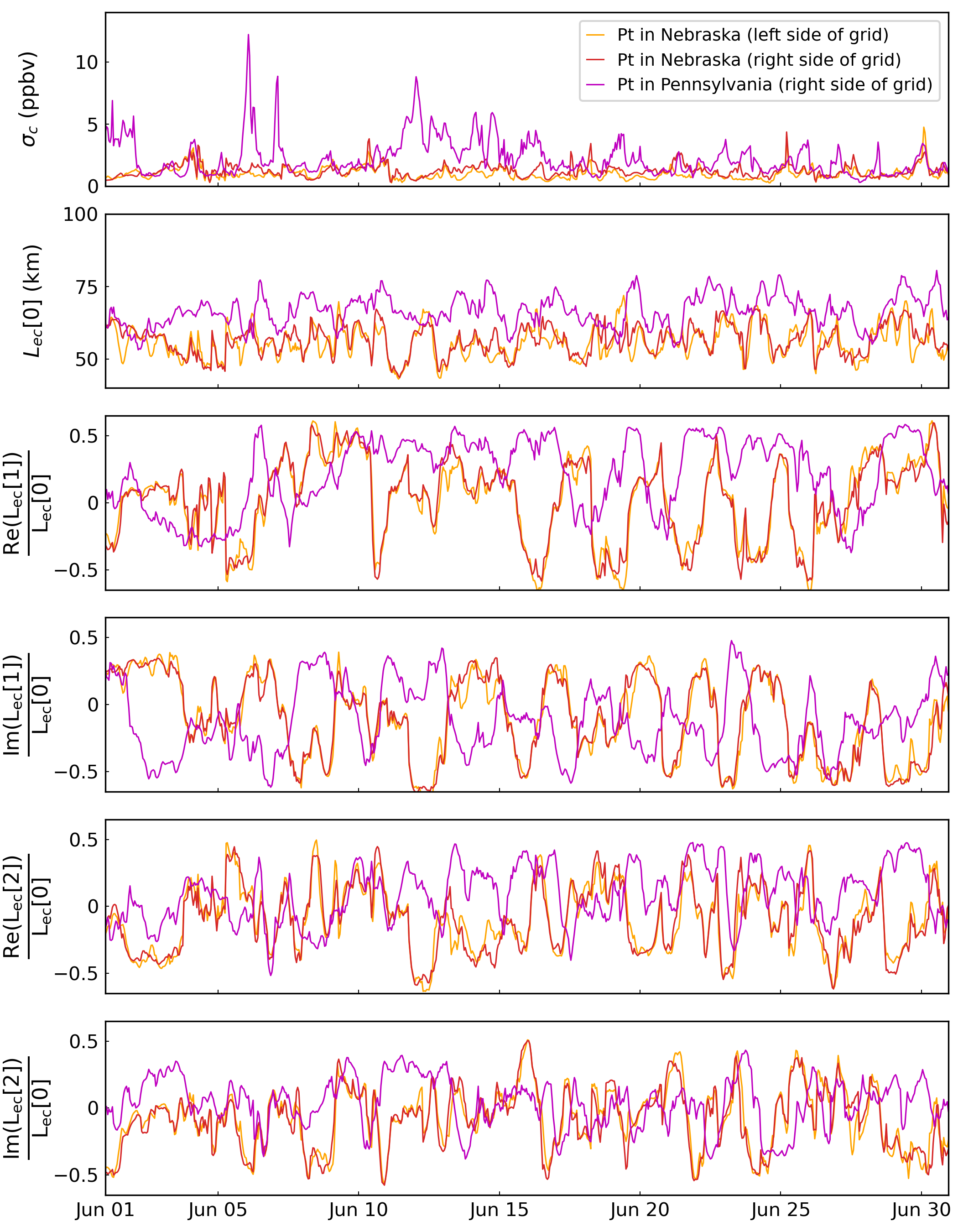}
\end{center}
\caption{Time series of components of the error covariances for EnsAI in the test of its translational invariance. The orange curves show the error covariance components of the original EnsAI run at a location in Nebraska, which is on the left side of EnsAI's grid. The red curves show the error covariance components for the location in Nebraska when the 90 leftmost values of the input fields are translated to the right side of the grid. In this translation, the location in Nebraska gets translated to a grid point in Pennsylvania. The curves in purple show the covariance components of the location in Pennsylvania in the original EnsAI run.}
\end{figure}

\begin{table}[ht]
\begin{center}
\begin{tabular}{lllll}
\hline
\hline
month & background covariance & bias (\%) & $STDE$ (\%) & $r$ \\
\hline
\hline
Jan & GEM-MACH &       -3.9 $\pm$ 18.0       &         43 $\pm$ 15         &       0.46 $\pm$ 0.44       \\
 & EnsAI &       -8.8 $\pm$ 15.7       &         40 $\pm$ 13         &       0.45 $\pm$ 0.49       \\
 & static &        8.6 $\pm$ 26.0       &         59 $\pm$ 22         &       0.44 $\pm$ 0.46       \\
\hline
Feb & GEM-MACH &        8.4 $\pm$ 6.9        &         87 $\pm$ 14         &       0.18 $\pm$ 0.06       \\
 & EnsAI &        7.9 $\pm$ 7.1        &         89 $\pm$ 15         &       0.18 $\pm$ 0.06       \\
 & static &       11.9 $\pm$ 7.0        &         87 $\pm$ 14         &       0.18 $\pm$ 0.06       \\
\hline
Mar & GEM-MACH &       -0.1 $\pm$ 4.9        &         70 $\pm$ 10         &       0.59 $\pm$ 0.05       \\
 & EnsAI &       -0.9 $\pm$ 4.8        &         70 $\pm$ 10         &       0.60 $\pm$ 0.05       \\
 & static &       -2.8 $\pm$ 4.8        &         69 $\pm$ 10         &       0.60 $\pm$ 0.06       \\
\hline
Apr & GEM-MACH &       32.0 $\pm$ 14.0       &        164 $\pm$ 22         &       0.47 $\pm$ 0.07       \\
 & EnsAI &       36.8 $\pm$ 14.8       &        174 $\pm$ 25         &       0.46 $\pm$ 0.08       \\
 & static &       38.4 $\pm$ 15.3       &        180 $\pm$ 27         &       0.44 $\pm$ 0.08       \\
\hline
May & GEM-MACH &       25.6 $\pm$ 15.4       &        182 $\pm$ 36         &       0.59 $\pm$ 0.09       \\
 & EnsAI &       23.7 $\pm$ 15.4       &        182 $\pm$ 37         &       0.59 $\pm$ 0.09       \\
 & static &       29.3 $\pm$ 15.4       &        183 $\pm$ 38         &       0.59 $\pm$ 0.09       \\
\hline
Jun & GEM-MACH &       -5.8 $\pm$ 9.1        &        105 $\pm$ 17         &       0.42 $\pm$ 0.07       \\
 & EnsAI &       -5.3 $\pm$ 9.3        &        106 $\pm$ 17         &       0.40 $\pm$ 0.07       \\
 & static &       -6.3 $\pm$ 9.2        &        106 $\pm$ 17         &       0.39 $\pm$ 0.07       \\
\hline
Jul & GEM-MACH &       21.1 $\pm$ 9.8        &        122 $\pm$ 20         &       0.56 $\pm$ 0.05       \\
 & EnsAI &       24.2 $\pm$ 9.7        &        120 $\pm$ 20         &       0.56 $\pm$ 0.05       \\
 & static &       24.7 $\pm$ 9.9        &        123 $\pm$ 21         &       0.56 $\pm$ 0.05       \\
\hline
Aug & GEM-MACH &       15.1 $\pm$ 9.5        &        116 $\pm$ 18         &       0.53 $\pm$ 0.07       \\
 & EnsAI &       18.5 $\pm$ 10.1       &        122 $\pm$ 18         &       0.52 $\pm$ 0.07       \\
 & static &       17.4 $\pm$ 9.9        &        120 $\pm$ 17         &       0.51 $\pm$ 0.07       \\
\hline
Sep & GEM-MACH &       27.0 $\pm$ 6.9        &        102 $\pm$ 15         &       0.60 $\pm$ 0.07       \\
 & EnsAI &       30.7 $\pm$ 7.1        &        104 $\pm$ 16         &       0.59 $\pm$ 0.07       \\
 & static &       32.9 $\pm$ 7.7        &        112 $\pm$ 18         &       0.58 $\pm$ 0.07       \\
\hline
Oct & GEM-MACH &       23.5 $\pm$ 7.9        &        100 $\pm$ 17         &       0.64 $\pm$ 0.06       \\
 & EnsAI &       19.9 $\pm$ 7.3        &         92 $\pm$ 15         &       0.64 $\pm$ 0.06       \\
 & static &       15.0 $\pm$ 6.9        &         86 $\pm$ 13         &       0.63 $\pm$ 0.06       \\
\hline
Nov & GEM-MACH &        9.9 $\pm$ 8.6        &        109 $\pm$ 20         &       0.47 $\pm$ 0.17       \\
 & EnsAI &       13.2 $\pm$ 9.1        &        115 $\pm$ 23         &       0.47 $\pm$ 0.17       \\
 & static &       16.7 $\pm$ 9.1        &        115 $\pm$ 23         &       0.47 $\pm$ 0.17       \\
\hline
Dec & GEM-MACH &       43.0 $\pm$ 11.7       &        138 $\pm$ 46         &       0.35 $\pm$ 0.14       \\
 & EnsAI &       62.0 $\pm$ 12.4       &        146 $\pm$ 44         &       0.34 $\pm$ 0.14       \\
 & static &       95.8 $\pm$ 14.6       &        170 $\pm$ 50         &       0.33 $\pm$ 0.14       \\
\hline
\hline
\end{tabular}
\caption{Comparison statistics between the posterior ammonia surface concentrations and ammonia surface observations from the AMoN and NAPS networks for 2015.}
\end{center}
\end{table}

\end{document}